\documentclass[11pt,twoside,a4paper,note,final]{cms-tdr}
\def\svnVersion{208949:208954}

\begin{document}

\hyphenation{had-ron-i-za-tion}
\hyphenation{cal-or-i-me-ter}
\hyphenation{de-vices}

\RCS$Revision: 187570 $
\RCS$HeadURL: svn+ssh://svn.cern.ch/reps/tdr2/notes/jolsen_001/trunk/jolsen_001.tex $
\RCS$Id: jolsen_001.tex 187570 2013-05-27 12:34:46Z alverson $
\newlength\cmsFigWidth
\ifthenelse{\boolean{cms@external}}{\setlength\cmsFigWidth{0.85\columnwidth}}{\setlength\cmsFigWidth{0.4\textwidth}}
\ifthenelse{\boolean{cms@external}}{\providecommand{\cmsLeft}{top}}{\providecommand{\cmsLeft}{left}}
\ifthenelse{\boolean{cms@external}}{\providecommand{\cmsRight}{bottom}}{\providecommand{\cmsRight}{right}}
\cmsNoteHeader{NOTE-13-002} 
\title{Projected Performance of an Upgraded CMS Detector at the LHC and HL-LHC: Contribution to the Snowmass Process}

\author{CMS Collaboration}

\date{\today}

\abstract{
   The physics reach of the CMS detector achievable with $300(0)\fbinv$ of proton-proton collisions
   recorded at $\sqrt{s}=14\TeV$ is presented.  Ultimate precision on measurements of Higgs boson
   properties, top quark physics, and electroweak processes are discussed, as well as the discovery
   potential for new particles beyond the standard model.  In addition, the potential for future heavy
   ion physics is presented.  This document has been submitted as a white paper to the Snowmass process,
   an exercise initiated by the American Physical Society's Division of Particles and Fields to assess
   the long-term physics aspirations of the US high energy physics community.
}

\hypersetup{%
pdfauthor={CMS Collaboration},%
pdftitle={CMS Physics Reach at High Luminosity: Snowmass report},%
pdfsubject={CMS},%
pdfkeywords={CMS, physics, future}}

\maketitle 
\section{Introduction}\label{intro}

The Division of Particles and Fields of the American Physical Society has
initiated a Snowmass process~\cite{snowmass} to assess the long-term physics 
aspirations of the US high energy physics community.  This exercise has been 
organized into several ``frontiers,'' of which the primary one relevant to 
the Compact Muon Solenoid (CMS) collaboration is the Energy Frontier.  The 
CMS Collaboration also performs measurements that are relevant to the Cosmic 
and Intensity frontiers, and there are frontiers devoted to forefront advances 
in Instrumentation and Computing that impact directly on the future capabilities 
of the detector.  The high energy physics landscape is studied with respect to 
future capabilities provided by both accelerator-based facilities and detector 
facilities distinct from accelerators, which have been assessed by the Frontier 
Capabilities working group.  Of the future accelerator options currently under 
study, the Large Hadron Collider (LHC) is the only facility currently operating.  
In this document we summarize the physics potential of the upgraded CMS detector 
operating during the future LHC running planned over the next two decades.

\begin{figure}[tbp]
  \begin{center}
    \includegraphics[width=0.48\textwidth]{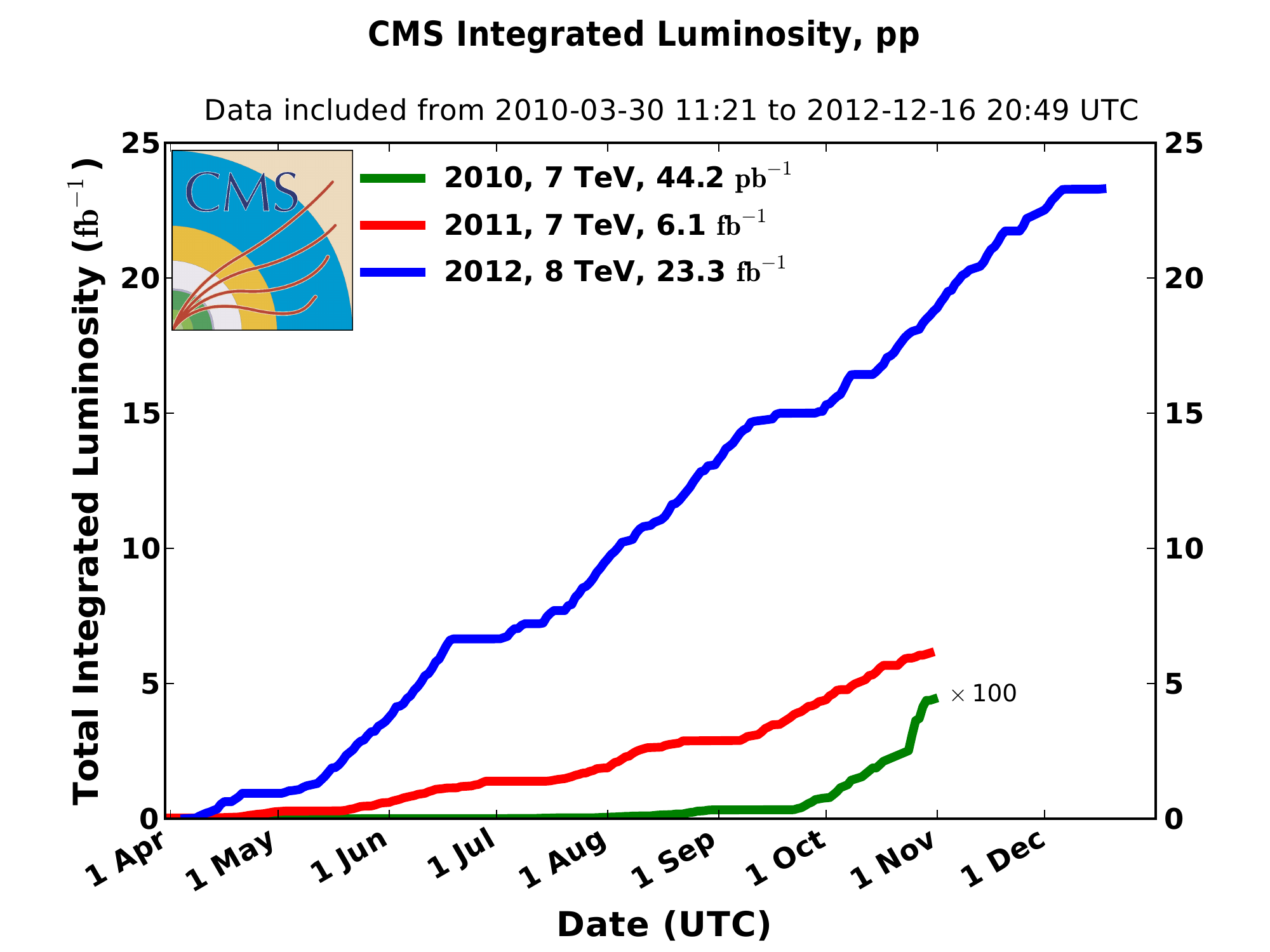}
    \includegraphics[width=0.50\textwidth]{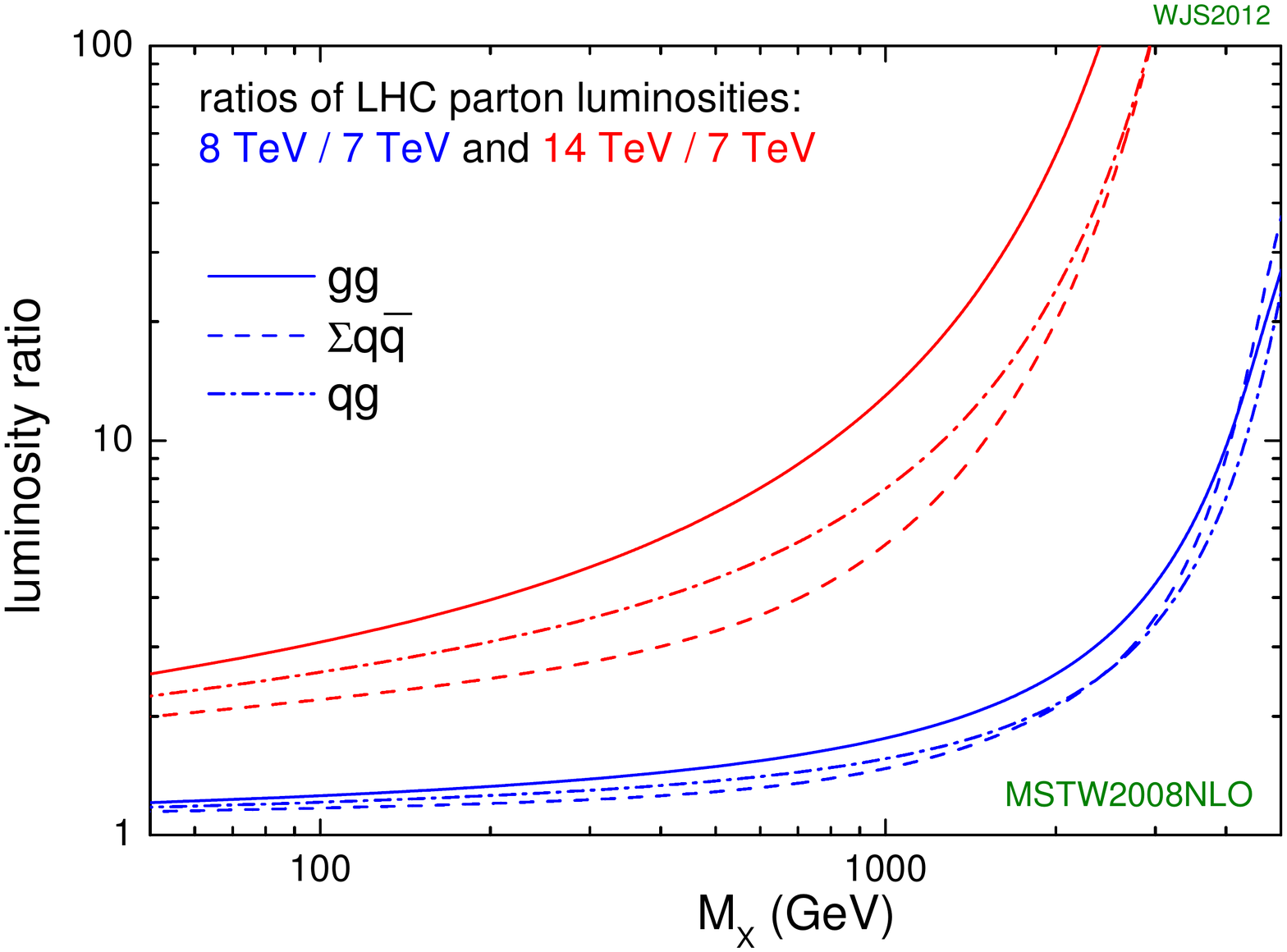}
    \caption{Left: LHC integrated luminosity delivered to CMS during the
    2010 (green), 2011 (red), and 2012 (blue) running periods.  Right:
    ratio of parton luminosities at the LHC for center-of-mass energies of 
    $8$ and $14\TeV$ relative to $7\TeV$.  Luminosities are shown separately
    for processes initiated by $gg$, $qg$, and $qq$ collisions~\cite{Sterling}.}
    \label{fig:LHC}
  \end{center}
\end{figure}

The LHC has performed flawlessly since initiating high-energy pp collisions at
$\sqrt{s}=7\TeV$ in early 2010, delivering $30\fbinv$ of data to CMS during the 
past three years (see Fig.~\ref{fig:LHC}).  In 2012 the energy was raised to a 
new record of $4\TeV$ per beam, while the instantaneous luminosity exceeded 
$7\times 10^{33}\,{\rm cm}^{-2}\,{\rm s}^{-1}$ and the average number of 
interactions per pp crossing (pile-up) reached $21$.  The CMS 
detector was able to operate effectively in this high-occupancy environment, 
recording $27\fbinv$ of high-quality pp data with efficiency in excess of $90\%$.  
This data has been used to discover a Higgs boson, to extend the search for particles 
beyond the standard model (SM) to the multi-TeV range, and to make measurements of 
electroweak processes and top-quark properties with a precision exceeding that 
achieved by the Tevatron.  In addition, CMS has collected $150~\mu{\rm b}^{-1}$ 
of lead-lead and $31\nbinv$ of proton-lead collisions that have fundamentally 
expanded our understanding of heavy ion physics.

Currently, the LHC is in the middle of its first long shutdown (LS1) 
in order to prepare for running at $\sqrt{s}=13\TeV$ in 2015, on the way to the 
design energy of $14\TeV$.  The bunch spacing will most likely be reduced to 
$25\,{\rm ns}$, the luminosity will reach the design value
($10^{34}\,{\rm cm}^{-2}\,{\rm s}^{-1}$) with $25$ pile-up interactions, and the 
goal will be to integrate $300\fbinv$ of pp data by the end of 2021.  A second 
long shutdown (LS2) in 2018 will be used to upgrade the detectors for running at 
double the design luminosity and an average pile-up of $50$.  The next phase of planned 
LHC operation, referred to as the High Luminosity LHC (HL-LHC), will begin with the 
third long shutdown (LS3) in the period 2022-2023, where the machine and detectors 
will be upgraded to allow for pp running at a luminosity of 
$5\times 10^{34}\,{\rm cm}^{-2}\,{\rm s}^{-1}$ and an average pile-up of $128$, with
the goal of eventually accumulating $3000\fbinv$.

The increase in LHC beam energy will have a significant impact on the physics reach 
of CMS beyond that gained by accumulating $10$ or $100$ times more data.   In 
addition to the increase in production cross section, a multi-$\TeV$ particle 
produced via gluon fusion will see an increase in the parton luminosity by one or 
two orders of magnitude relative to $7\TeV$ collisions (Fig.~\ref{fig:LHC}).  The 
jump in energy will lead to a doubling of the mass reach for discovery of new 
particles early in the next run, while enabling precision measurements of Higgs 
boson properties and SM processes that will either help to elucidate 
the nature of newly discovered particles or exclude a large set of possible 
alternatives to the standard model.

The purpose of this document is to summarize, in the context of the 
Snowmass process, the future physics potential of the CMS detector at the LHC 
operating with protons and heavy ions at design energy and luminosities up to 
$5\times 10^{34}\,{\rm cm}^{-2}\,{\rm s}^{-1}$.  The methodology used to make
projections is based on the assumption that the planned upgrades of the CMS detector 
will achieve the goal of mitigating the increased radiation damage and complications
arising from higher luminosity and higher pile-up.  With this primary assumption,
existing public results based on current data are extrapolated to higher energy and
luminosities.  In most cases, the analyses are assumed to be unchanged, which is a
conservative assumption given the fact that all analyses will be reoptimized to maximally 
exploit the higher energy and luminosity.  This white paper updates and extends the 
conclusions summarized in the CMS report~\cite{CMS-NOTE-2012-006} submitted to the 
European Strategy Preparatory Group in October, 2012, and is organized as follows.  
Section~2 summarizes the current physics landscape at the Energy Frontier, while Sec.~3 
describes the CMS upgrade plans for LHC Phases 1 and 2.  Section~4 presents the projected 
measurement sensitivity of Higgs boson properties, while Secs.~5 and 6 summarize the 
discovery reach for supersymmetry and exotic resonances, respectively.  Sections 
$7$-$9$ summarize the physics potential for top-quark, electroweak, and heavy-ion 
physics, respectively, and concluding remarks are given in Sec.~10.

\section{LHC Physics Landscape (2013)}\label{landscape}

By the end of the 2010 LHC data-taking period at $7\TeV$, all of the SM 
particles had been rediscovered by both CMS and ATLAS (neutrinos through missing 
energy).  By the end of 2011 the search for the SM Higgs boson had 
excluded a wide range of masses, leaving only a narrow allowed region around 
$125\GeV$ where an indication of a signal had appeared.  Increasingly precise 
measurements of top quark and electroweak processes continued to confirm the standard 
model, and the absence of any signals in the search for new particles beyond the 
standard model (BSM) motivated a new class of simplified supersymmetric (SUSY) models 
to test in the $8\TeV$ data.

In July of 2012 the landscape changed fundamentally when the ATLAS~\cite{Aad:2012tfa} 
and CMS~\cite{Chatrchyan:2012ufa} collaborations announced the discovery of a new 
particle with a mass near $125\GeV$ possessing properties consistent with that of 
the long-sought Higgs boson.  Since that time, both experiments have analyzed the 
full $8\TeV$ dataset, comprising approximately $20\fbinv$ of proton-proton collision 
data, and reported preliminary results for the main boson decay 
channels~\cite{CMS-HIG-13-001,CMS-HIG-13-002,CMS-HIG-13-003,Aad:2013wqa}.  
CMS has also shown preliminary results on the full dataset for the primary fermion 
decay channels~\cite{CMS-HIG-13-004,CMS-HIG-13-012}, where an indication of a signal 
has started to materialize, while ATLAS has presented preliminary results on the
full dataset in the bottom-quark channel~\cite{ATLAS-CONF-2013-079}.  
All of the main production channels accessible at the 
LHC have now been investigated experimentally, including gluon fusion, vector-boson 
fusion, and associated production with vector bosons and top quarks.  Within the 
existing experimental and theoretical precision, all measurements of the coupling of 
the new boson to photons, W and Z bosons, tau leptons, and bottom quarks are consistent 
with the expected couplings of the SM Higgs 
boson~\cite{ATLAS-CONF-2013-034,CMS-HIG-13-005}.  In addition, the hypothesis that the 
new boson is a scalar has been tested against alternative spin-parity hypotheses by 
CMS~\cite{Chatrchyan:2012jja} and ATLAS~\cite{Aad:2013xqa}, with results disfavoring 
all but the SM prediction.  One year after the initial discovery it is 
now clear that this particle is a Higgs boson, but whether it is the single particle 
predicted in the standard model, or only one of many Higgs bosons, remains a central 
open question.  This question will be the focus of intense research in particle 
physics for the foreseeable future.

As of the time of writing, no new particles have been discovered apart from a Higgs 
boson.  Exclusions for gluinos are now extended up to $\sim 1.3\TeV$, while first- and 
second-generation squarks are excluded up to $\sim 0.8\TeV$ assuming an eightfold 
squark-mass degeneracy.  Third-generation squarks are excluded up to $\sim 650\GeV$, 
and elecroweak gauginos are excluded up to $\sim 300\GeV$ when sleptons are decoupled.  
The search for new gauge bosons with SM couplings has yielded lower limits 
of $3\TeV$, while more exotic models such as black holes and string resonances are 
excluded up to $5\TeV$.  The increase in energy of the LHC will have a significant 
effect on the discovery reach for new particles, and the full exploitation of this new 
phase space will be a primary goal of the next run after LS1.

Electroweak processes, including diboson production and W and Z production in 
association with up to four jets, have been measured with increasing precision and 
no deviations from the SM predictions have been found.  Such processes 
represent important reducible and irreducible backgrounds to Higgs boson production 
and will be critical to improve upon in future running to achieve the target precision 
on its properties.  CMS has measured the top-quark mass to a precision of better 
than $1\%$, while the uncertainties on the top-pair and single-top cross sections in 
the t-channel are less than $10\%$.  Searches for new particles produced in association
with, or decaying to, top quarks have so far resulted in only exclusion limits.

Given this current landscape, the physics goals of the future LHC running are clear.  
The properties of the new boson must be measured to the highest achievable precision, 
including the Higgs self-coupling, while additional Higgs bosons and exotic decays must 
be either found or excluded.  At the same time, the existence of a Higgs boson intensifies 
the search for supersymmetric particles, and the upgraded CMS detector must be able to 
simultaneously trigger on electroweak-scale physics while remaining sensitive to broadband 
particle searches in the multi-TeV regime.  The precision of top and electroweak 
measurements must continue to improve, both as a way to reduce the systematic uncertainty 
on Higgs boson measurements and as potential probes of subtle new physics effects.  The 
CMS upgrades are designed to enable this physics by not only mitigating the effects of 
radiation damage and higher luminosity, but by improving the performance of the detector 
in key areas relative to the existing performance at $8\TeV$.

\section{CMS Upgrades}\label{upgrades}

A number of modifications to the LHC experiments are required to deal with
the increased instantaneous luminosity that the LHC will deliver in Run 2
and beyond.  The improvements to the LHC experiments that will be installed
through the period encompassing LS1 and LS2 are referred to as ``Phase 1''
upgrades, while those planned for installation in LS3 are referred to as
``Phase 2.''  This section describes the CMS upgrade program and its goals.

\subsection{Phase 1 Upgrades to the CMS Experiment}

For the CMS experiment, the planned Phase 1 upgrades involve the
innermost tracking detector composed of silicon pixels, the hadron calorimeter,
and the first level of the CMS trigger.  The pixel detector will be replaced
with an improved device that adds a fourth layer to its design. In addition to
improving the overall quality and robustness of track reconstruction, the
upgraded pixel system will provide substantially improved b-tagging capability.
The readout of the hadron calorimeter will be replaced to exploit newly
available silicon phototransducer technology. In addition to addressing
shortcomings of the existing readout system, this will allow for longitudinal
segmentation of the calorimeter, allowing for improved ``particle flow"
reconstruction in high-occupancy events.  The L1 Trigger system will be
upgraded to allow use of the full granularity of the calorimeter. The upgraded
system will also permit the combination of cluster information from the
different muon subsystems directly at the L1 muon track reconstruction level,
as well as improved capability to exploit the intrinsic cluster position resolution.

Planning for each of these upgrades is at an advanced stage, with the technical description and expected performance of these upgraded detectors documented
in detail in Technical Design Reports approved by the LHCC~\cite{CMS:2012sda, CMS:2012tda, CMS:2013l1tdr}.  Salient aspects of these details with respect
to the CMS physics program are summarized in the following sections, wherein the vital necessity of each component of the Phase 1 upgrades is demonstrated.

\subsubsection{Pixel Detector Upgrade}

The CMS pixel detector has efficiently recorded data since the first LHC collisions in 2009. It provides high track reconstruction efficiency and precise measurement of track origin, as needed for online event selection and offline physics analyses. CMS analyses are based on the particle flow technique to provide a globally optimized reconstruction of physics objects (isolated leptons and photons, jets and missing transverse energy). With the 2012 LHC dataset, it has been observed that the pixel performance is essential to mitigate the effect of numerous interactions in the same bunch crossing to maintain efficient identification of all physics objects.  This performance, however, is not maintainable with the current pixel detector when LHC beam conditions significantly exceed their original design specifications and it must be replaced with a new device.

\begin{figure}[th]
  \begin{center}
    \includegraphics[width=0.57\textwidth]{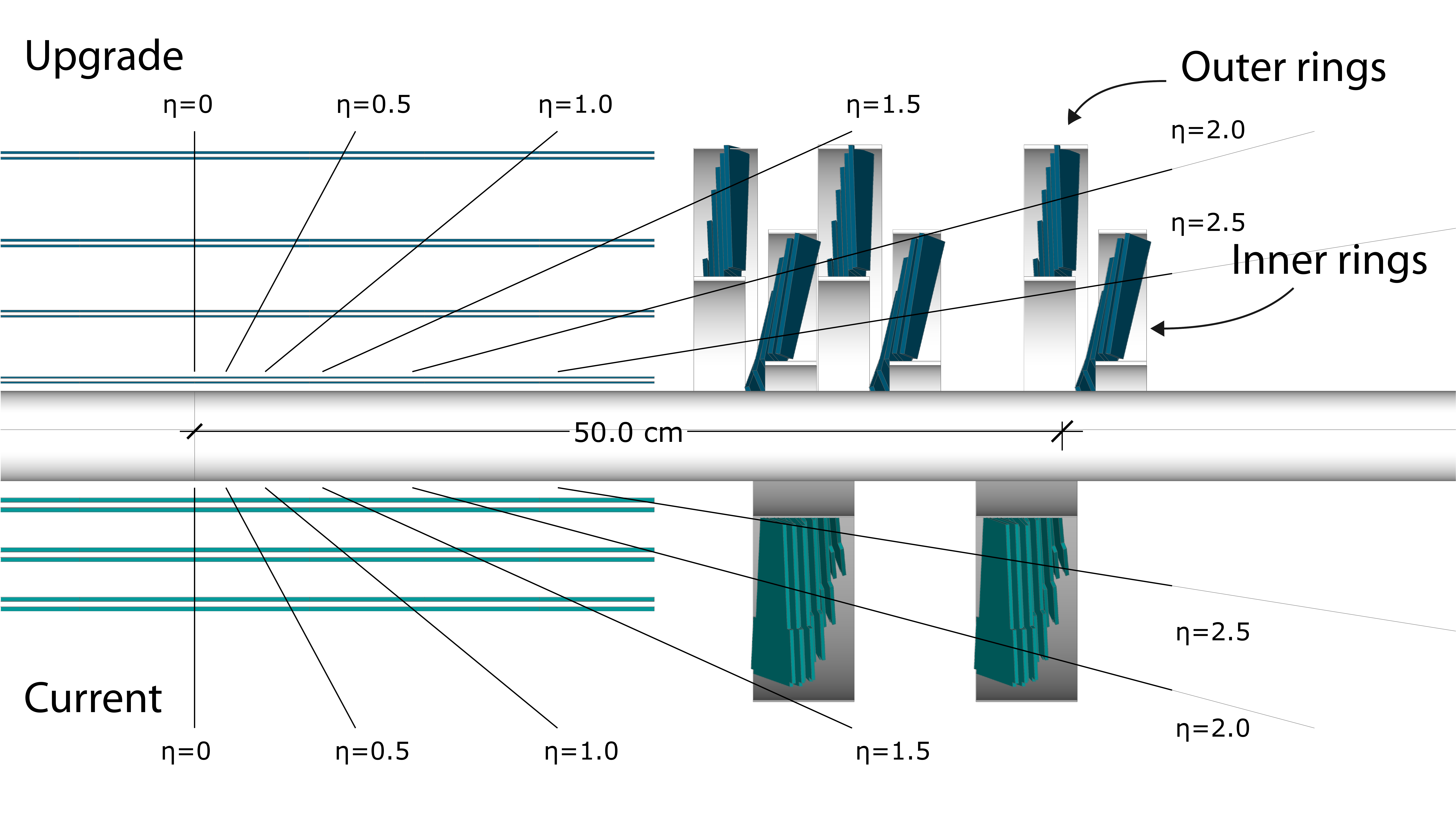}
    \includegraphics[width=0.42\textwidth]{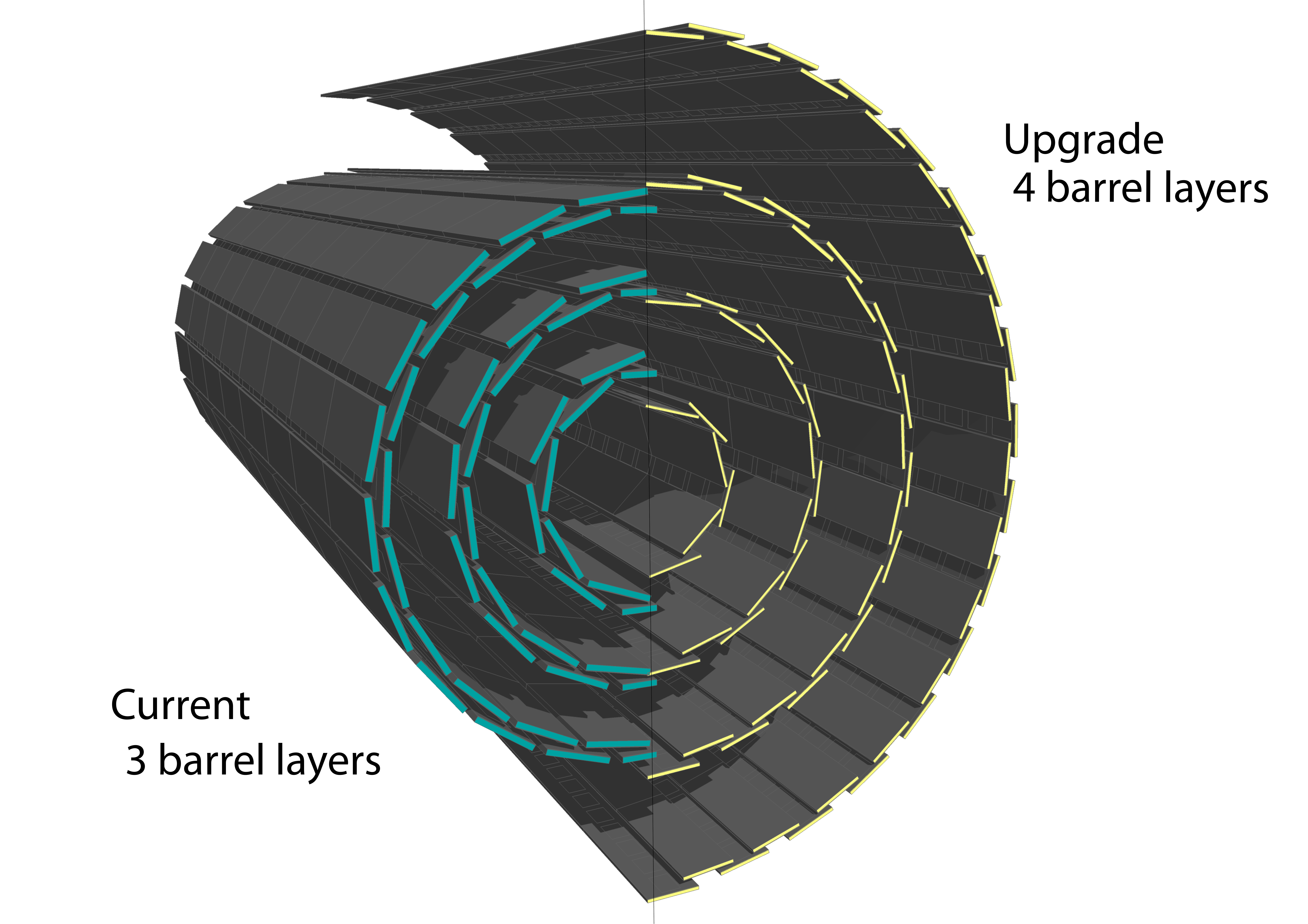}
    \caption{Left: Conceptual layout comparing the different layers and disks
      in the current and upgrade pixel detectors. Right:
      Transverse-oblique view comparing the pixel barrel layers in the
      two detectors.}
    \label{fig:ConceptualLayout}
  \end{center}
\end{figure}

The performance of the present pixel detector at high luminosity is limited by reduced hit efficiency in the readout chip for the innermost layers, and bandwidth limitations in the optical links.  A new detector has been designed that overcomes these limitations, improves the track impact parameter resolution, and reduces the effect of particle interactions in the detector materials. It contains an optimized configuration for 4-hit coverage over the pseudo-rapidity range up to $\eta=2.5$, with 4 layers in the barrel and 3 disks in the end-caps, a new read-out chip (ROC) architecture with high hit rate capability, a significantly reduced material budget, new optical links and data acquisition system with a higher output bandwidth, and a modified power supply system using DC-DC converters on the detector to reduce power losses.  The new pixel detector, with these characteristics, will survive the radiation dose expected through LS3, with a single exchange of the innermost barrel layer at mid-term.

\begin{figure}[bh]
\begin{center}
\includegraphics[width=0.49\textwidth]{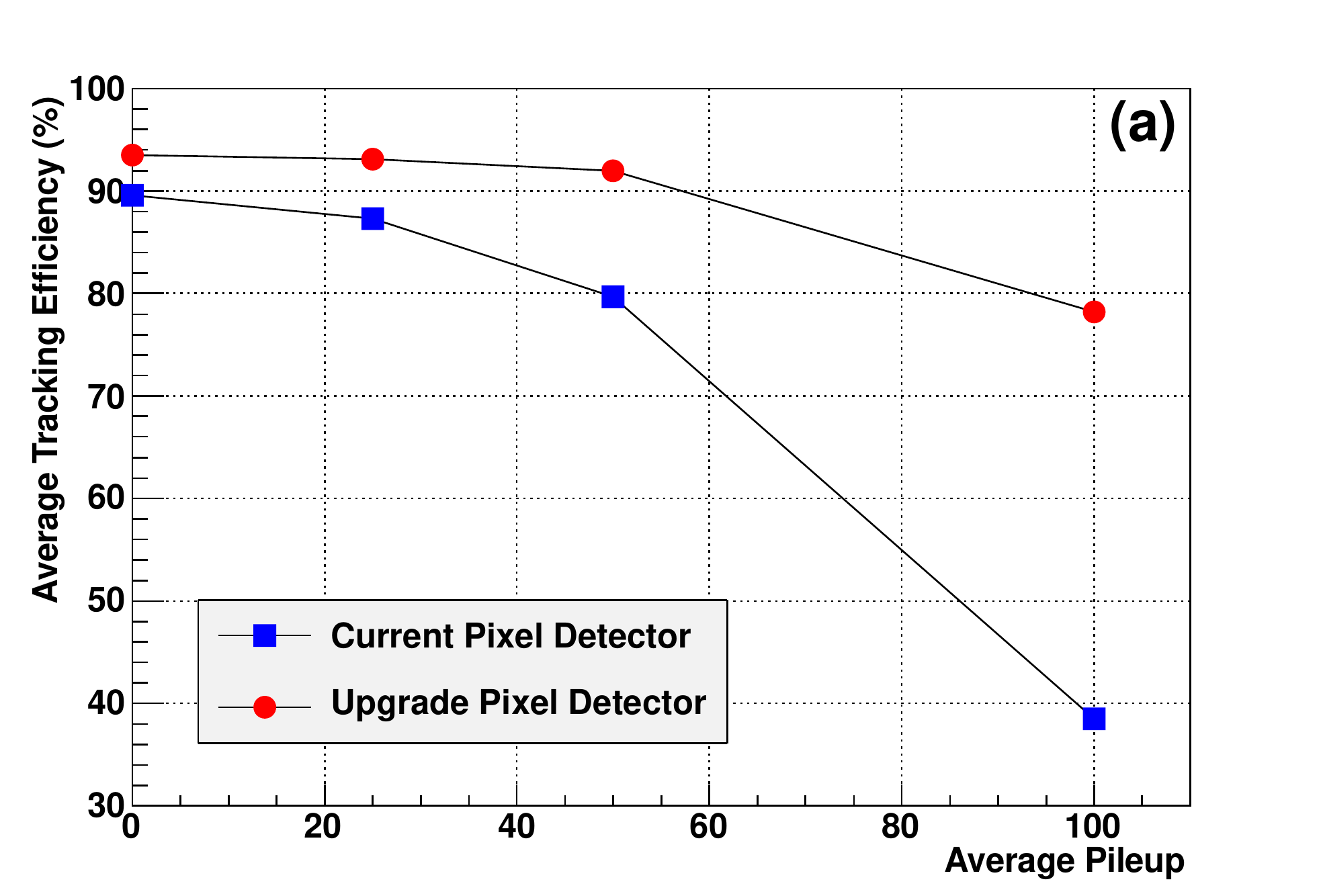}
\includegraphics[width=0.49\textwidth]{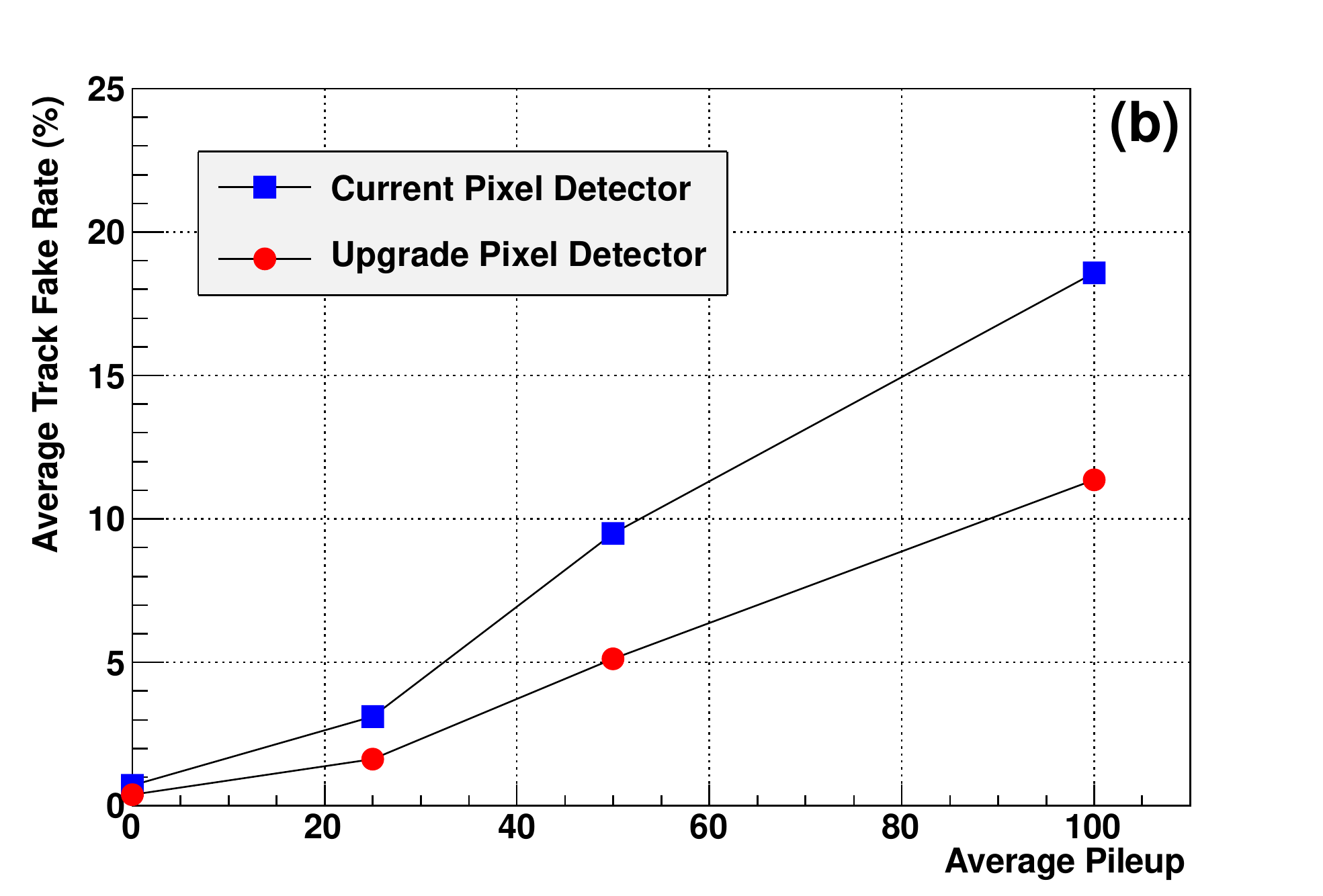}
\caption{Average tracking efficiencies (a) and fake rates (b) as a function of
pile-up for the $t\bar{t}$ event selection.}
\label{fig:TrackingEfficiencies}
\end{center}
\end{figure}

The overall configuration of the new detector is shown in Fig.~\ref{fig:ConceptualLayout}. The first barrel layer is moved closer to the interaction point, by 14~mm, at a radius of
30~mm; this will improve the track impact parameter (IP) resolution. The radius of the outermost layer, now the fourth layer, increases to 160~mm, closer to the Tracker Inner Barrel (TIB) layers; this will reduce the rate of fake tracks and mitigate future inefficiencies in the TIB. The new detector will have $\sim$123M~pixels, almost twice the present system.

The performance of the proposed upgrades to the pixel detector has been studied with a full GEANT simulation of the CMS detector, using complete descriptions of the detector and beam pipe geometries and materials. Both the present and new detectors have been simulated, including emulation of the ROC signal thresholds and data loss.  In these studies, CMS track reconstruction has not been re-optimized for the new detector nor have the track selection and the algorithm used for the b-tagging been tuned to the upgrade conditions. The performance presented for the new pixel detector is therefore likely conservative.  Studies have been performed for luminosities of 10$^{34}$~cm$^{-2}$~s$^{-1}$ with 25~ns bunch spacing, used as a reference for the present detector, 2x10$^{34}$~cm$^{-2}$~s$^{-1}$ with a 25~ns bunch spacing (pile-up of 50) and for the extreme case of a 100 pile-up corresponding to a 50~ns bunch spacing at the same luminosity.  The performance comparison of the current and new detectors is presented in Fig.~\ref{fig:TrackingEfficiencies}. It shows the average efficiency and average rate of fake tracks.  Major improvements in the track reconstruction efficiency are achieved with the
new design, resulting from the increased number of layers (disks) and the
mitigation of the ROC data loss.

\begin{figure}[tbph]
\begin{center}
\includegraphics[width=0.85\textwidth]{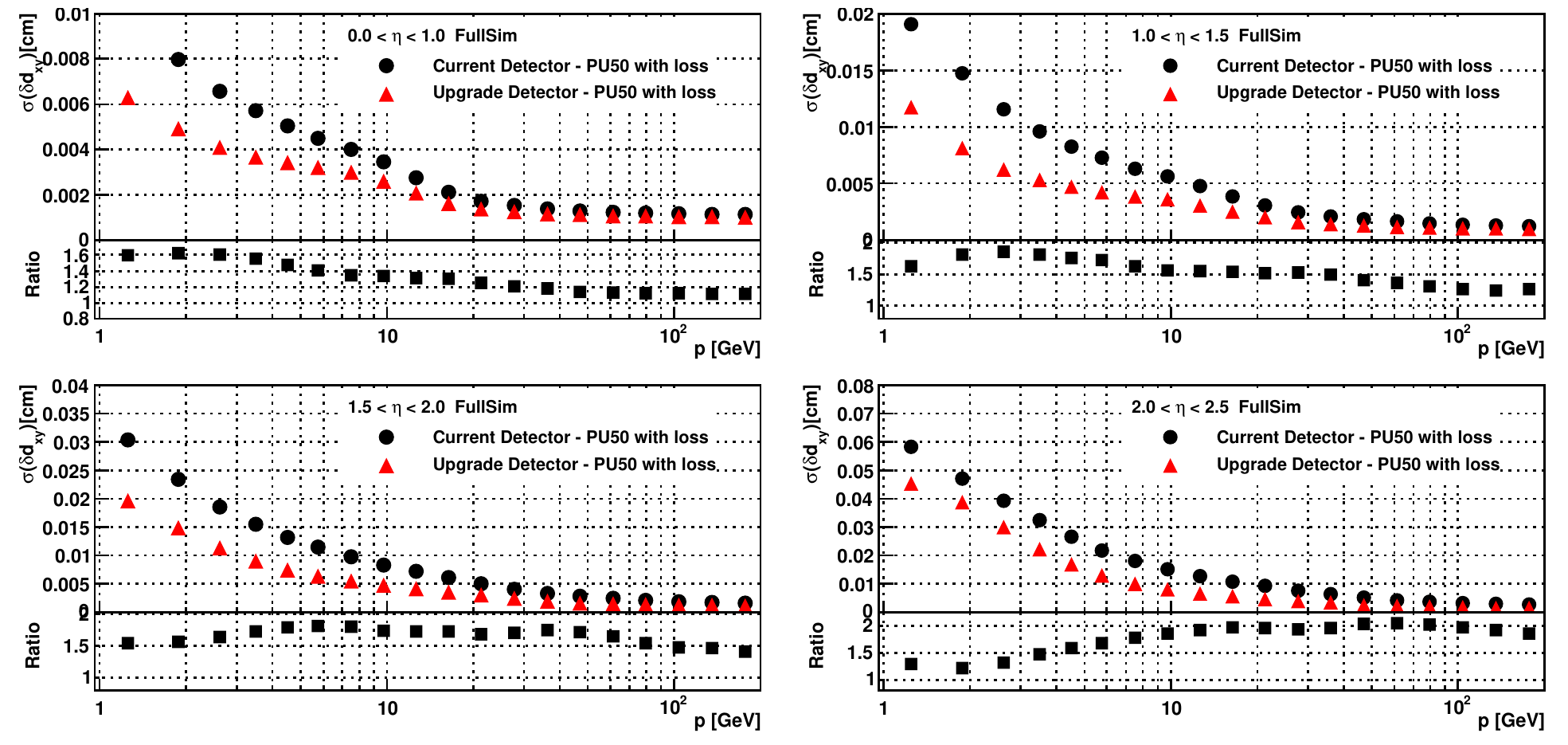}
\caption{Transverse IP resolution for muon tracks as a function of momentum for different pseudorapidity regions. The current and new detectors arerepresented with black dots and red triangles, respectively.}
\label{fig:ResolutionMuonTracks}
\end{center}
\end{figure}

\begin{figure}[tbph]
\begin{center}
\includegraphics[width=0.85\textwidth]{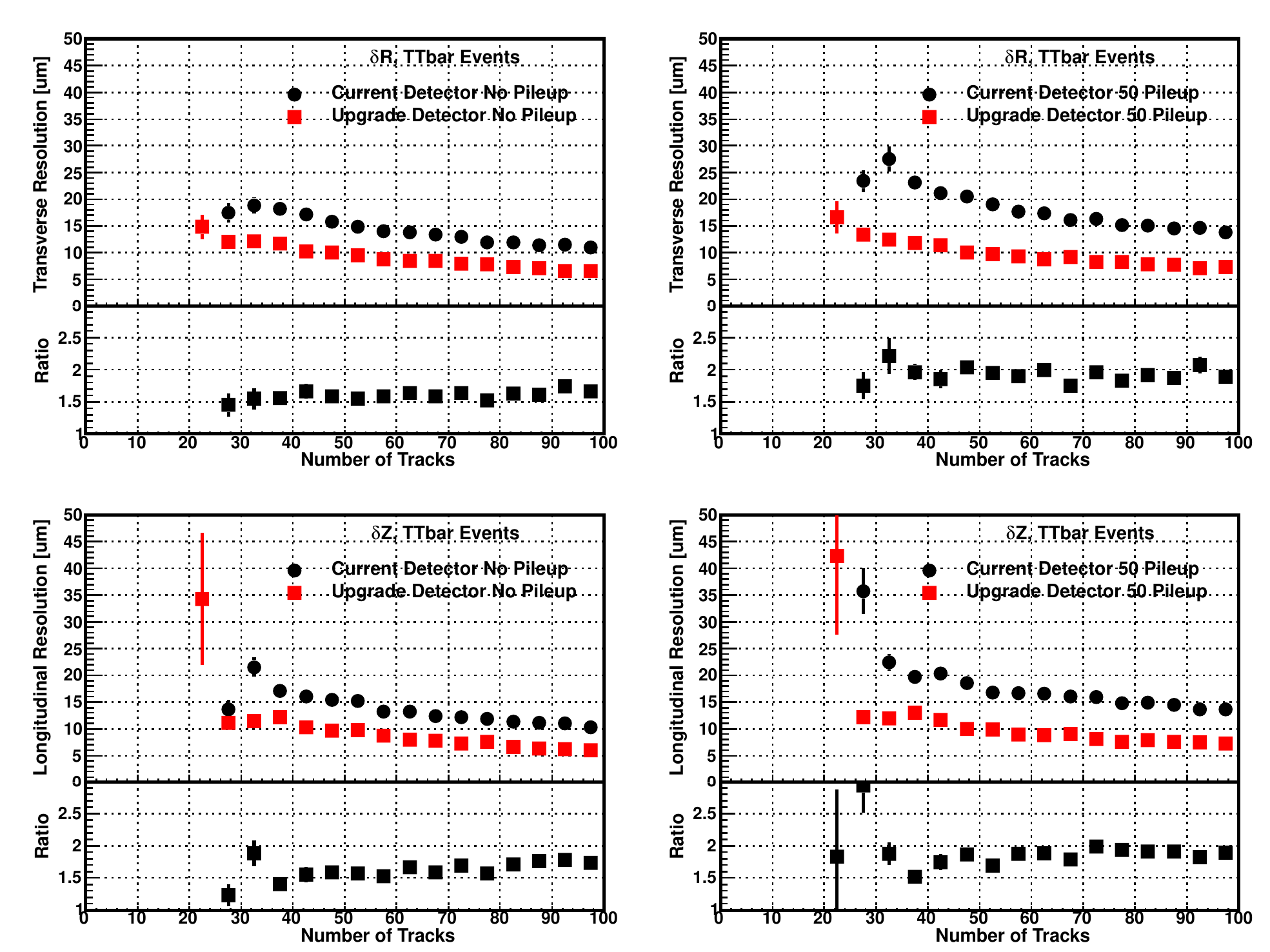}
\caption{Transverse (top) and longitudinal (bottom) primary vertex position resolutions as a function of the number of tracks; without pile-up (left) and at a 50 pile-up (right). The current and new detectors are respectively represented with black dots and red squares.}
\label{fig:VertexPositionResolutions}
\end{center}
\end{figure}

\begin{figure}[tbph]
\begin{center}
\includegraphics[width=0.49\textwidth]{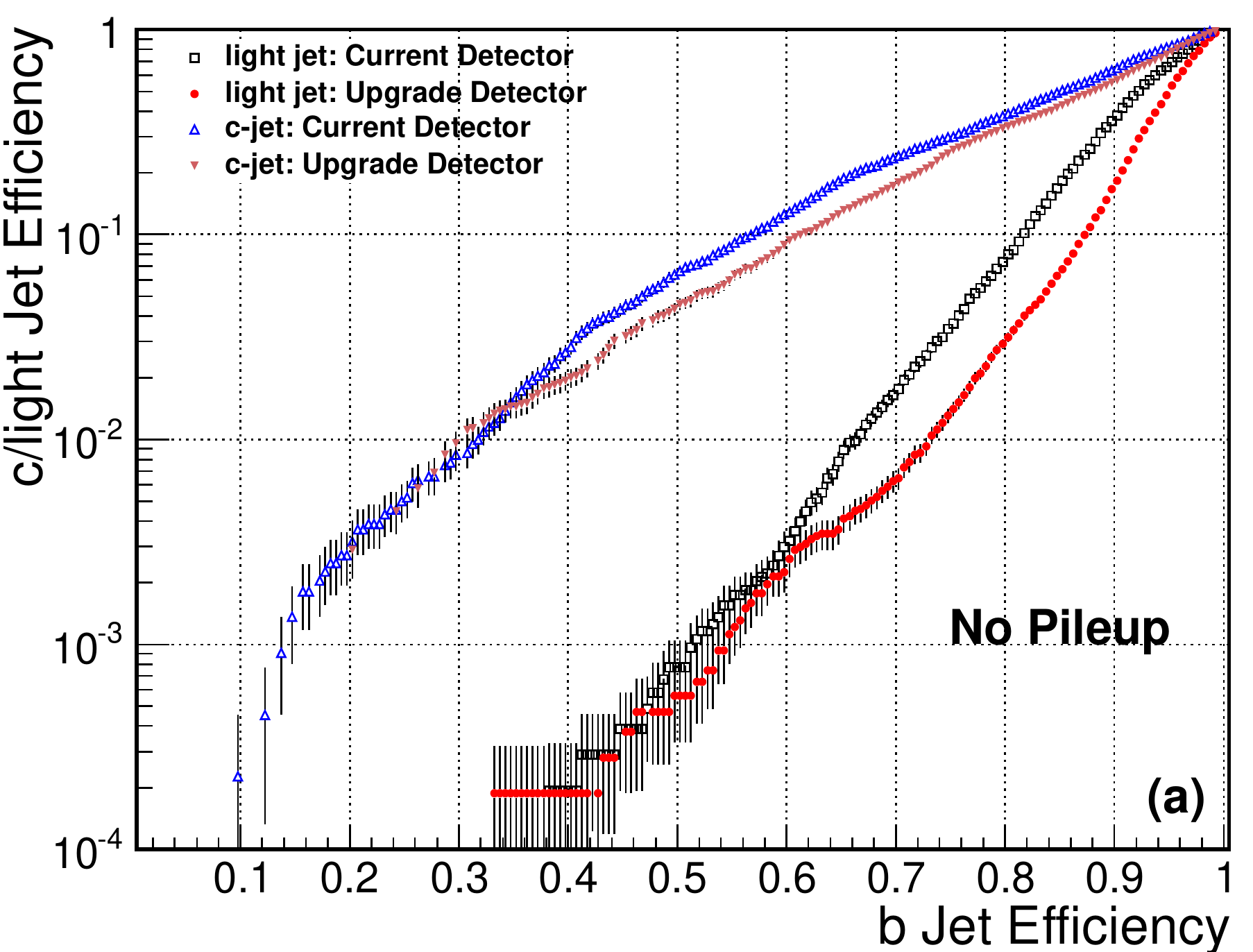}
\includegraphics[width=0.49\textwidth]{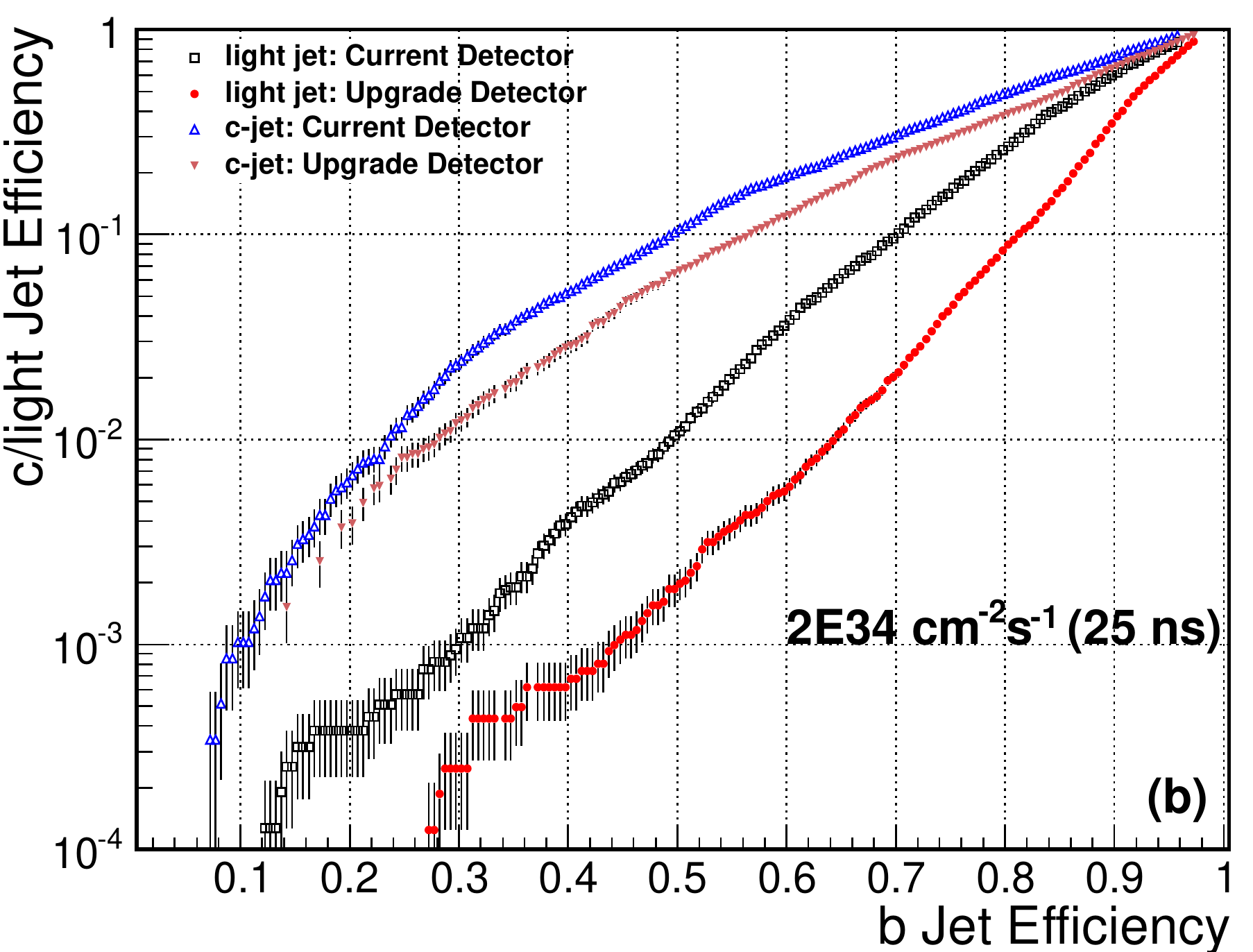}
\caption{Fraction of c jets or light-quark jets misidentified as b jets for the
current (blue/black) and upgraded (red) detectors as a function of the efficiency
to tag true b jets in the absence of pile-up (left), and with 50 pile-up (right).}
\label{fig:Fraction}
\end{center}
\end{figure}

\begin{figure}[tbph]
\begin{center}
\includegraphics[width=0.49\textwidth]{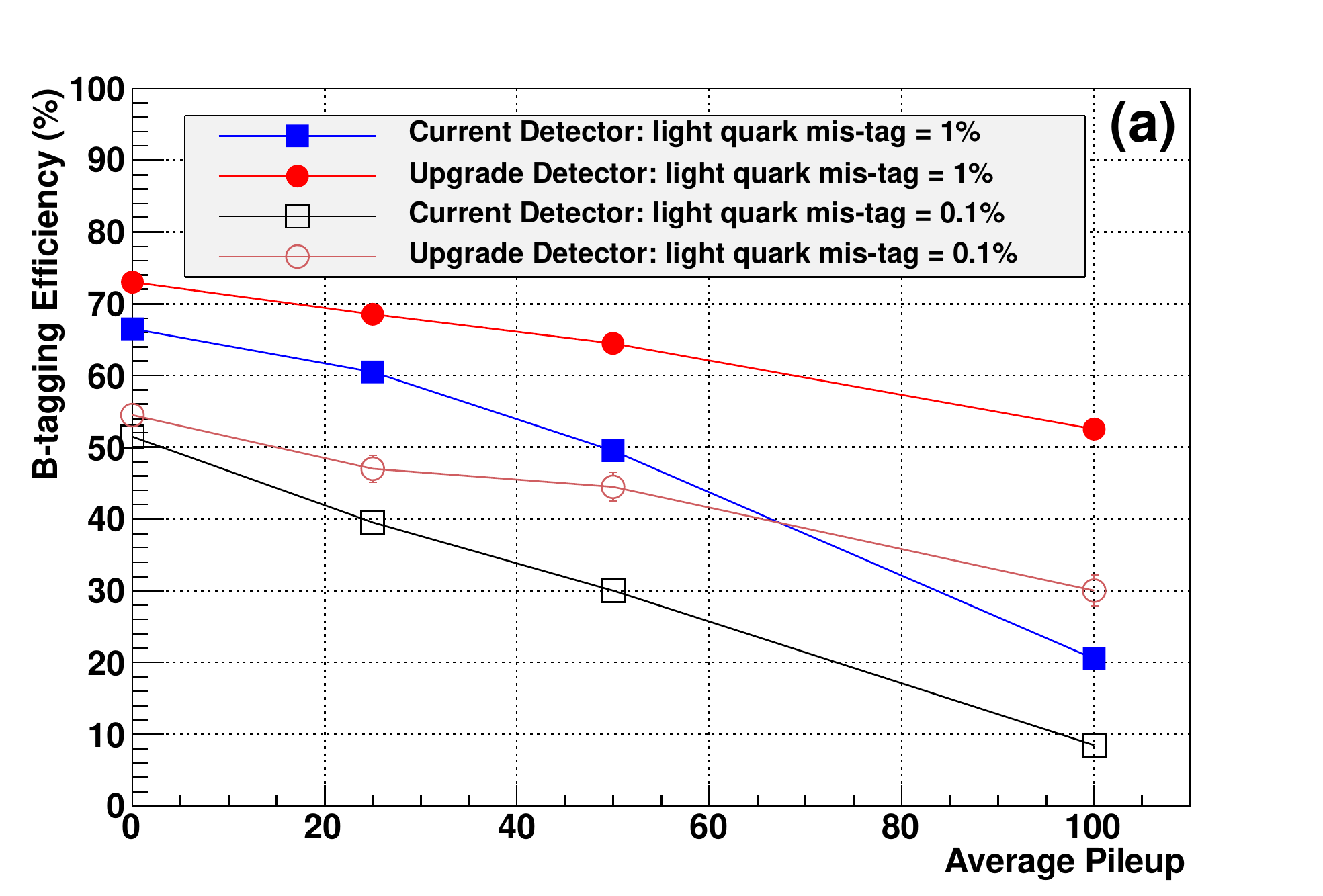}
\includegraphics[width=0.49\textwidth]{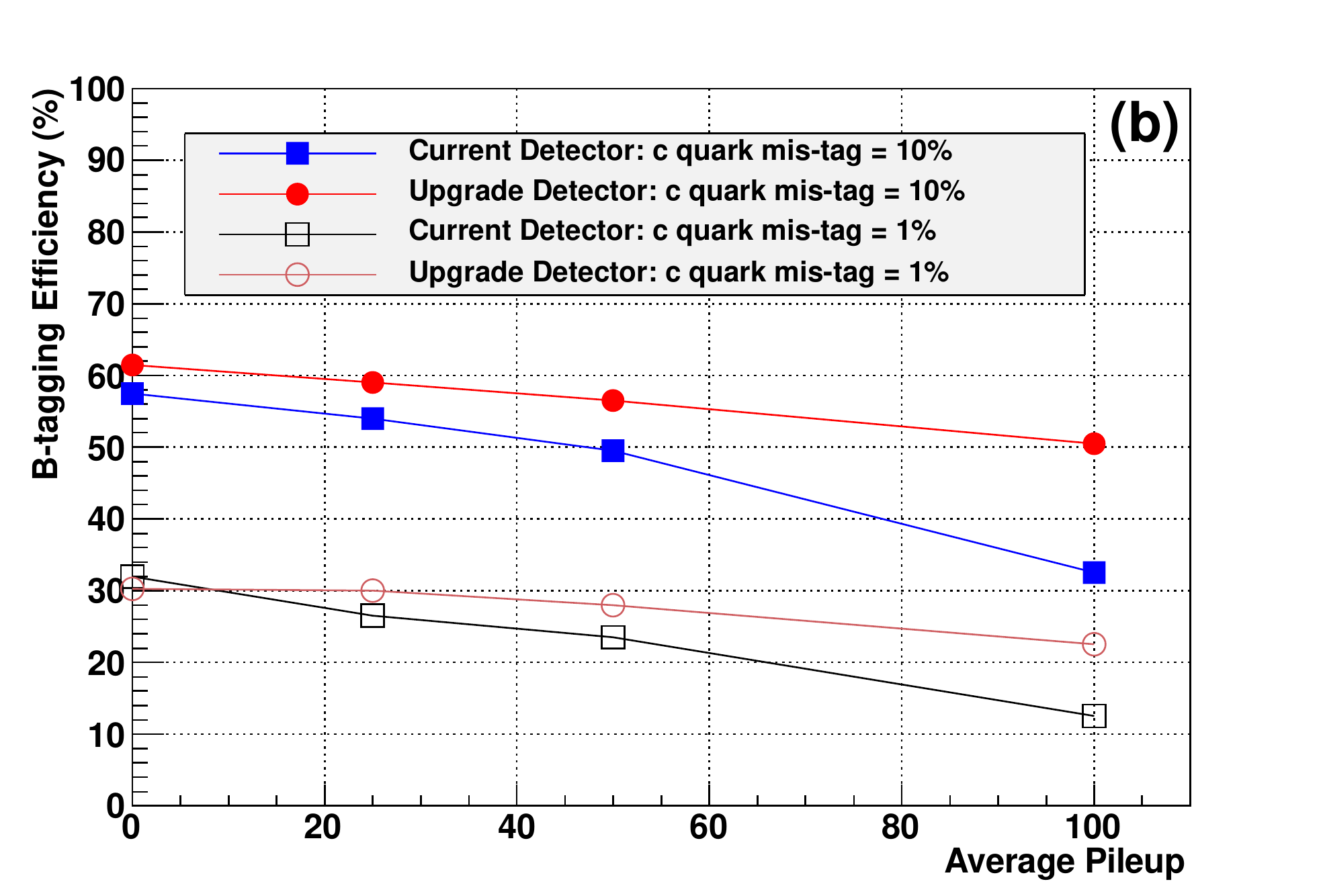}
\caption{The b-tagging efficiency of the current (blue/black) and upgraded
(red) detectors as a function of pile-up for typical values of mistagging fractions
for light-quark jets (left) and c jets (right).}
\label{fig:Efficiency}
\end{center}
\end{figure}

Another benefit with the new detector is improved resolution on measurements of the transverse and longitudinal components of the track impact parameter (IP), and of the primary vertex position.  These resolutions are key elements in b-tagging efficiency. The improvements result from the increase in the number of space points, the lower radius of the first layer, the lower ROC data loss and signal thresholds and the reduced detector mass. Expected IP resolutions estimated with simulated muon tracks in different pile-up conditions are shown in Fig.~\ref{fig:ResolutionMuonTracks}.  The expected primary-vertex resolution is estimated using the $t\bar{t}$  event sample. It is presented in Fig.~\ref{fig:VertexPositionResolutions} for different pile-up conditions. The b-tagging performance of the new pixel detector has been studied for the "Combined Secondary Vertex" (CSV) algorithm with the $t\bar{t}$ event sample. The CSV is a multivariate method using both track IPs and secondary vertex reconstruction.  Fig.~\ref{fig:Fraction} shows the fraction of c jets or light-quark jets (u,d,s) misidentified as b jets as a function of the efficiency to tag the true b jets. Figure~\ref{fig:Efficiency} shows this efficiency as a function of pile-up for typical mis-tagging fractions.

The results of these studies indicate that while performance would be seriously compromised without the proposed upgrade, with the new detector there is no degradation, and in several cases significant improvement, to the performance of the reconstruction of the objects that are relevant for physics.  These studies give us confidence in the validity (as concerns tracking and b-tag performance) of the extrapolations to 300 fb$^{-1}$ of physics results presented in this report.

\subsubsection{Hadron Calorimeter Upgrade}

The CMS hadron calorimetry system (HCAL) has
four major sections: the HCAL Barrel (HB), HCAL Endcap (HE), HCAL
Outer (HO), and HCAL Forward (HF). The HCAL upgrade takes advantage of new technologies that have become available since the design and construction of the original calorimeters and improves the performance of the calorimeters as built, primarily through the replacement of the phototransducers and
electronics.  It will also address and mitigate weaknesses that have been identified in the current systems.

The upgrades of the HF and HB/HE systems are based on the replacement of the phototransducers for these calorimeters.  For the HF, the
current single-anode PMTs will be replaced by multi-anode PMTs with increased quantum efficiency.  As discussed below, the upgrade uses
dual-anode readout of the PMTs and a new TDC capability to suppress background from spurious signals and recover the calorimeter response
where possible.  For the HB/HE systems, a newly-proven technology the Silicon Photomultipler (SiPM) will be used.  These new photodetectors
will be followed by a common chain of upgraded electronics which provides increased robustness to channel or link failures and a
greatly-enhanced capacity for calculating and delivering inputs to the calorimeter trigger.   The upgraded electronics should tolerate the full radiation dose
anticipated through HL-LHC operation, for 3000 \fbinv total luminosity.

The performance gains of the upgraded detector originate from the superior characteristics of the new photodetectors to be installed in the barrel, endcap, and forward calorimeters and from the expanded functionalities of the proposed front-end and back-end electronics
upgrades.  The SiPM photodetectors with newly designed readout chips provide an order of magnitude higher signal-to-noise ratio in the barrel and endcap calorimeters.  In the forward calorimeter the multi-anode signals provide redundant sampling of the light from a single calorimeter cell and high quantum efficiency for measuring Cerenkov signals while the new TDC capability allows rejection of background.

\begin{figure}[tbph]
\begin{center}
\includegraphics*[width=0.65\linewidth]{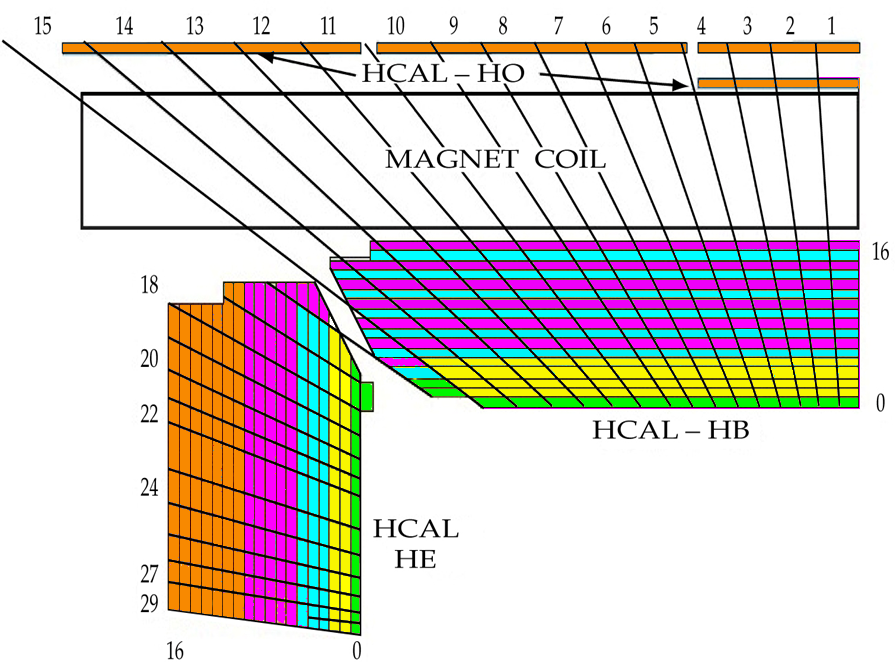}
\caption{Current proposed depth segmentation structure for the HB and HE calorimeters, made possible by the use of
SiPM photodetectors.
\label{fig:hbhesegment_new}}
\end{center}
\end{figure}

In the HB and HE, the switch to the SiPM with its much-decreased noise
level and higher gain allows longitudinal segmentation of the calorimeter without introducing excessive noise.  Figure~\ref{fig:hbhesegment_new} shows the current proposed depth segmentation.  This depth segmentation allows better determination of hadronic shower development.  Studies in simulation of particle-flow behavior at high pile up show large numbers of anomalous hits and large clusters of energy as the particle flow algorithms can no longer distinguish individual high-energy particle showers.  The addition of depth segmentation eliminates the anomalous hits and improves the association of clusters and tracks which is crucial for particle-flow techniques.

Depth segmentation improvements, along with the high photon-detection-efficiency of the SiPMs, will also allow better management of the radiation damage which will occur with the high-$\eta$ region of the HE calorimeter.  The longitudinal segmentation of the hadron calorimeter will also provide shower profile information to be used to verify that electromagnetic particles identified in the ECAL have little energy in the HCAL. In particular, the segmentation suppresses the influence of pileup particles that contribute to the first layer of HCAL but not to deeper layers.  Similarly, the deepest segment of HCAL can be useful for efficient identification of prompt muons and rejection of muons produced in the decay of hadrons in flight.

\subsubsection{L1 Trigger Upgrade}
The first level (L1) trigger for the CMS experiment must facilitate a physics program wherein comparable sensitivity to electroweak scale physics and TeV scale searches is maintained with respect to the pre-LS1 program. It must do so with an increased LHC center-of-mass energy near 14~\TeV, with pile-up of about 50 interactions per crossing, and with a luminosity in excess of $2\times 10^{34}$~cm$^{-2}$s$^{-1}$.  With the increase in energy, luminosity and pile-up, a substantial increase in trigger thresholds would be required to fit within the nominal limit of 100 kHz, especially for pile-up sensitive multi-object triggers. This would have a detrimental impact on the physics acceptance of the CMS experiment, in particular at the electroweak scale where the study of the couplings of the newly discovered Higgs boson is a priority.  An upgrade to the L1 trigger hardware that expands both the capability and flexibility of the system is needed.  The L1 trigger upgrade will provide the following:

\begin{itemize}
\item Improved electromagnetic object isolation using calorimeter
  energy distributions with pile-up subtraction;
\item Improved jet finding with pile-up subtraction;
\item Improved hadronic tau identification with a smaller fiducial area;
\item Improved muon transverse momentum (\PT) resolution in difficult regions;
\item Isolation of muons using calorimeter energy distributions with
  pile-up subtraction;
\item Improved global Level-1 trigger menu with a greater number of
  triggers and with more sophisticated relations involving the input objects.
\end{itemize}

The L1 trigger upgrade will also be flexible and scalable to accommodate
future upgrades to the experiment beyond the present ones such as those required
for the HL-LHC physics program. The upgraded system will
form a basis to which additional inputs can be easily incorporated and to which
additional processing power can be added in a straightforward way.

To meet these requirements, CMS will upgrade the electronics for the calorimeter trigger, the muon trigger, and the global trigger.  Additional interconnections between
these systems will also be provided, to implement algorithms such as muon isolation. As noted above, a key feature of this upgrade is that it will offer
a large increase in flexibility beyond that provided by the current trigger system. Flexibility has been important in adapting to the rapidly evolving running conditions since LHC start-up, and will continue to be needed in order to implement further rate reduction and efficiency improvements as algorithms improve. This increased flexibility will be accomplished by using high bandwidth optical links for most of the data communication between trigger cards, and by using modern, large FPGAs and large memory resources for the trigger logic. The use of optical links allows the architecture to be readily changed, while large FPGAs allow for algorithms to evolve as needed.

The current L1 makes use of a large number of different electronics cards. CMS plans to use the upgrade as an opportunity to reduce this diversity, basing the upgrade on a small number of general-purpose designs. It is planned to pursue a similar path of consolidation on the software and firmware side of the
project, increasing that which is common to all components of the trigger. The electronic systems will be implemented in the telecommunications
standard $\mu$TCA, evolving away from the VME framework used previously, to take advantage of additional flexibility and higher bandwidth.  The proposed upgrade maintains the overall architecture and functionality of the present trigger system, with upgrades to each of the main areas of calorimeter, muon and global trigger systems.
The schedule for the upgrade foresees parallel commissioning during the LHC restart after LS1 in 2015. The upgraded trigger system is planned to be available for CMS data taking in 2016.

\begin{figure}[tbh!]
   \begin{center}
     \includegraphics[width=0.65\textwidth]{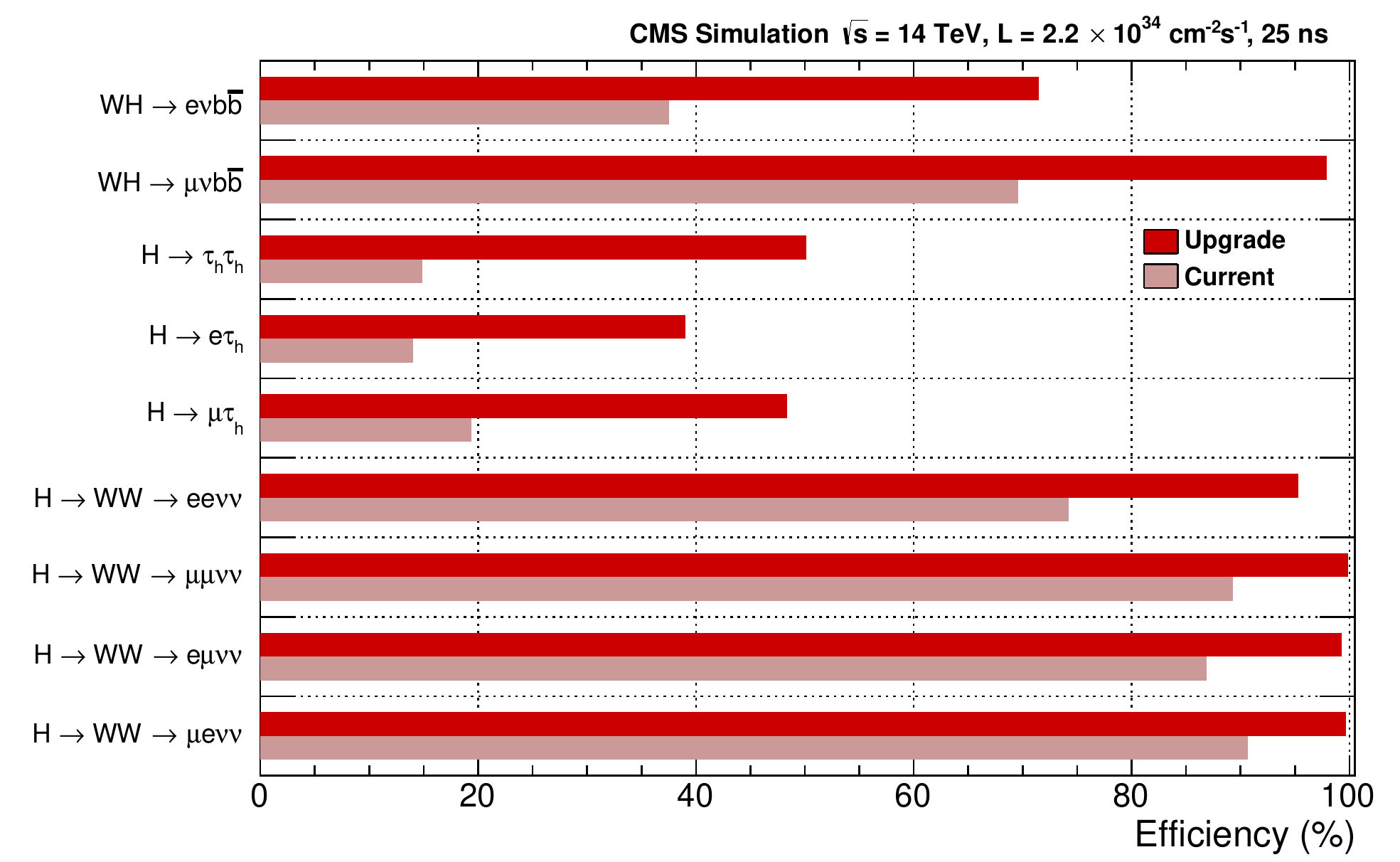}\\
     \vspace{1cm}
     \includegraphics[width=0.65\textwidth]{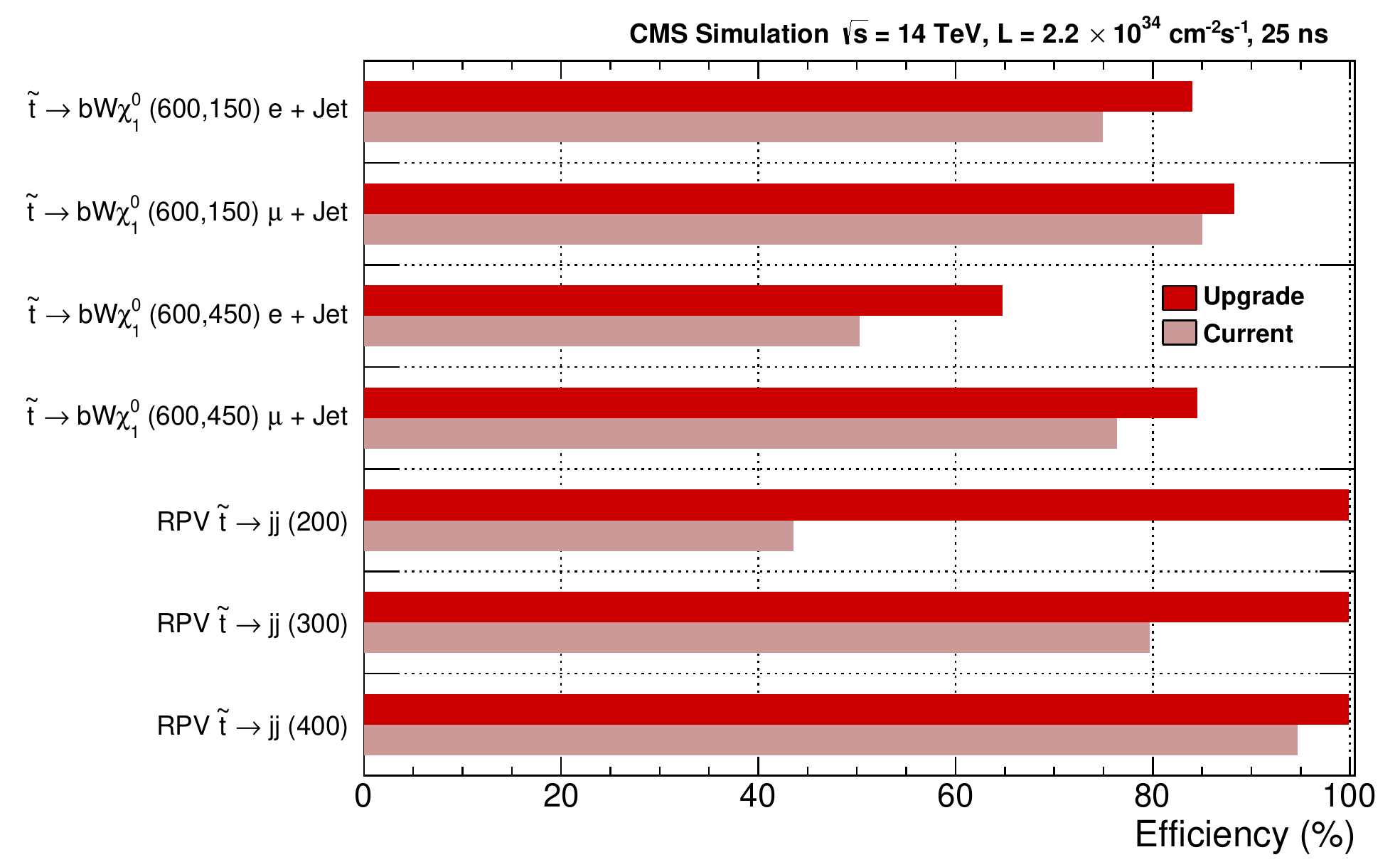}
     \caption{Signal efficiency obtained for the current and an upgraded L1 trigger system at $L = 2\times 10^{34}$~cm$^{-2}$s$^{-1}$ for selected Higgs (top) and SUSY (bottom) channels.
              \label{fig:trigsummary}}
     \end{center}
\end{figure}

To study the performance of the upgraded L1 trigger, we prepared simplified trigger menus for the present CMS and trigger upgrade scenarios.  These simplified menus contain a small sample of algorithms that capture much of the important physics processes and account for about 80\% of the L1 rate in 2012 operations. These menus are used to illustrate typical L1 trigger thresholds that will be attainable with the upgrade.  The thresholds are determined not only by the individual rate for each algorithm but also to the total rate of the full menu, which is required to remain below 100 kHz.  For any given event, several trigger criteria may be satisfied.  Such overlaps are accounted for when estimating the rate of the full menu.  Using fully simulated Monte Carlo samples of signal and background processes, the performance of key analyses in the CMS physics program outlined above was evaluated using the plateau efficiencies of the triggers in these simplified menus.  An assessment was done by comparing the analysis efficiencies that result from the use of the current L1 menus with those that result from using the upgraded L1 menus.  The studies presented applied the analysis strategies used on the 7/8\TeV pp collision data as closely as possible.  The results are summarized in Fig.~\ref{fig:trigsummary}.

\subsection{Phase 2 Upgrades to the CMS Experiment}

Due to radiation damage, aging, and the challenges of even higher instantaneous luminosities for the HL-LHC, a number of additional major upgrades to the CMS detector will be required in order to preserve the ability to carry out the diverse physics program of the CMS experiment. The performance of the tracking system will significantly degrade with radiation aging, and a new tracker will be needed. The new tracker will have substantially less material and in addition to providing improved tracking capability in a dense charged particle environment, will also provide tracks to the L1 trigger, allowing a substantial increase in trigger functionality.   We are also examining a possible upgrade to the front-end electronics and high-level trigger systems, which would allow up to 1~MHz L1 readout and up to 10~kHz event storage rate.

The upgrade must ensure that precision electromagnetic calorimetry and robust jet and missing transverse energy reconstruction capability are maintained at the HL-LHC.  Issues are particularly severe in the endcap region, where the present calorimeters will suffer radiation damage and where event pile-up is more pronounced. This region has critical importance in major parts of the physics program (e.g. vector-boson-fusion Higgs production and vector-boson-scattering studies).  Both the electromagnetic and hadronic endcap calorimeters must be replaced. Studies are underway to determine the optimal choice for technology and design. An option is under consideration to extend tracking beyond $\eta=2.5$ with additional pixel disks to improve particle flow reconstruction and pile-up mitigation in this region.  The use of precision timing measurement which could be integrated into an electromagnetic preshower detector is also being investigated. Such a system could provide further pile-up mitigation.

Several Phase 2 upgrade scenarios are under study using both parametrized detector resolutions/responses with which different configurations can be easily compared, and full GEANT simulations for investigations that require more complex treatment.  The result of these studies will be included along with a description of the design considerations in a Technical Proposal for the Phase 2 upgrade, anticipated in 2014.

Since the studies to optimize the design choices for Phase 2 are ongoing, it is challenging to extrapolate the performance of physics results with the 3000 fb$^{-1}$ dataset. This complication is in addition to the inherent difficulty of extrapolating to a dataset more than 100 times larger than the current dataset on which the studies are based and whose systematic uncertainties (both experimental and theoretical) will evolve over the next decade.  The present CMS expectation of this evolution is discussed in subsequent Sections.

\section{Higgs Boson Properties}\label{higgs}

The year 2012 will be remembered as the year of the Higgs boson
discovery~\cite{Aad:2012tfa, Chatrchyan:2012ufa}. Hypothesized more than 40 years
ago, the Higgs boson holds the key to understanding how fundamental particles
acquire their mass and how the electroweak symmetry is broken. The observation of a Higgs
boson by the ATLAS and CMS experiments at the LHC has opened a new era for particle
physics, namely precision consistency tests of the SM Higgs boson.

\begin{figure*}[h]
\centering
\includegraphics[width=0.5\textwidth]{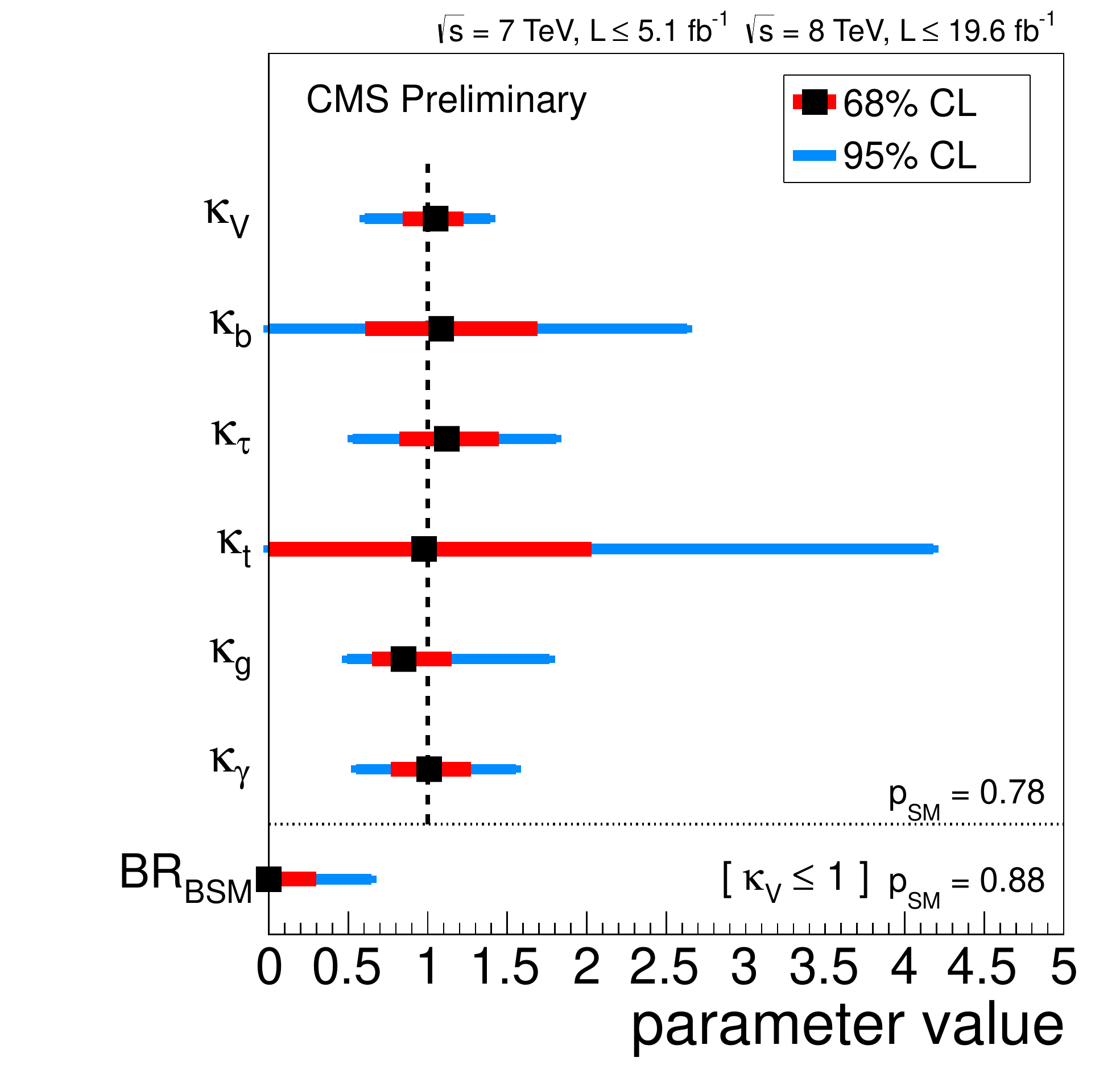}
\caption{The best fit of the Higgs boson coupling parameters are shown,
with the corresponding $68\%$ and $95\%$ CL intervals, and the overall
p-value (pSM) of the SM Higgs hypothesis is given. The result of the
fit when extending the model to allow for BSM decays, while
restricting the effective coupling to vector bosons to not exceed
unity ($\kappa_V \le$  1.0), is also shown.}
\label{fig:currentHiggsFit}
\end{figure*}

Five main decays have been studied: $\gamma\gamma$, $\cPZ\cPZ$, $\PW\PW$,
$\tau\tau$, and $\cPqb\cPqb$, using data samples corresponding to integrated
luminosities of up to 5.1
fb$^{-1}$ at 7~\TeV and up to 19.6 fb$^{-1}$ at 8~\TeV.
The mass of the new boson has been measured from the $\gamma\gamma$ and  $\cPZ\cPZ \to 4\ell$ channels
to be $125.7\pm 0.4\GeV$~\cite{CMS-HIG-13-005}.  The event yields obtained by the different analyses
targeting specific decay modes and production mechanisms are within
current uncertainties consistent with
those expected for the SM Higgs boson.
The best-fit signal strength for all channels combined,
expressed in units of the SM Higgs boson cross section,
is $0.80 \pm 0.14$~\cite{CMS-HIG-13-005}. Additional channels have
been studied including decays to two muons and $Z\gamma^*$.
The consistency of the couplings
with those predicted for the SM Higgs boson is tested in various ways, and no significant deviations
have been found.  Figure~\ref{fig:currentHiggsFit}  shows the results of a coupling fit assuming no BSM decays. A
second fit allowing for BSM decays restricting the effective
couplings to vector bosons is used to extract a 95\% confidence level
upper limit of 52\% on the branching fraction for
undetected Higgs decays~\cite{CMS-HIG-13-005}. Direct searches for a standard-model-like
Higgs boson produced in association with a Z boson or in vector boson
fusion production and decaying to invisible particles yields an observed (expected) 95\% confidence
level upper limit on the branching fraction of the Higgs boson to
invisible particles is 54\% (46\%)~\cite{CMS-HIG-13-013, CMS-HIG-13-018}. The limits can be translated in
the context of dark matter models, strongly constraining them.

The spin-parity of the boson has been studied, and the pure scalar hypothesis found to be consistent with the observation when compared to six
other spin-parity hypotheses. The data disfavor the pure pseudoscalar
hypothesis $0^{-}$ with a $CL_s$ value of 0.16\%, and disfavor the pure
spin-2 hypothesis of a narrow resonance with the minimal couplings to
the vector bosons with a $CL_s$ value of 1.5\%. The spin-1 hypotheses are
disfavored with an even higher confidence.
The measurement of the fraction of a
CP-violating contribution to the decay amplitude expressed through the
fraction of the corresponding decay rate is $f_{a3} =
0.00^{+0.23}_{-0.00}$, or equivalently $f_{a3} < 0.58$ at 95\%
CL~\cite{CMS-HIG-13-002}.

Precise measurements of the new boson's properties are of
utmost importance. More data are needed to check whether the properties of
this new state imply BSM physics.
The key properties are the couplings to each fermion and boson, which
are predicted by the standard model. Perhaps the most important measurement after the discovery of the
Higgs boson is the measurement of the Higgs potential itself. This can
be probed with the study of multiple Higgs boson production. Previous
studies have shown that a measurement~\cite{di-higgs} of
multiple Higgs boson production is possible with a dataset of
3000~fb$^{-1}$ at the LHC.
Promising final states are those which allow a
precision measurement of the mass of one of the two Higgs bosons or
have large branching fraction, e.g. the
bb$\gamma\gamma$ or bb$\tau\tau$ final states, which allow to reduce the experimental backgrounds.
It is interesting to note that  multiple Higgs boson production can be increased in BSM scenarios like the MSSM.

\subsection{Extrapolation Strategy}
In this summary only measurements that have been
made public by CMS as measurements applied to the 7 and 8~TeV data are used.
The results are extrapolated to larger
datasets of 300 and 3000 fb$^{-1}$ and a center-of-mass energy of
14~TeV by scaling signal and background event yields accordingly.
In order to study the precision of future measurements, a number of
assumptions are made. As stated in the introduction, the underlying assumption of
the extrapolations is that future CMS upgrades will provide the same level of
detector and trigger performances achieved with the current detector
in the 2012 data taking period.
The extrapolations do not take into consideration those channels that
were not utilized in the currently available dataset, and there is no
attempt to optimize the measurement in order to minimize the uncertainties
on Higgs coupling measurements.
Extrapolations are presented under two
uncertainty scenarios. In Scenario 1, all systematic uncertainties
are left unchanged. In Scenario 2, the theoretical uncertainties
are scaled by a factor of 1/2, while other systematic uncertainties
are scaled by the square root of the integrated luminosity. The
comparison of the two uncertainty scenarios indicates a range of
possible future measurements. The extrapolation
without theoretical uncertainties is also presented, to illustrate the
importance of reducing those uncertainties in the future.  Systematic
uncertainties are inputs to the fits. They can be further constraint
by the data when extracting the signal strength, coupling modifier or ratios of such. Similar extrapolations have been discussed in~\cite{CMS-NOTE-2012-006}.

\subsection{Search channels}
Higgs cross sections and coupling measurements are obtained by combining information from many Higgs
production and decay channels. Table~\ref{tab:channelsSum} lists the main features of these channels,
namely the exclusive final state and the approximate instrumental mass resolution.
The simultaneous analysis of the data selected by all individual
analyses accounts for all statistical and systematic uncertainties and their correlations.

\begin{table}[h]
\begin{center}
\small
\caption{Summary of the information on the analyses used as input in this combination, including
decay mode, production channel (tag), final states, analysis categories, mass resolution, and documentation.}
\label{tab:channelsSum}
\begin{tabular}{|l|c|l| c|c|c|}
\hline
\PH decay & prod. tag& exclusive final states & cat. & res. & ref. \\
\hline \hline
\multirow{4}{*}{$\gamma\gamma$}  & untagged & $\gamma\gamma$ (4 diphoton classes)
& 4 & 1-2\% & \multirow{4}{*}{\cite{CMS-HIG-13-001}}  \\
& VBF-tag    &  $\gamma\gamma + (jj)_{\rm VBF}$ &  2  & $<$1.5\%
&   \\
& VH-tag &  $\gamma\gamma + (\Pe,\mu,\mathrm{MET})$ & 3
& $<$1.5\%       &  \\
\cline{2-6}
& $\mathrm{tt}\PH$-tag & $\gamma\gamma$  (lep. and had. top decay) & 2  & $<$1.5\%       &  \cite{CMS-HIG-13-015} \\
\hline
\multirow{2}{*} {$\cPZ\cPZ \to 4\ell$}             & $N_{\rm jet}<2$   &
\multirow{2}{*} {$4\Pe, \, 4\mu, \, 2e2\mu $}         & 3
& \multirow{2}{*} {1-2\%}       & \multirow{2}{*} {\cite{CMS-HIG-13-002}}  \\
                                                   & $N_{\rm jet}\ge 2$ &
                                                   & 3         &      &   \\
\hline
\multirow{3}{*} {$\PW\PW \to \ell\nu\ell\nu$}      & 0/1-jets  &  (DF or SF
dileptons) $\times$ (0 or 1 jets)          & 4             & 20\%
                   & \cite{CMS-HIG-13-003}  \\
                   & VBF-tag    &  $ \ell\nu\ell\nu + (jj)_{\rm VBF}$ (DF or SF
                   dileptons)      & 2             & 20\%          & \cite{CMS-HIG-12-042}  \\
                   & WH-tag &  $3\ell 3\nu$ (same-sign SF and otherwise)
                   & 2                  &
                   & \cite{CMS-HIG-13-009}  \\
\hline
\multirow{5}{*}{$\tau\tau$} & 0/1-jet & $(\Pe\tau_h, \, \mu\tau_h, \, \Pe\mu, \,
\mu\mu) \times $ (low or high $p_T^{\tau}$)  & 16 &
\multirow{3}{*}{15\%}
& \multirow{3}{*}{\cite{CMS-HIG-13-004}}  \\

                            & 1-jet & $\tau_h \tau_h$  & 1 & &  \\

                                    & VBF-tag                & $(\Pe\tau_h, \,
                                    \mu\tau_h, \, \Pe\mu, \, \mu\mu, \, \tau_h
                                    \tau_h) + (jj)_{\rm VBF}$      & 5 &&\\
\cline{2-6}
                                    & ZH-tag &  $(\Pe\Pe, \, \mu\mu)
                                    \times(\tau_h\tau_h, \, \Pe\tau_h, \,
                                    \mu\tau_h,\, \Pe\mu)$                & 8 &  &
                                    \multirow{2}{*}{\cite{CMS-HIG-12-053}}  \\
                                    & WH-tag &  $\tau_h\mu\mu,\tau_h \Pe\mu, \Pe
                                    \tau_h\tau_h,\Pgm\tau_h \tau_h$
                                    & 4 & &\\
\hline
\multirow{3}{*}{bb}              & VH-tag &  $(\nu\nu, \, \Pe\Pe, \, \mu\mu,
\,\Pe\nu, \, \mu\nu$ with 2 b-jets$) \times$x  & 13   & 10\%  & \cite{CMS-HIG-12-044}  \\
\cline{2-6}
         &       \multirow{2}{*}{$\mathrm{tt}\PH$-tag} &  (\(\ell\) with
         4, 5 or \(\ge\)6 jets) $\times$ (3 or \(\ge\)4 b-tags);    &
         6 &   & \multirow{2}{*}{\cite{CMS-HIG-12-035}}  \\
         &                                &  (\(\ell\) with 6 jets with 2 b-tags); (\(\ell \ell\) with 2 or \(\ge\)3 b-jets)
         &  3         &                             &
         \\
\hline
$\cPZ\Pgg$ & inclusive & $(\Pe\Pe, \, \mu\mu) \times (\gamma)$ & 2 &
& \cite{CMS-PAPERS-HIG-13-006}\\
\hline
$\Pgm\Pgm$ & 0/1-jets & $\Pgm\Pgm$ & 12 &  \multirow{2}{*} {1-2\%}
& \multirow{2}{*}{\cite{GianottiManganoVirdee,dimu1,dimu2}} \\
    & VBF-tag & $\Pgm\Pgm+(jj)_\mathrm{VBF}$ & 3 & & \\
\hline
invisible & ZH-tag & $(\Pe\Pe, \, \mu\mu) \times (\mathrm{MET})$ & 2 &
&
\cite{CMS-HIG-13-018} \\
\hline

\end{tabular}
\end{center}
\end{table}

\subsection{Signal Strength}
\begin{figure*}[h]
\centering
\includegraphics[width=0.48\textwidth]{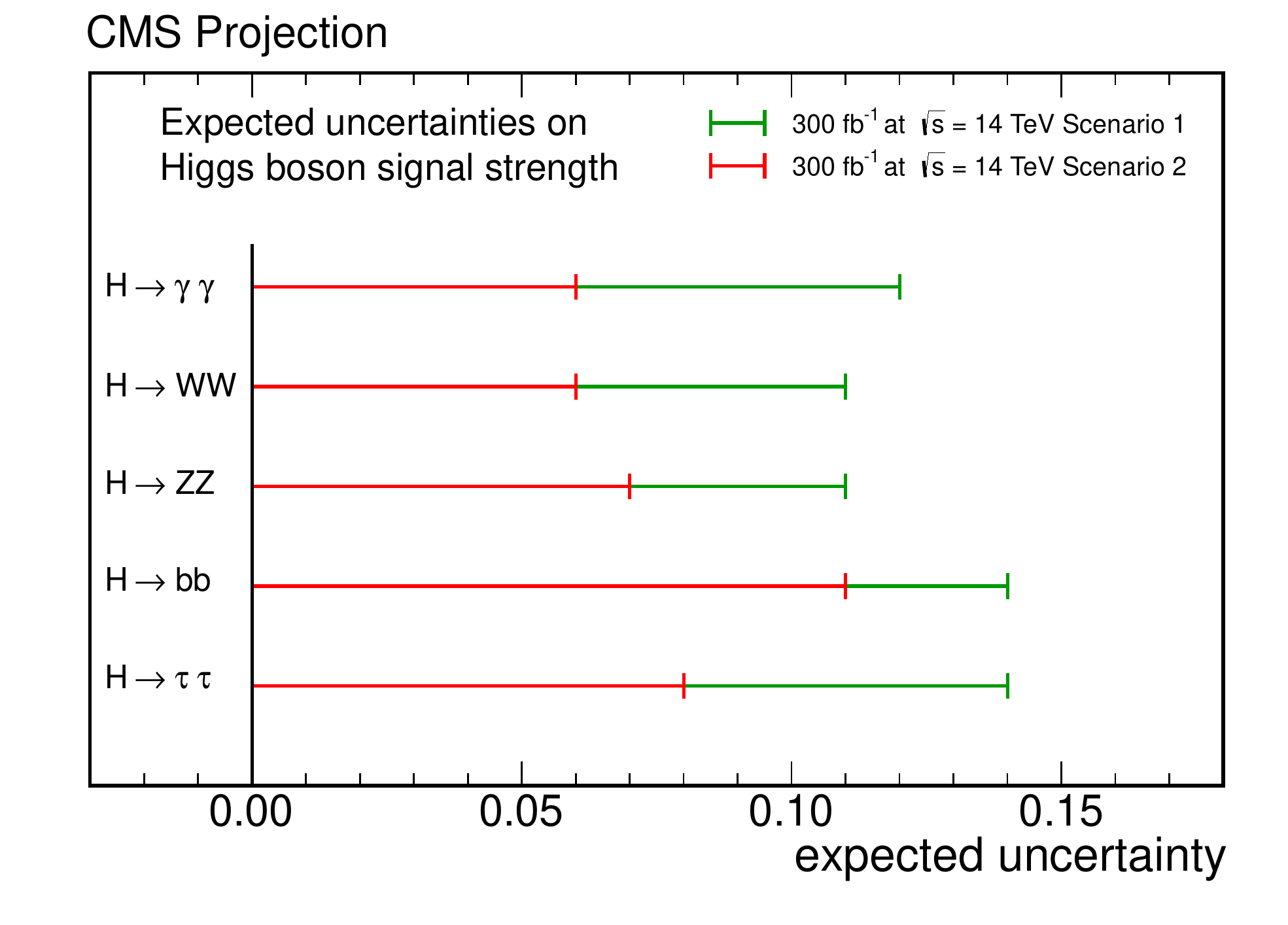}
\includegraphics[width=0.48\textwidth]{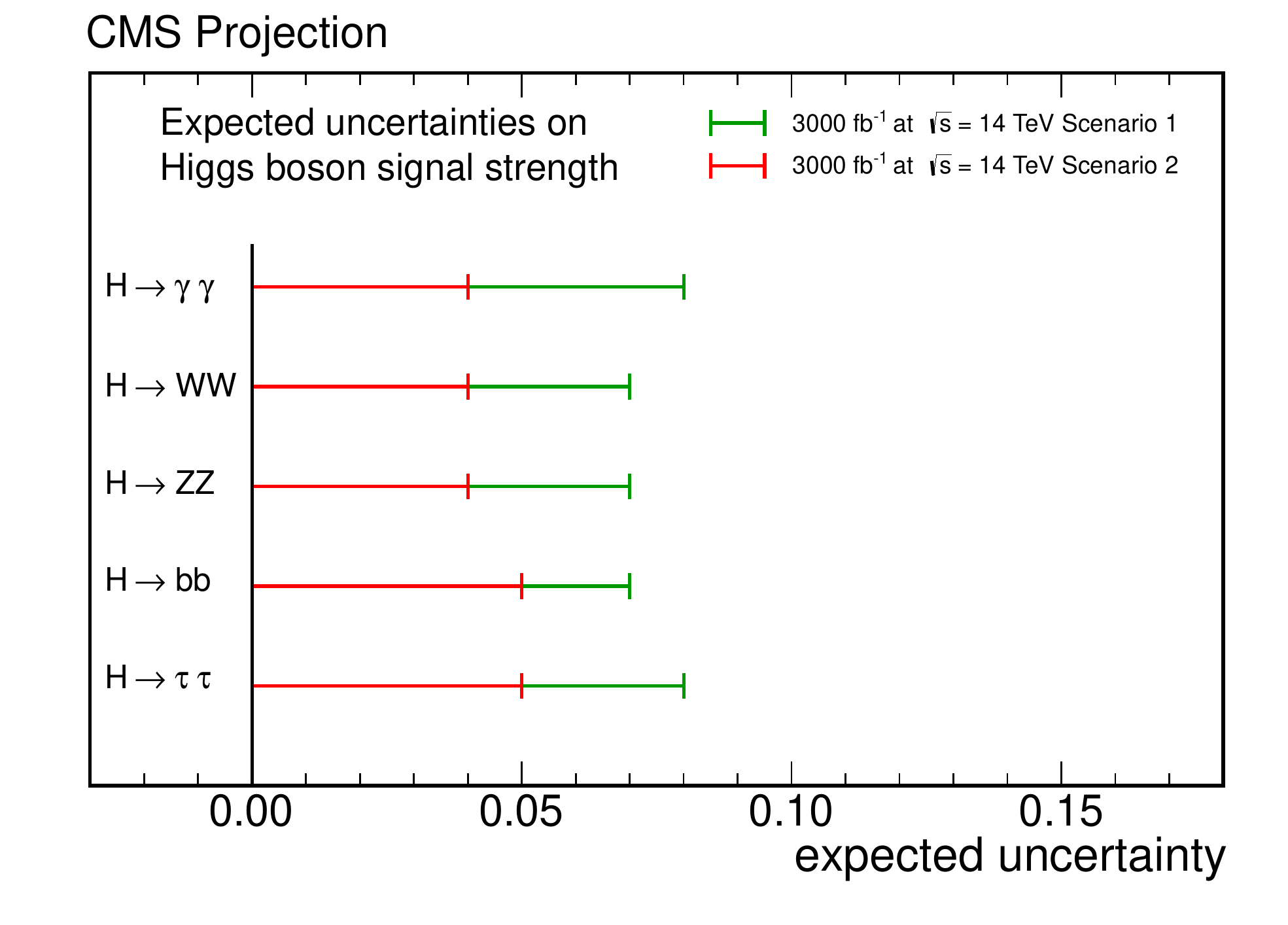}
\caption{Estimated precision on the measurements of the signal strength for a
SM-like Higgs boson.  The projections assume $\sqrt{s} = 14$~TeV and an integrated dataset of 300 fb$^{-1}$ (left) and 3000 fb$^{-1}$ (right). The projections are obtained with the two uncertainty scenarios described in the text.}
\label{fig:mufit}
\end{figure*}

The signal strength modifier $\mu = \sigma / \sigma_{\mathrm{SM}}$,
obtained in the combination of all search channels, provides a first
compatibility test. Figure~\ref{fig:mufit} and Table~\ref{tab:mufit} show the $\mu$ uncertainties obtained in different sub-combinations of search channels, organized by decay mode for an integrated dataset
of 300 fb$^{-1}$ and 3000 fb$^{-1}$.  We predict a precision $6$--$14\%$ for
300 fb$^{-1}$ and $4$--$8\%$ for a dataset of 3000 fb$^{-1}$. Studies
show that  future measurements of the signal strength will be limited by
theoretical uncertainty of the signal cross
section, which is included in the fit. Figure~\ref{fig:cfitnotheo} (left)
shows the uncertainty on the signal strength omitting the
uncertainties from QCD scale and PDFs for signal and background.

\begin{table}[h]
\centering
\caption {Precision on the measurements of the signal strength per decay mode for a
SM-like Higgs boson.
These values are obtained at $\sqrt{s} = 14$~TeV using an integrated dataset of 300 and 3000 fb$^{-1}$.
Numbers in brackets are \% uncertainties on the measurements estimated under [Scenario2, Scenario1],
as described in the text.  For the direct search for invisible Higgs
decays the $95\%$ CL on the branching fraction is given.}
\begin{tabular} {|c|c|c|c|c|c|c|c|c|}
\hline
L (fb$^{-1}$)  &  $\Pgg\Pgg$ &  $\PW\PW$ &  $\cPZ\cPZ$ & $\cPqb\cPqb$ &  $\Pgt\Pgt$ &  $\cPZ\Pgg$ & $\Pgm\Pgm$ & inv. \\ \hline
300               & [6, 12] &  [6, 11] &  [7, 11] &  [11, 14] &  [8, 14] &
[62, 62] & [40,42] & [17, 28] \\ \hline
3000             & [4, 8]   &  [4, 7] &  [4, 7] &  [5, 7] &  [5, 8]
& [20, 24] & [20,24] & [6, 17] \\ \hline
\end{tabular}
\label {tab:mufit}
\end{table}

The direct search for invisible Higgs decays in events
produced in association with a Z boson yields a $95\%$ confidence level
upper limit on the branching fraction of $28\,(17)\%$ for Scenario 1 and
$17\,(6.4)\%$ for Scenario 2 for $300\,(3000)\fbinv$.

\subsection{Coupling-Modifier Fit}

The event yield for any (production)$\times$(decay) mode
is related to the production cross section and the partial and
total Higgs boson decay widths via the narrow-width approximation:

\begin{equation}
\left(\sigma\cdot\text{BR}\right)(\mathit{x}\to\PH\to\mathit{ff}) = \frac{\sigma_{\mathit{x}}\cdot\Gamma_{\mathit{ff}}}{\Gamma_{\mathrm{tot}}},
\end{equation}
where $\sigma_{\mathit{x}}$ is the production cross section through the
initial state $\mathit{x}$, $\Gamma_{\mathit{ff}}$ is the partial decay width into
the final state $\mathit{ff}$, and $\Gamma_{\mathrm{tot}}$ is the total
width of the Higgs boson. In particular, $\sigma_\mathrm{gg\PH}$, $\Gamma_{\mathrm{gg}}$, and
$\Gamma_{\gamma\gamma}$ are generated by quantum loops and are directly
sensitive to the presence of new physics.
The possibility of Higgs boson decays to BSM particles,
with a partial width $\Gamma_{\mathrm{BSM}}$, is
accommodated by keeping $\Gamma_{\mathrm{tot}}$ as a dependent parameter so that
$\Gamma_{\mathrm{tot}} = \sum \Gamma_{ii} +
\Gamma_{\mathrm{BSM}}$, where the $\Gamma_{ii}$ stand for the partial width of
decay to all SM particles.
The partial widths are proportional to the square of the effective Higgs boson
couplings to the corresponding particles.
To test for possible deviations in the data from the rates expected in the different channels
for the SM Higgs boson, factors $\kappa_{i}$ corresponding to the
coupling modifiers are introduced and fit to the data~\cite{LHCHiggsCrossSectionWorkingGroup:2012nn}.

\begin{figure*}[h]
\centering
\includegraphics[width=0.48\textwidth]{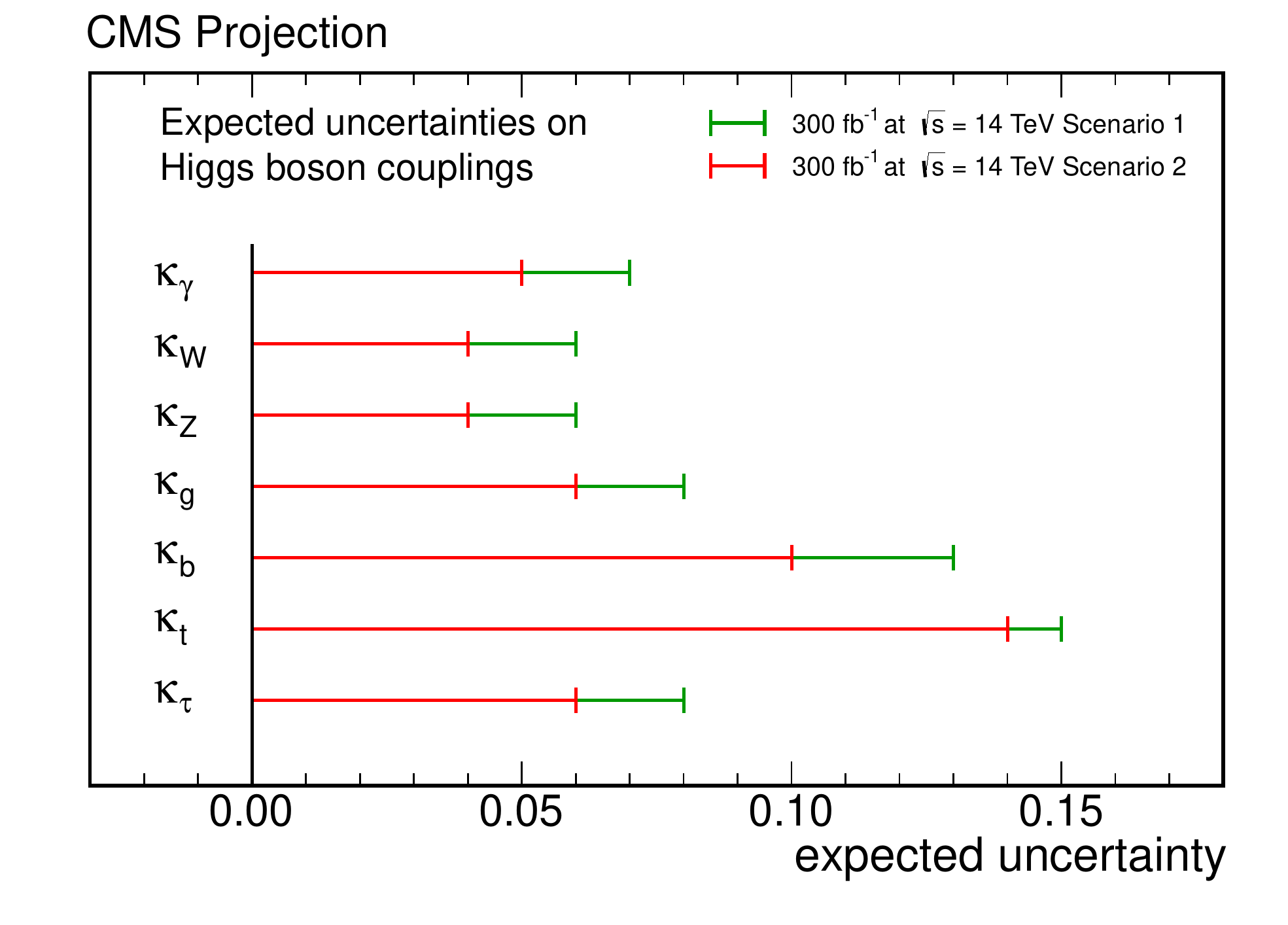}
\includegraphics[width=0.48\textwidth]{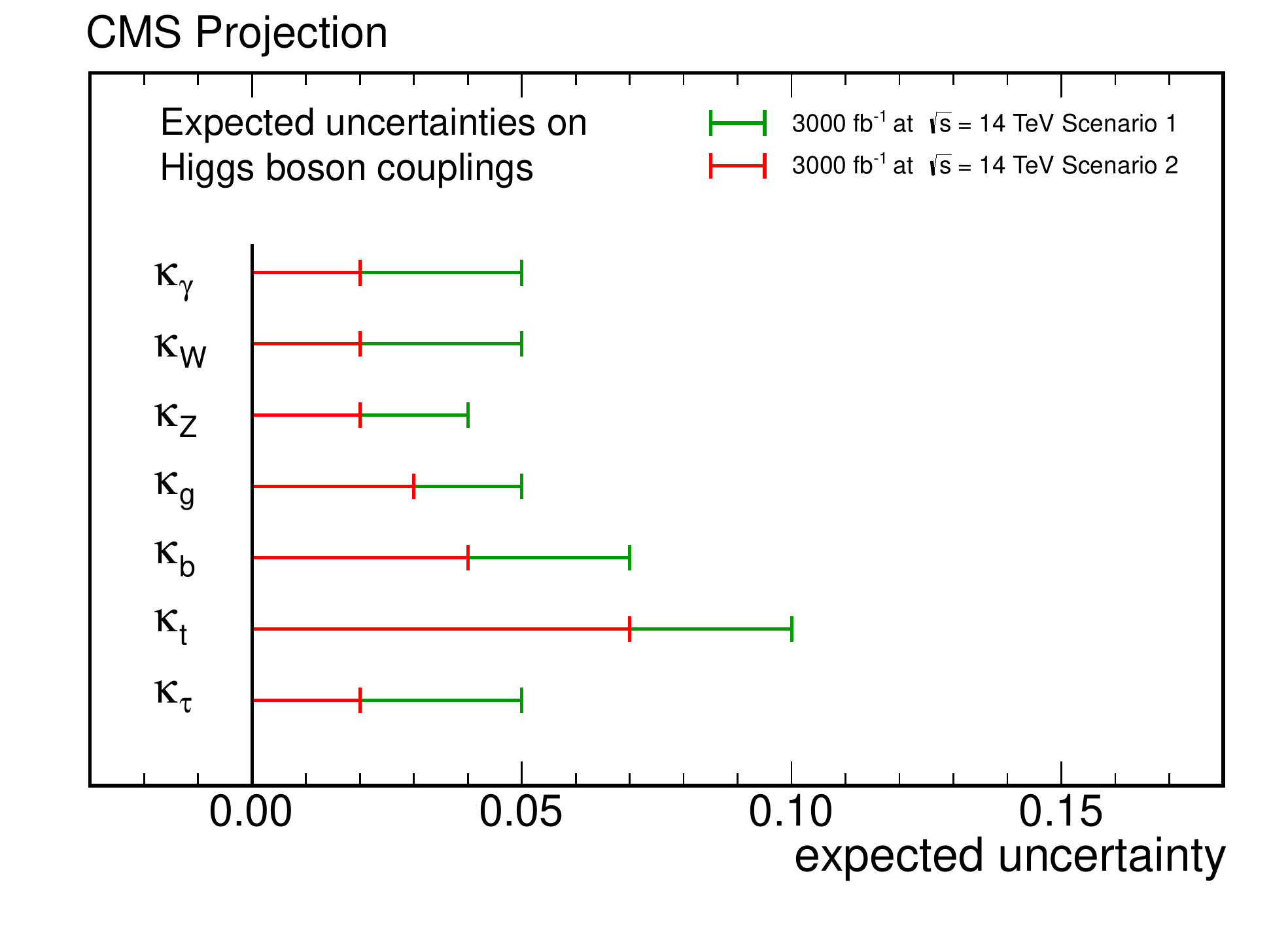}
\caption{Estimated precision on the measurements of $\kappa_\gamma$, $\kappa_W$, $\kappa_Z$, $\kappa_g$, $\kappa_b$, $\kappa_t$ and $\kappa_\tau$.
The projections assume $\sqrt{s} = 14$~TeV and an integrated dataset of 300 fb$^{-1}$ (left) and 3000 fb$^{-1}$ (right). The projections are obtained with the two uncertainty scenarios described in the text.}
\label{fig:cfit}
\end{figure*}

\begin{figure*}[h]
\centering
\includegraphics[width=0.48\textwidth]{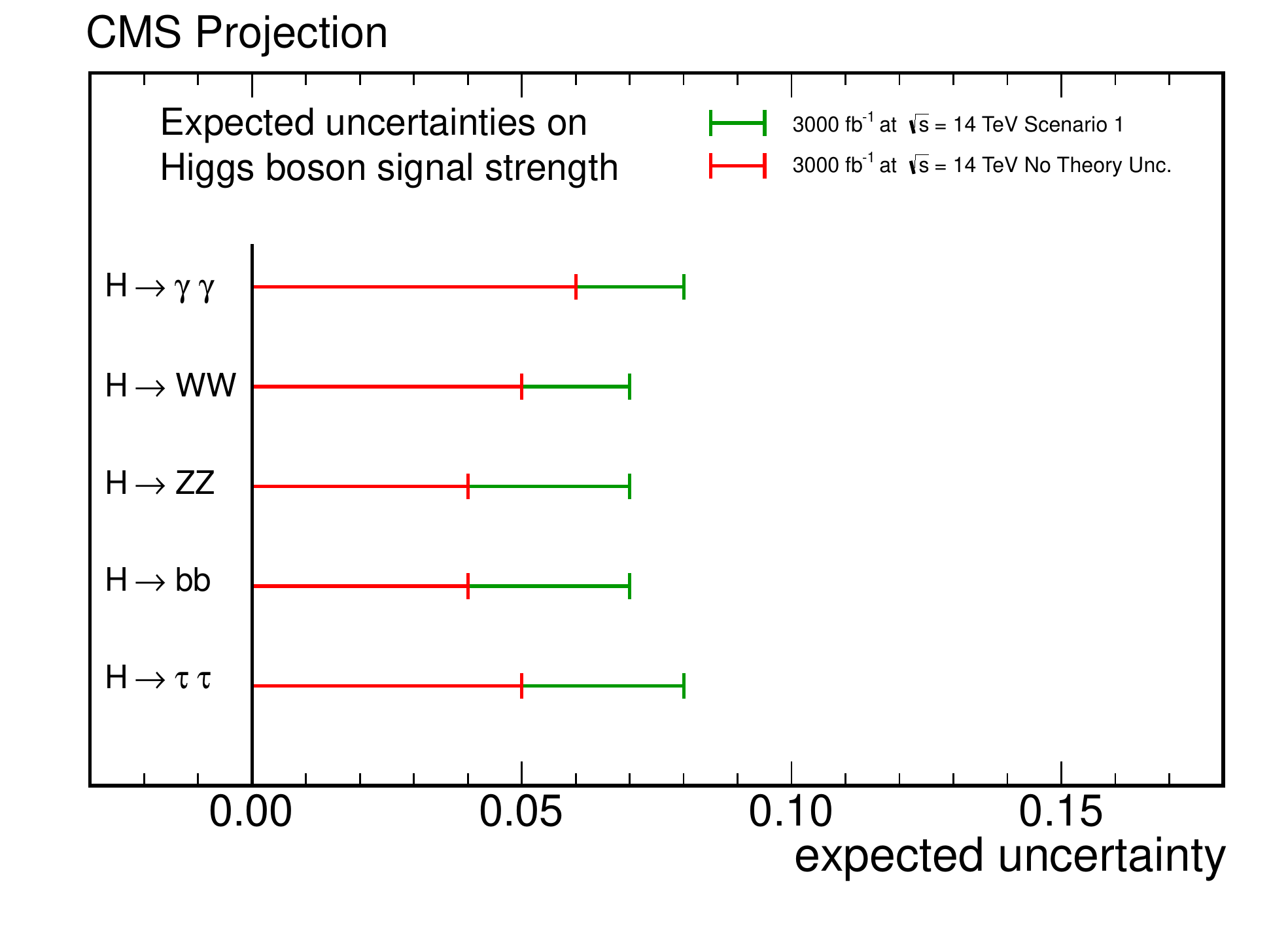}
\includegraphics[width=0.48\textwidth]{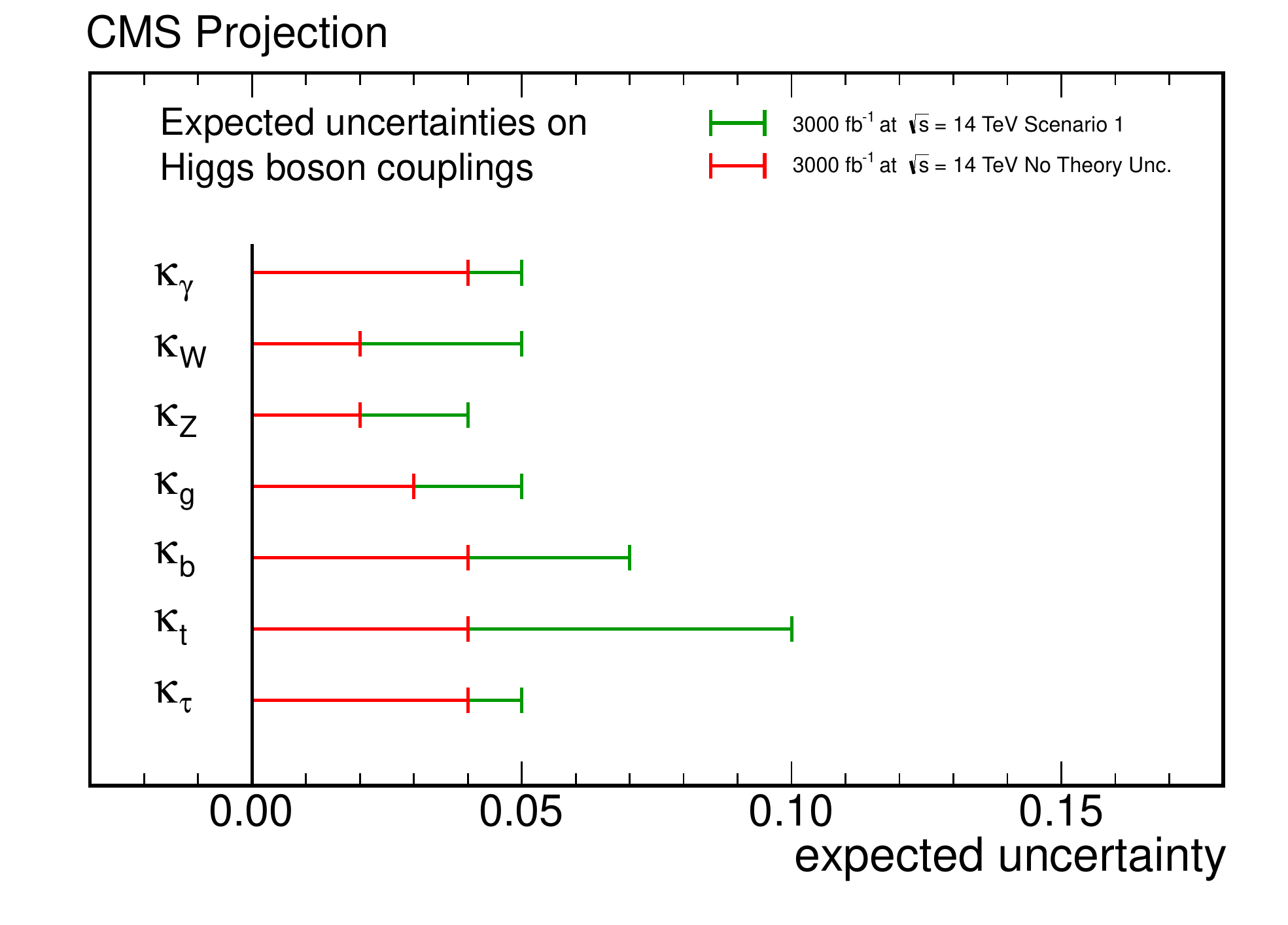}
\caption{Estimated precision on the signal strengths (left) and coupling modifiers (right).
The projections assuming $\sqrt{s} = 14$~TeV, an integrated dataset
of 3000 fb$^{-1}$ and Scenario 1 are compared with a projection neglecting
theoretical uncertainties.}
\label{fig:cfitnotheo}
\end{figure*}

Figure~\ref{fig:cfit} and Table~\ref{tab:cfit} show the uncertainties
obtained on $\kappa_{i}$ for an integrated dataset
of $300\fbinv$ and $3000\fbinv$.  The expected precision ranges from
$5$--$15\%$ for $300\fbinv$ and $2$--$10\%$ for a dataset of $3000\fbinv$.
The measurements will be limited by systematic uncertainties on the cross
section, which is included in the fit for the signal strength. The
statistical uncertainties on $\kappa_{i}$ are below one percent. As
for the results on the signal strength, to
illustrate the importance of theoretical uncertainties, a
fit was performed without considering theoretical systematics.
The results are shown in Fig.~\ref{fig:cfitnotheo}.

\begin{figure*}[h]
\centering
\includegraphics[width=0.48\textwidth]{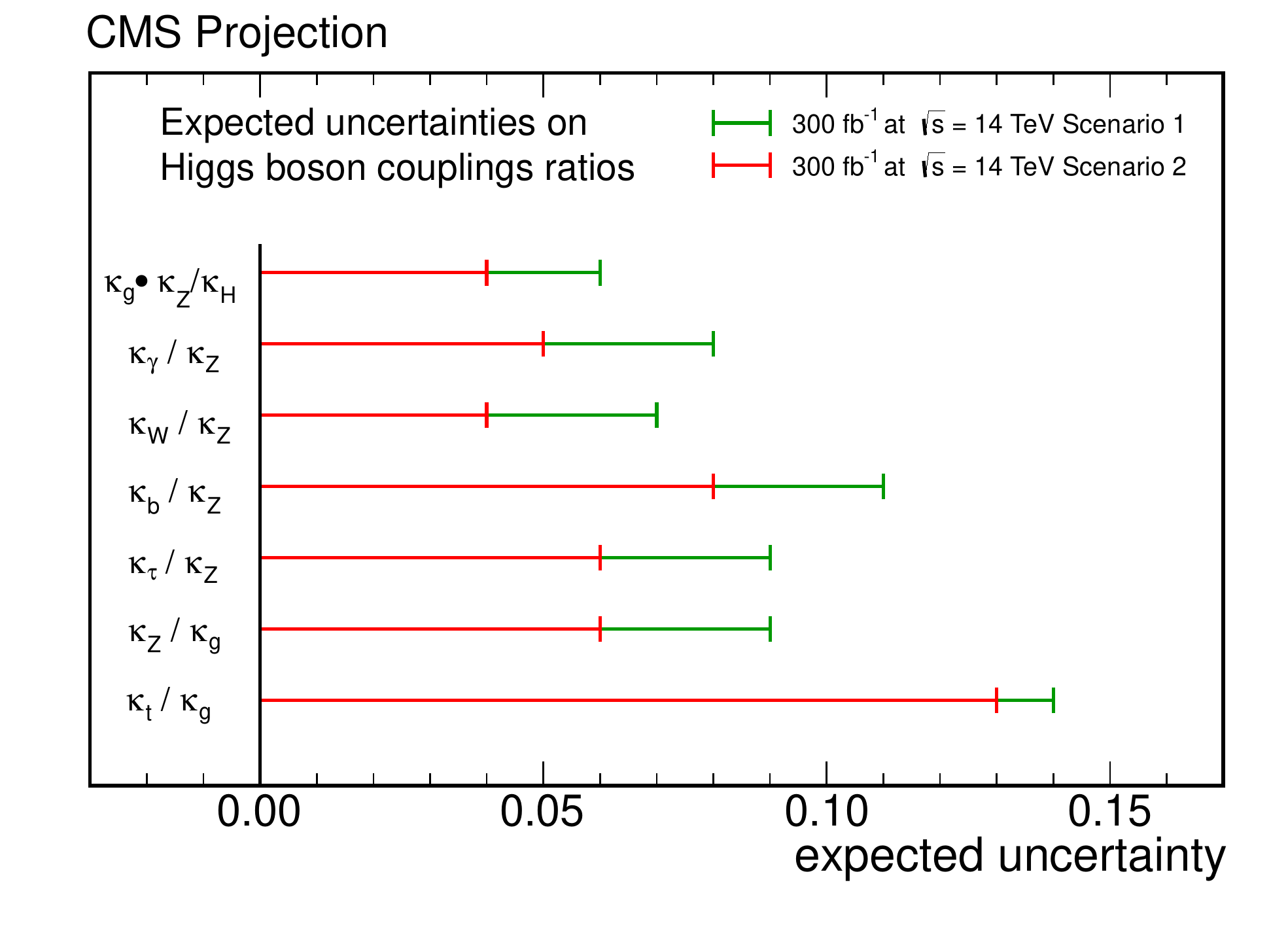}
\includegraphics[width=0.48\textwidth]{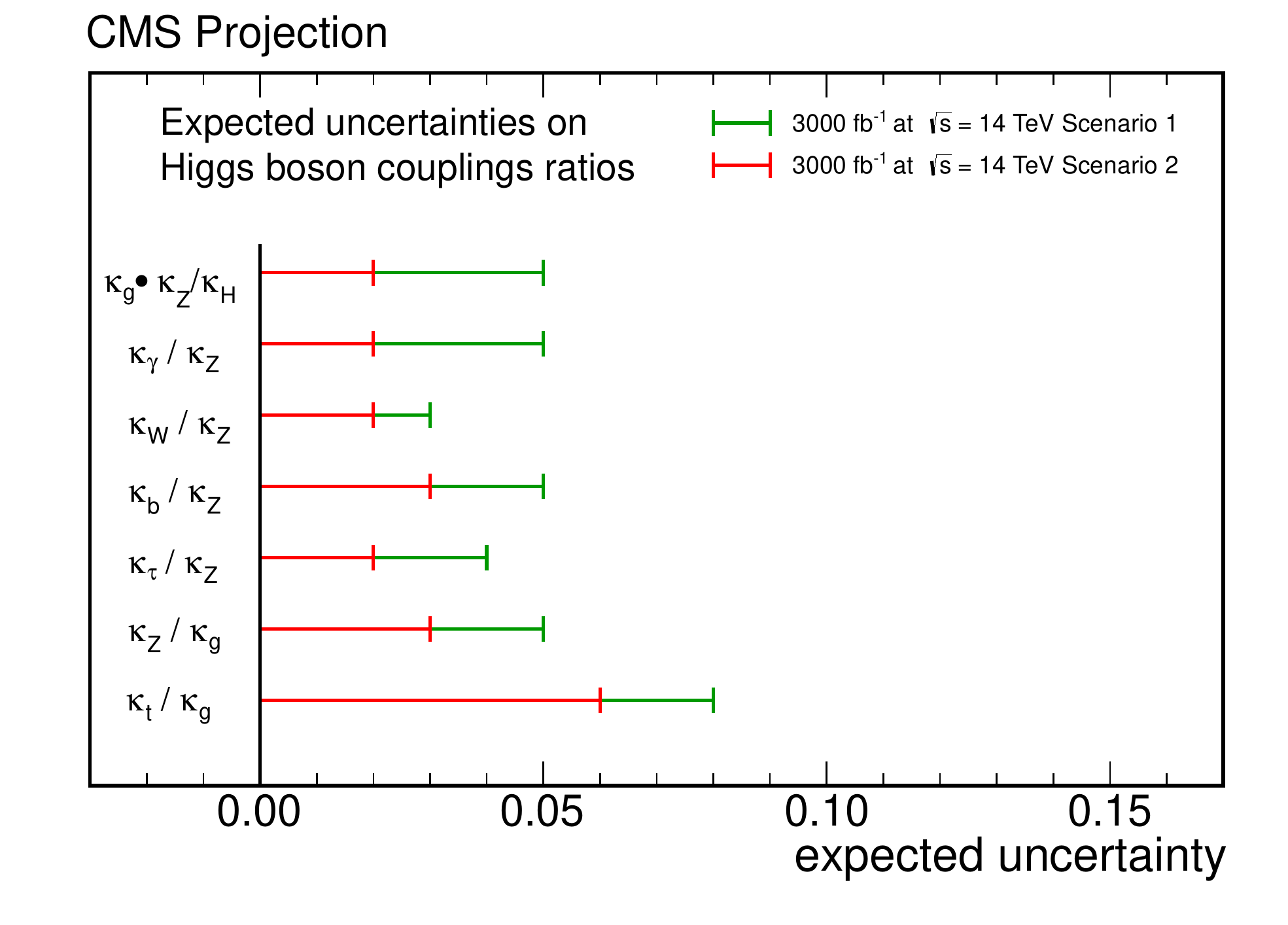}
\caption{Estimated precision on the measurements of ratios of Higgs
  boson couplings (plot shows ratio of partial width. It will be
  replaced by a plot of ratio of couplings by the time of the
  pre-approval. Uncertainties are 1/2). The projections assume $\sqrt{s} = 14$~TeV and an integrated dataset of 300 fb$^{-1}$ (left) and 3000 fb$^{-1}$ (right). The projections are obtained with the two uncertainty scenarios described in the text.}
\label{fig:cfratio}
\end{figure*}

The likelihood scan versus
$\mathrm{BR}_{\mathrm{BSM}}=\Gamma_{\mathrm{BSM}}/\Gamma_{\mathrm{tot}}$
yields a 95\% CL of the invisible BR of 18 (11) \% for Scenario 1 and
14 (7) \% for Scenario 2 for 300 (3000) fb$^{-1}$. This scan assumes
that the coupling to the W and Z boson are equal to or smaller than the SM values.
Fits for ratios of Higgs boson couplings do not require assumptions on
the total width or couplings to the W and Z boson. The results are
shown in Figure~\ref{fig:cfratio} and Table~\ref{tab:ratio}.

The measurement of couplings can be extended to first- and
second-generation fermions. Previous studies have shown that the Higgs
decay to a pair of muons can be observed in gluon-gluon fusion and via
vector-boson fusion production~\cite{GianottiManganoVirdee,dimu1,dimu2}.
The dimuon events can be observed as a narrow resonance over a falling background
distribution. The shape of the background can be parametrized and
fitted together with a signal model. Assuming the current performance
of the CMS detector, we confirm these studies and estimate a measurement of the
h${\mu\mu}$ coupling with a precision of 8\%, statistically limited in
3000 fb$^{-1}$.\\
\begin{table}[h]
\centering
\caption {Precision on the measurements of $\kappa_\gamma$, $\kappa_W$, $\kappa_Z$, $\kappa_g$, $\kappa_b$, $\kappa_t$ and $\kappa_\tau$.
These values are obtained at $\sqrt{s} = 14$~TeV using an integrated
dataset of 300 and 3000 fb$^{-1}$. Numbers in brackets are \%
uncertainties on couplings for [Scenario 2, Scenario 1] as described in
the text. For the fit including the possibility of Higgs boson decays to BSM particles
d the $95\%$ CL on the branching fraction is given.}
\begin{tabular} {|c|c|c|c|c|c|c|c|c|c|c|}
\hline
L (fb$^{-1}$)  &  $\kappa_\gamma$ & $\kappa_W$ & $\kappa_Z$ &
$\kappa_g$ & $\kappa_b$ & $\kappa_t$ &$ \kappa_\tau$ &$
\kappa_{\cPZ\Pgg}$ & $\kappa_{\Pgm\Pgm}$ & BR$_\mathrm{SM}$ \\ \hline
300               & [5, 7] &  [4, 6] &  [4, 6] &  [6, 8] &  [10, 13] &
[14, 15] &  [6, 8]  & [41, 41] & [23, 23] & [14, 18] \\ \hline
3000             & [2, 5] &  [2, 5] &  [2, 4] &  [3, 5] &
[4, 7]   &  [7, 10] &  [2, 5]  & [10, 12] & [8, 8] & [7, 11] \\ \hline
\end{tabular}
\label {tab:cfit}
\end{table}

\begin{table}[h]
\centering
\caption {Estimated precision on the measurements of ratios of Higgs
  boson couplings (plot shows ratio of partial width. It will be
  replaced by a plot of ratio of couplings by the time of the
  pre-approval. Uncertainties are 1/2).
These values are obtained at $\sqrt{s} = 14$~TeV using an integrated dataset of 300 and 3000 fb$^{-1}$.
Numbers in brackets are \% uncertainties on the measurements estimated under [scenario2, scenario1],
as described in the text. }
\begin{tabular} {|c|c|c|c|c|c|c|c|c|c|}
\hline
L (fb$^{-1}$)  &  $\kappa_{g}\cdot\kappa_{\cPZ} / $ $\kappa_{H}$ &  $\kappa_{\gamma} / \kappa_{\cPZ}$ &  $\kappa_W
/ \kappa_{\cPZ}$ &  $\kappa_{b} / \kappa_{\cPZ}$ &  $\kappa_{\tau} /
\kappa_{\cPZ}$ &  $\kappa_{\cPZ} / \kappa_{g}$ &  $\kappa_{t} /
\kappa_{g}$ & $\kappa_{\mu} / \kappa_{\cPZ}$ & $\kappa_{\cPZ\gamma} / \kappa_{\cPZ}$\\ \hline
300               & [4,6] &  [5,8] &  [4,7] &  [8,11] &  [6,9] & [6,9] & [13,14]  & [22,23] & [40,42] \\ \hline
3000             & [2,5] &  [2,5] &  [2,3] &  [3,5]   &  [2,4] & [3,5] & [6,8] & [7,8] & [12,12] \\ \hline
\end{tabular}
\label {tab:ratio}
\end{table}

\subsection{Spin-parity}
Besides testing Higgs couplings, it is important to determine the spin and
quantum numbers of the new particle as accurately as possible.  The full case
study has been presented by CMS with the example
of separation of the SM Higgs boson model and the pseudoscalar
($0^-$)~\cite{CMS-HIG-13-002}. Studies on the prospects of measuring
CP-mixing of the Higgs boson are presented using the  H$\rightarrow ZZ^{*} \rightarrow 4l$ channel.
The decay amplitude for a spin-zero boson defined as
\begin{equation}
A(H \to ZZ) = v^{-1} \left (
  a_1 m_{Z}^2 \epsilon_1^* \epsilon_2^*
+ a_2 f_{\mu \nu}^{*(1)}f^{*(2),\mu \nu}
+ a_3  f^{*(1)}_{\mu \nu} {\tilde f}^{*(2),\mu  \nu}
\right )\,.
\label{eq:fullampl-spin0}
\end{equation}
where $f^{(i),{\mu \nu}}$ ($ {\tilde f}^{(i),{\mu \nu}}$) is
the (conjugate) field strength tensor of a $Z$ boson with polarization
vector $\epsilon_i$ and $v$ the vacuum expectation value of the Higgs
field. The spin-zero models $0^+$ and $0^-$ correspond to the terms with $a_1$ and $a_3$,
respectively.

Four independent real numbers describe the process in Eq.~(\ref{eq:fullampl-spin0}),
provided that  the overall rate is treated separately and one overall complex phase is not measurable.
For a vector-boson coupling, the four independent parameters can be represented by two fractions
of the corresponding cross-sections ($f_{a2}$ and $f_{a3}$)
and two phases ($\phi_{a2}$ and $\phi_{a3}$).
In particular, the fraction of $C\!P$-odd contribution is defined under the assumption $a_2=0$ as
\begin{eqnarray}
&& f_{a3} =  \frac{|a_{3}|^2\sigma_3}{| a_{1}|^2\sigma_1+ |a_{3}|^2\sigma_3},
\nonumber
\label{eq:fractions1}
\end{eqnarray}
where $\sigma_i$ is the effective cross section of the process corresponding to $a^{}_{ i}=1, a_{j \ne i}=0$.
Given the measured value of  $f_{a3}$, the coupling constants can be extracted in any parameterization.
For example, following Eq.~(\ref{eq:fullampl-spin0}) the couplings will be
\begin{eqnarray}
&&  \frac{|a_3|}{|a_1|}=
\sqrt{\frac{f_{a3}}{(1-f_{a3})}} \times \sqrt{\frac{\sigma_1}{\sigma_3}},
\nonumber
\label{eq:fractions2}
\end{eqnarray}
where ${\sigma_1}/{\sigma_3}=6.240$ for a Higgs boson with a mass of $126\GeV$.

\begin{figure}[htbp]
\begin{center}
\centerline{
\includegraphics[width=0.5\linewidth]{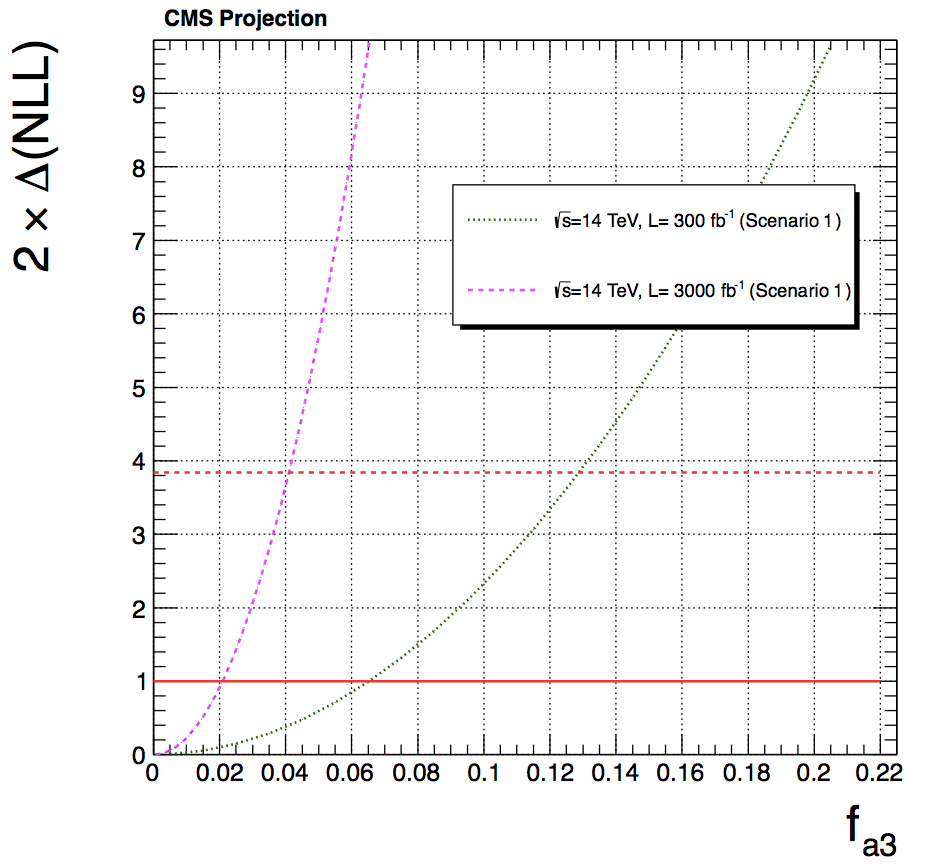}
}
\caption{
Distribution of expected $-2\ln{\cal L}$ for $f_{a3}$ for the projection to 300 fb$^{-1}$ (green, dotted) and 3000 fb$^{-1}$ (magenta,
dot-dashed).
}
\label{fig:fa3}
\end{center}
\end{figure}

A fit is performed on the parameter $f_{a3}$, which is
effectively a fraction of events (corrected for reconstruction efficiency) corresponding to
the $0^-$ contribution in the (${\cal D}_{0^-}, {\cal D}_{\rm bkg} $) distribution.
Projections of the expected $-2\ln{\cal L}$ values from the fits assuming $300\fbinv$ and
$3000\fbinv$ are shown in Fig.~\ref{fig:fa3}.
A 68\% (95\%) CL limit on the contribution of $f_{a3}$ can be achieved at the level of $0.07\,(0.13)$
with $300\fbinv$ and $0.02\,(0.04)$ with $3000\fbinv$.
The analysis is limited by statistical uncertainties up to a high luminosity, but all sources of systematic
uncertainties are preserved in the projections.

\section{Discovery Potential: Supersymmetry}\label{susy}

After the observation of a Higgs boson at the LHC, the question about the large 
quantum corrections to its mass are more pressing than ever. A natural solution to this
hierarchy problem would be the cancellation of these corrections from new particles
predicted by supersymmetry (SUSY), which have the same quantum numbers as their
SM partners apart from spin.  No evidence for supersymmetric particles has been 
found with the data taken at the LHC with $\sqrt{s}=8\TeV$, but the energy upgrade to 
$14\TeV$ together with higher luminosities will open the possibility to access a new 
interesting mass window in the next years.

Extrapolations of several searches for SUSY by
CMS~\cite{SUS-13-007,SUS-12-024,SUS-13-011,SUS-13-013,ss-hcp,SUS-13-006} are performed
by scaling the luminosity and taking into account the change of cross section with higher 
energy accordingly. The projections are made based on 8\TeV Monte Carlo samples and
without optimizing the selections for searches at higher energies and higher luminosities.
In ``Scenario A'' the signal and background yields, and the uncertainty on the
background, are scaled by the ratio of 
the luminosities ($20\fbinv$ for $8\TeV$ and $300\fbinv$ for $14\TeV$) and by the ratio of 
the cross sections for signal and background $(\sigma_{\rm sig}$ and $\sigma_{\rm bkg}$):
\begin{equation} \label{eq:susy_extrapolation}
R_{\rm sig(bkg)} = \frac{300\fbinv}{20\fbinv} \times 
\frac{\sigma_{\rm sig(bkg)}(14\TeV)}{\sigma_{\rm sig(bkg)}(8\TeV)}.
\end{equation}
For some analyses, a less conservative scenario, called ``Scenario B,'' is defined
where the relative uncertainty on the background is reduced. 
These are similar to the Scenarios 1 and 2 used in the Higgs projections
discussed in Sec.~\ref{higgs}, but not the same in detail and
have different implications for SUSY searches where higher mass regions will be progressively
searched in the future.
The exact procedures differ slightly in projections for different SUSY models and are 
described in detail in the following sections.

\begin{figure}[htbp]
  \centering
    \includegraphics[width=0.6\textwidth]{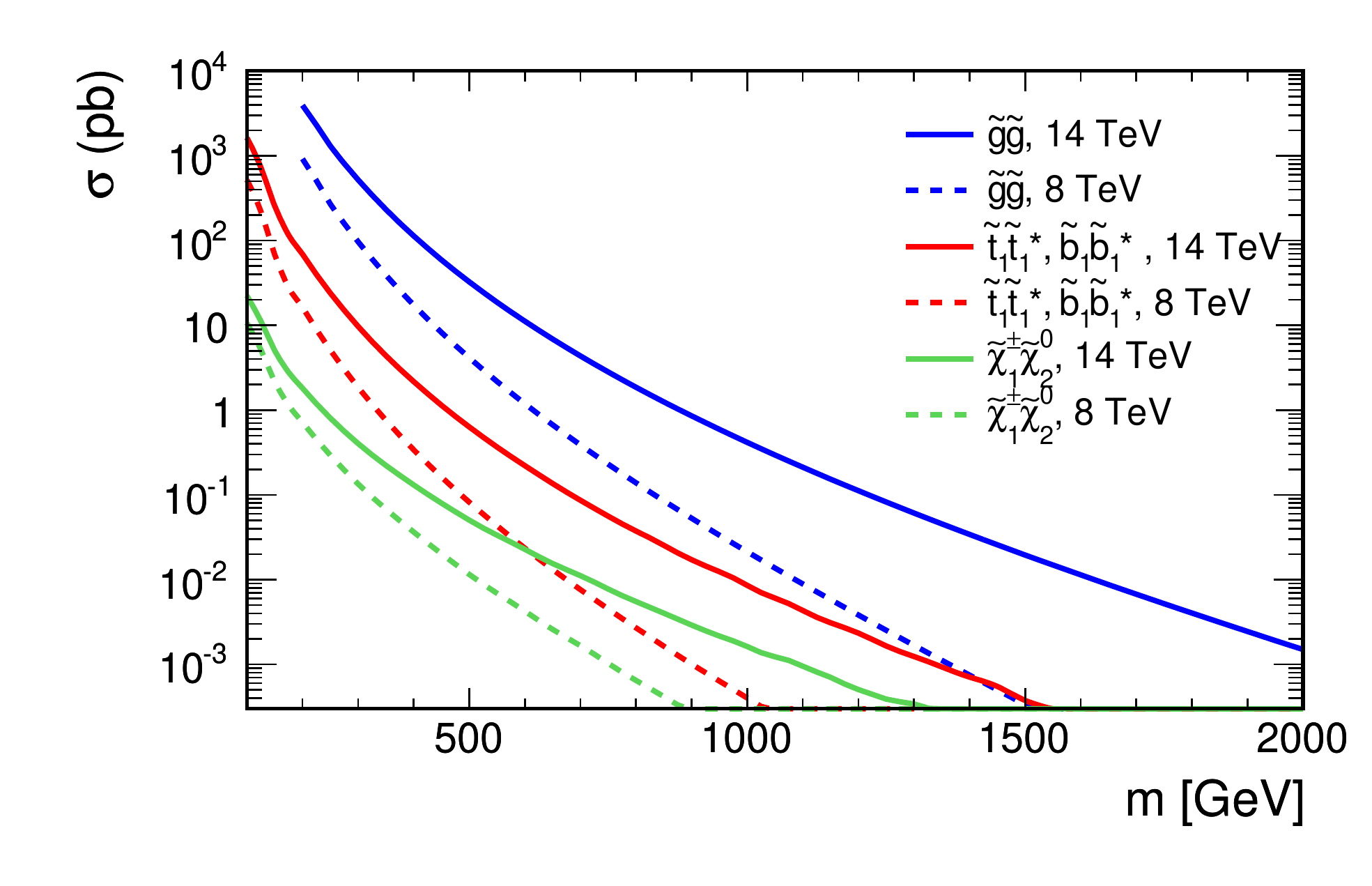}
  \caption{Next-to-leading order cross sections for gluino-pair production,
    stop-pair (sbottom-sbottom) production, and chargino-neutralino
    production versus the mass of the pair-produced SUSY particles.
    The chargino-neutralino production cross section is presented for common
    $m_{\chi_1^\pm}=m_{\chi_2^0}$ masses.
    }
  \label{fig:susy_xsec}
\end{figure}

The following models, assuming $100\%$ branching fractions, are considered: gluino-pair production
with each gluino decaying to the lightest SUSY particle (LSP) and either a $\ttbar$ or $\bbbar$ pair; 
direct stop production with each stop decaying to a top quark and LSP; chargino-neutralino production 
with final states containing W and Z bosons and missing transverse energy; and direct sbottom 
production with decay to chargino and top quark.  The cross sections for these SUSY particle production 
processes, computed at the next-to-leading-order accuracy in $\alpha_s$ using 
Prospino2~\cite{Beenakker:1996ch,Beenakker:1997ut,Beenakker:1999xh},
are shown in Fig.~\ref{fig:susy_xsec}.

\subsection{Gluino-Pair Production with Four Top Quarks in the Final State}
\label{sec:ra4b_projection}
\begin{figure}[htbp]
\centering
\subfigure[\label{fig:t1tttt_topology}]{
        \includegraphics[width=0.38\textwidth, viewport=1 -50 220 200]{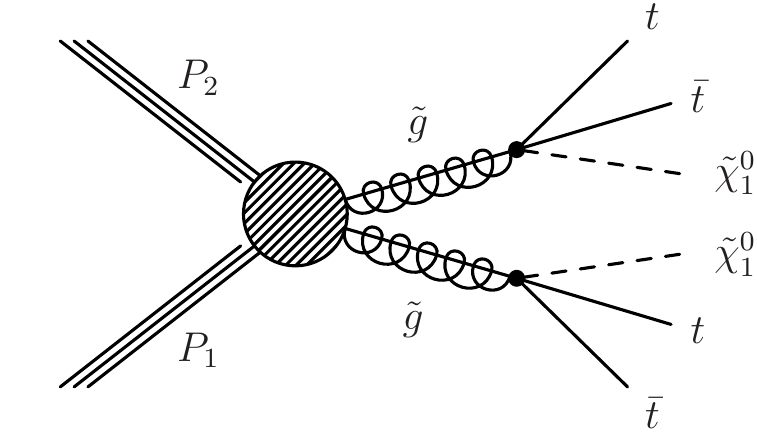}}
\subfigure[\label{fig:ra4_projection_plot}]{
        \includegraphics[width=0.58\textwidth]{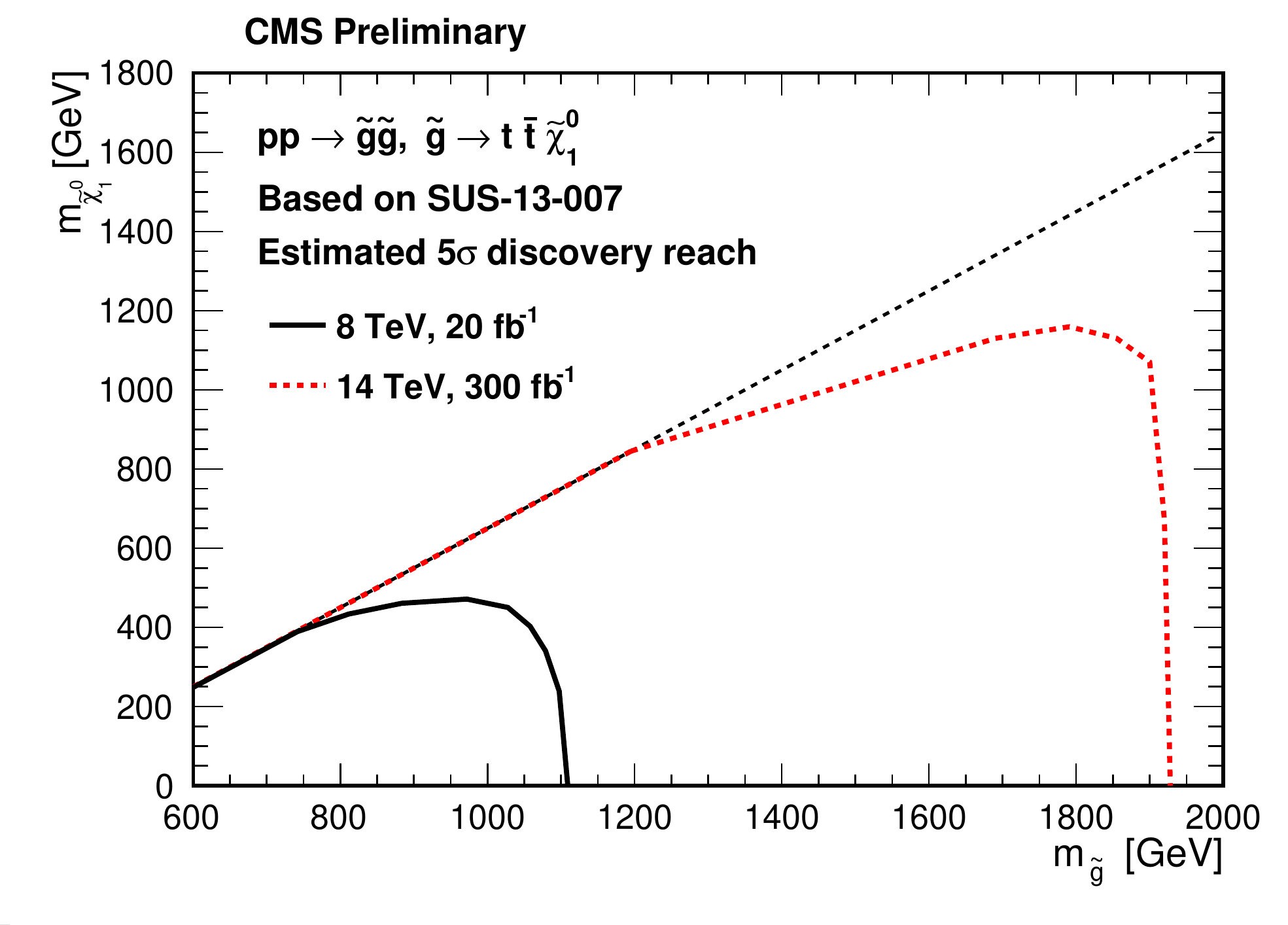}} \\
\caption{(a) The simplified model topology for gluino production,
             where the gluinos decay to two top quarks and an LSP each,
             and (b) the projected 5$\sigma$ discovery reaches for this model.}
\label{fig:ra4_projections}
\end{figure}

Naturalness predicts not only light third-generation masses, but also gluinos that are not
much heavier than a$\TeV$. In this case they could decay to third-generation 
squarks. This section focuses on gluino-pair production, where each gluino
decays to a top and a stop quark that then decays to a top quark and the LSP.
This is described by a simplified model, where pair-produced gluinos
each decay to a \ttbar pair and the LSP (see Fig.~\ref{fig:t1tttt_topology}).
Due to the presence of four W bosons in the final state, a search in the single lepton 
final state has a large branching fraction ($\sim 40\%$) and good sensitivity. 
Hence, the sensitivity to this simplified model topology is projected to $14\TeV$ 
based on the results obtained in the SUSY search in the single-lepton channel~\cite{SUS-13-007},
performed in pp collisions at a center-of-mass energy of $\sqrt{s}=8\TeV$ and
corresponding to an integrated luminosity of $19.4\fbinv$.

The numbers of signal and background events are scaled from the $8\TeV$ analysis
based on Eq.~\eqref{eq:susy_extrapolation}.  As the background is dominated by $\ttbar$ 
production, it is scaled up based on the $\ttbar$ cross section ratio between 
$14\TeV$ and $8\TeV$, which is about a factor of $3.9$.
For Scenario A, the same relative systematic uncertainties as for the $8\TeV$ analysis are kept,
which is a conservative assumption.  Nevertheless, the dominant uncertainty of the 
analysis is the statistical uncertainty from the control regions used for 
the background estimation, which is reduced by $1/\sqrt{R_{\rm bkg}}$.
Thus, even a more aggressive treatment of the systematic uncertainties
would not lead to a sizable improvement on the sensitivity.

The expected significance is calculated using the profile likelihood method
and the signal Monte Carlo samples generated with \PYTHIA~6~\cite{Sjostrand:2006za}
with a CMS custom underlying event tuning~\cite{pythia-z2}. 
Figure~\ref{fig:ra4_projection_plot} shows the 5$\sigma$ significance line
in the 2-dimensional plane of neutralino versus gluino mass for the different scenarios investigated.
Gluino masses up to $\sim1.9$\TeV for neutralino masses around 0.9\TeV or less 
can be discovered at 14 TeV with an integrated luminosity of 300~fb$^{-1}$.
It should be noted that the current results are obtained without performing any optimization 
on the current analysis~\cite{SUS-13-007}, and further improvements in the sensitivities
are expected by re-optimizing the analysis selection for the different scenarios.

\subsection{Gluino-Pair Production with Four Bottom Quarks in the Final State}
\label{sec:ra2b_projection}
\begin{figure}[htbp]
\centering
\subfigure[\label{fig:t1bbbb_topology}]
  {\includegraphics[width=0.38\textwidth, viewport=1 -50 220 200]{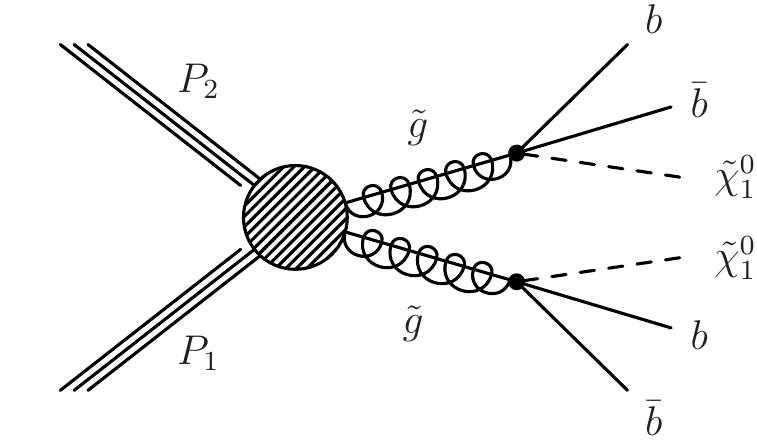}}
\subfigure[\label{fig:ra2_projection_plot}]         
  {\includegraphics[width=0.58\textwidth]{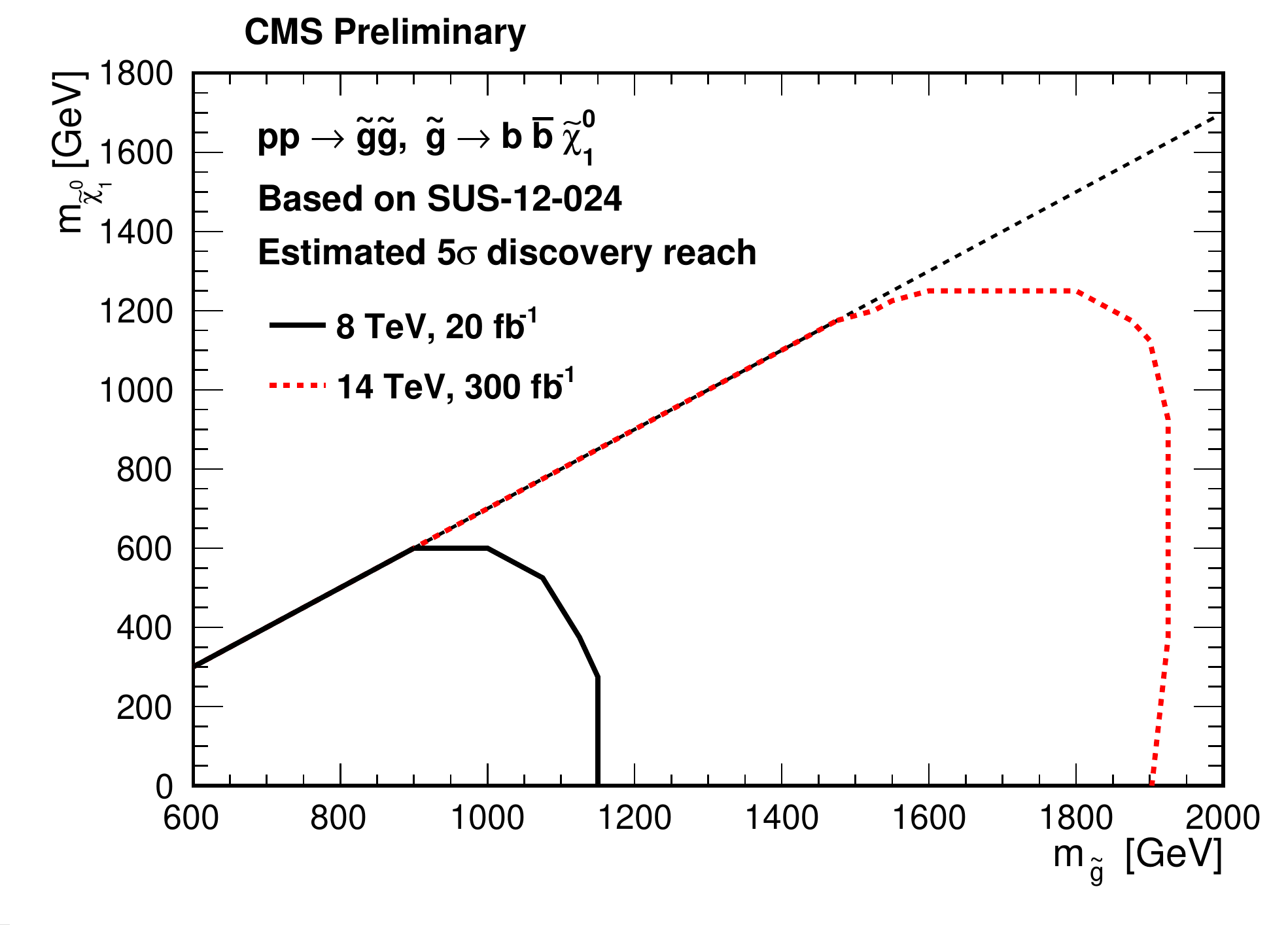}} \\
\caption{The simplified model topology for gluino production, 
where the gluinos decay to two bottom quarks and an LSP each (a),
and the projected 5$\sigma$ discovery reaches for this model (b).}
\label{fig:ra2_projections}
\end{figure}

Similar to the gluino decay to four top quarks and two LSPs in the previous section, 
one can also investigate a model for gluino-pair production, where each gluino decays 
to \bbbar and the LSP (see Fig.~\ref{fig:t1bbbb_topology}).  The projection of the 
sensitivity for 14\TeV is studied based on the results of the search in events with 
multiple jets, large missing transverse energy, and b tags~\cite{SUS-12-024}.

The signal and background yields are scaled based on the cross section ratios for
the different center-of-mass energies, and the luminosity increase.  The systematic 
uncertainty is conservatively kept the same as for the $8\TeV$ analysis, 
corresponding to the Scenario A described above. 
The 
signal samples produced with \PYTHIA~6~\cite{Sjostrand:2006za} are 
used for this projection. 
Figure~\ref{fig:ra2_projection_plot} shows the projected 
5$\sigma$ discovery reaches
for this model. This analysis is 
sensitive to the discovery of gluinos with masses up to $1.9\TeV$ for LSP masses of $1.2\TeV$ or below.

For a center-of-mass energy of $14\TeV$ and an integrated luminosity of $300\fbinv$,
we expect more than 100 background events in the highest sensitivity bin of the analysis 
. An actual analysis would
be designed in a way that the background is kept smaller, enhancing the sensitivity to a
possible signal. Therefore, the given limit can be considered as conservative.

\subsection{Stop-Pair Production}
\label{sec:t2tt_projection}
One of the most pressing questions for the next run of the LHC is whether third-generation squarks can 
be observed. 
Light stop quarks, with masses less than $\sim1$\TeV, are required to
cancel the large radiative corrections to the Higgs mass from the top
quark without large fine-tuning.
One possible production mechanism is the decay of (light) gluinos to stops and sbottoms,
if they are lighter than the gluinos and the gluinos are within the LHC reach with $13$--$14\TeV$. 
These models are studied in the previous Secs.~\ref{sec:ra4b_projection}--\ref{sec:ra2b_projection}.
Here, we study the model where the stops are the lightest squarks and are directly produced in pairs.
The extrapolation is based on the result obtained from a search in final states with a muon or 
electron~\cite{SUS-13-011}.  This analysis has a discovery reach for stop masses of $300$--$500\GeV$ 
and a maximum neutralino mass of $75\GeV$ for a center-of-mass energy of $8\TeV$ and an integrated 
luminosity of $20\fbinv$.

The projections to higher energy and luminosity are based on the $8\TeV$ Monte Carlo simulated samples
produced with the \MADGRAPH~5~\cite{MadGraph} simulation program.  For Scenario A, the signal and background 
yields, as well as the uncertainty on the background, are scaled by the ratios $R_{\rm sig}$ and $R_{\rm bkg}$, 
respectively (Eq.~\eqref{eq:susy_extrapolation}).
The cross sections 
for direct stop production are enhanced for $14\TeV$ by a factor of $\sim 4$--$20$ for
stop masses of $200$--$1000\GeV$.  The main background consists of $\ttbar$ events, which are 
scaled by the cross section ratio. 
The ratio of the cross sections for the second highest background, W+jets, is smaller than $\ttbar$, 
leading to a conservative background estimation.
The signal extrapolation is done in the same way for the less conservative Scenario B, but the 
uncertainty on the background is reduced by $1/\sqrt{R_{\rm bkg}}$, as it is assumed that the 
uncertainty is largely driven by the statistical precision from the control samples, which will improve
with more data. Nevertheless, a fixed lower limit on the relative uncertainty of at least $10\%$ is kept.

\begin{figure}[htbp]
\begin{center}
  \subfigure[\label{fig:t2tt_topology}]{ \includegraphics[width=0.38\textwidth, viewport=1 -50 210 200]{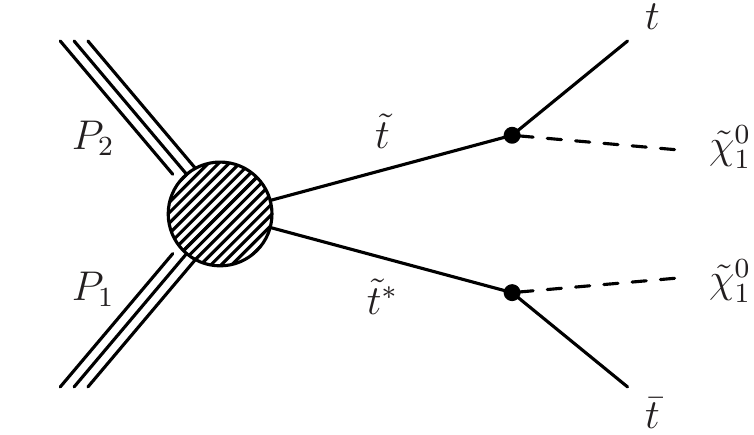} }
  \subfigure[\label{fig:t2tt_projection}]{ \includegraphics[width=0.58\textwidth]{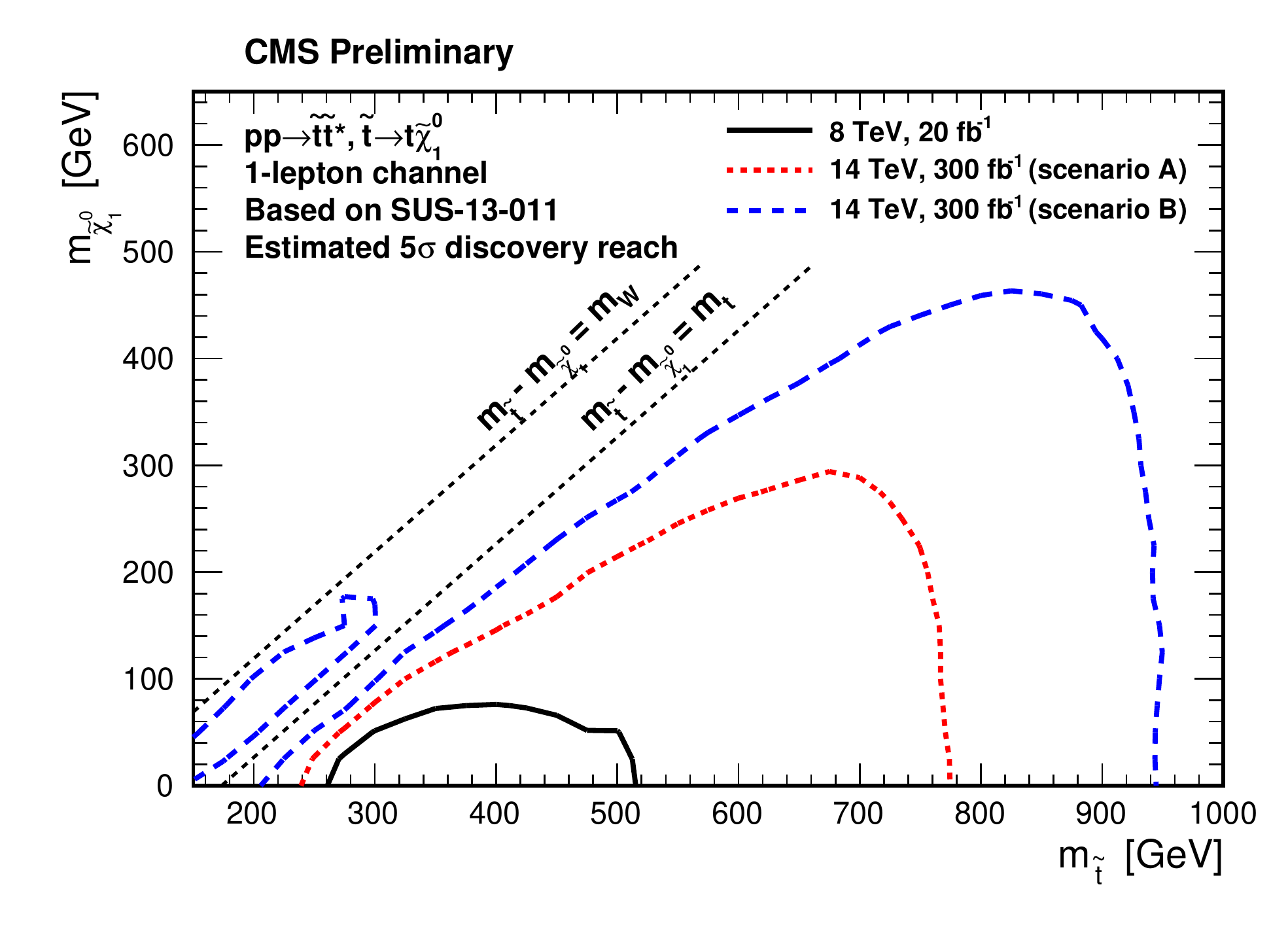} }
\caption{The simplified model topology direct stop production, where the stops decay to a top quark and an LSP each (a),
and the projected 5$\sigma$ discovery reaches for this model (b).}
\label{fig:t2tt_14TeV}
\end{center}
\end{figure}

The results are summarized in Fig.~\ref{fig:t2tt_14TeV}. 
A discovery reach for stop masses of $750$--$950\GeV$, and LSP masses of $300$--$450\GeV$, is expected.
More stringent selection requirements could suppress the background further, 
leading to an improvement of the signal-to-background ratio and discovery potential.
Also, when searching for stop signals at higher masses, many top quarks from stop decays 
are highly boosted, but the use of the boosted top taggers are not yet
explored to gain extra sensitivity.

\subsection{Sbottom-Pair Production with Four W Bosons and Two Bottom Quarks in the Final State}
\label{sec:ssb_projection}
Here, a model is considered where sbottom quarks are relatively light and are directly produced in pairs.  
The corresponding simplified model assumes that a sbottom quark decays solely to a top quark 
and a chargino, with the chargino subsequently decaying to a W and the LSP.  The model 
considered here additionally assumes mass splittings such that the top and W are on-shell.
The extrapolation is based on the result obtained from a search in a final state with a same-sign lepton pair, 
jets, b-tagged jets, and missing transverse energy~\cite{SUS-13-013}.

The background is considered to be composed of two components --- one from rare SM processes 
producing genuine same-sign lepton pairs and another consisting of processes where at least one lepton 
comes from a jet, hereafter referred to as a fake isolated lepton.  These two components comprise over $95\%$ of the 
background to searches for strongly produced new physics in the same-sign dilepton final state, 
with rare SM processes contributing $50$--$80\%$ depending on the search region.
The rare SM background consists mainly of processes producing multiple weak bosons or top quarks in the final state, 
with the largest contribution coming from the production of a \ttbar pair in association with a W boson.  
The background containing fake isolated leptons arises mostly from \ttbar events, where one prompt lepton originates from a W boson
and the other lepton comes from the decay of a b quark.

We scale the number of signal events $N_{\rm sig}$ by the ratio $R_{\rm sig}$ as defined in Eq.~\eqref{eq:susy_extrapolation}.
The signal is simulated with \MADGRAPH. 
The signal cross section increases from 8 to 14\TeV
approximately by a factor of 5 to 12 for sbottom masses between 300 and $700\GeV$.
The fake background yield $N_{\rm fake}$ 
and the rare SM background yield $N_{\rm rare}$
are also scaled by Eq.~\eqref{eq:susy_extrapolation}.
The scaling of $N_{\rm fake}$ is based on the \ttbar cross section ratio, and the scaling of $N_{\rm rare}$ 
is based on the \ttbar{}W cross section ratio of 3.3 between 14 and
8\TeV~\cite{Campbell:2012dh}.

\begin{figure}[htbp]
\begin{center}
  \subfigure[]{ \includegraphics[width=0.38\textwidth, viewport=1 -50 210 200]{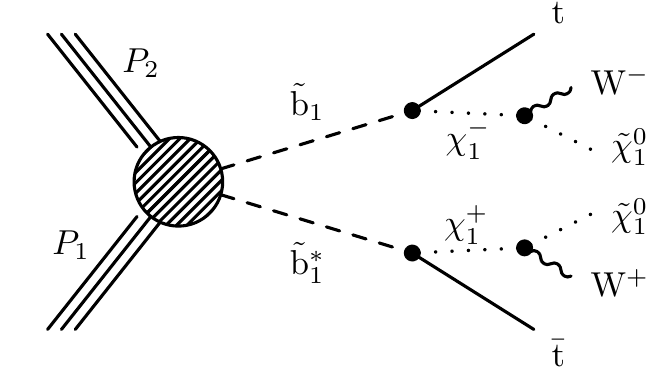} }
  \subfigure[]{ \includegraphics[width=0.58\textwidth]{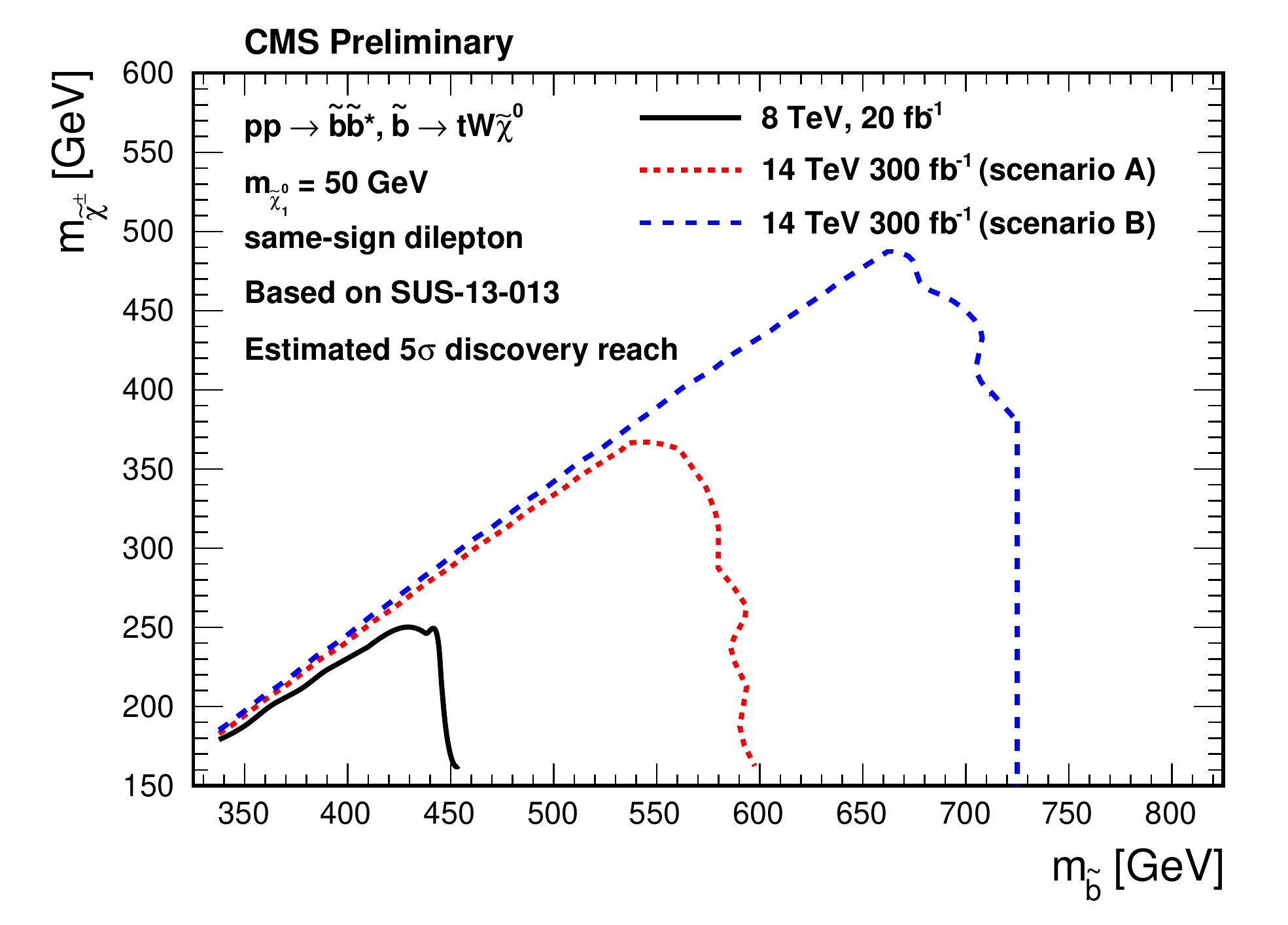} }
\caption{The simplified model topology for direct sbottom production, where the sbottoms decay
to a top quark and a chargino each, and the chargino
decays to a W boson and a LSP (a),
and the projected 5$\sigma$ discovery reaches for this model (b).}
\label{fig:t6ttww}
\end{center}
\end{figure}

The uncertainty on each component of the background, $\sigma_{\rm rare}$ and $\sigma_{\rm fake}$, is comprised of a 50\% 
systematic uncertainty and a statistical component.  For Scenario A, the uncertainties remain the 
same as for the $8\TeV$ analysis, except for the statistical uncertainty on the fake prediction, which is scaled down by the
square-root of the luminosity and cross section increase, as this uncertainty is driven purely by the fakeable object count
in the isolation sideband.
For Scenario B,
the signal extrapolation is done in the same way, but the 
systematic uncertainty on the rare SM background is reduced from 50\% to 30\%, as it can be assumed 
that the cross sections and kinematic properties of these processes will be measured and better understood.  
The systematic uncertainty on the fake background is reduced from 50\% to 40\%.

Figure~\ref{fig:t6ttww} shows the topology of the investigated simplified model and the $5 \sigma$ discovery region, 
which is extended up to sbottom masses of $600$--$700\GeV$ and LSP masses up to $350\GeV$.

\subsection{Chargino-Neutralino Production with Decays to a Z Boson}
\label{sec:tchiwz_projection}
With higher luminosities, the searches for the electroweak SUSY particles may become increasingly more important.
Charginos and neutralinos can be produced in cascade decays of gluinos and squarks
or directly via electroweak interactions, and,
in the case of heavy gluinos and squarks, gauginos would be produced dominantly via electroweak interactions.
Depending on the mass spectrum,
the charginos and neutralinos can have significant decay branching fractions to leptons or on-shell vector bosons,
yielding multilepton final states.
Here the projections of the discovery reach for
direct production of $\tilde{\chi}_1^{\pm}$ and $\tilde{\chi}_2^0$, which decay via 
W and Z bosons into the LSP ($\tilde{\chi}_1^0$)~\cite{SUS-13-006}, are presented. 
This production becomes dominant if sleptons are too massive and $\tilde{\chi}_1^{\pm}$ and $\tilde{\chi}_2^0$
are wino-like, which suppresses neutralino-pair production relative to neutralino-chargino production.

\begin{figure}[htbp]
\begin{center}
  \subfigure[]{ \includegraphics[width=0.38\textwidth, viewport=1 -50 210 200]{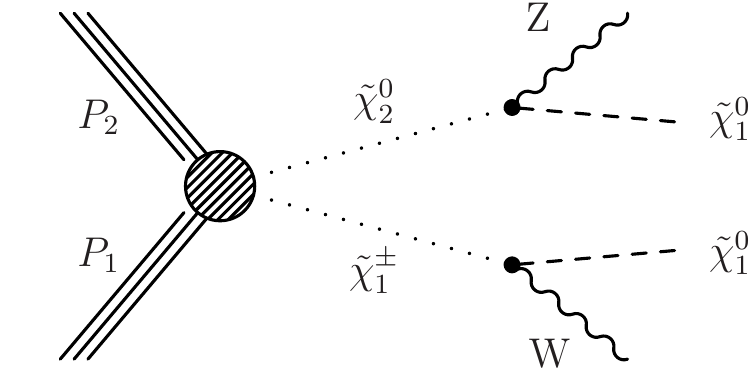} }
  \subfigure[]{ \includegraphics[width=0.58\textwidth]{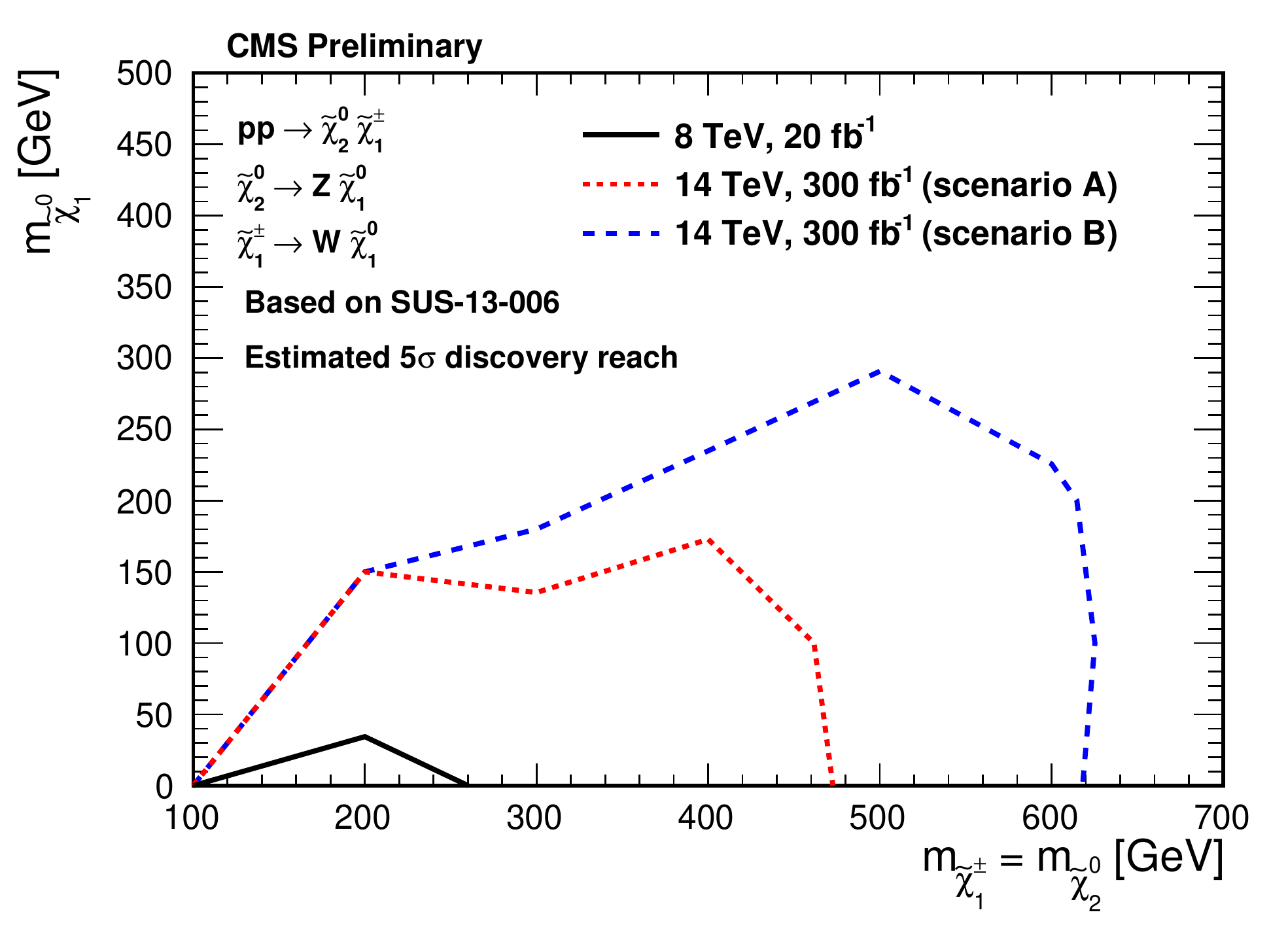} }
\caption{The simplified model topology for direct $\tilde{\chi}_1^{\pm}\tilde{\chi}_2^0$ production decaying to the WZ+\MET final state (a),
and the projected 5$\sigma$ discovery projections for this model (b).}
\label{fig:Tchiwz}
\end{center}
\end{figure}

The analysis is based on a three-lepton search, with electrons, muons, and at most one hadronically
decaying $\tau$ lepton. 
In order to get an estimate for the sensitivity at $14\TeV$ two different Scenarios (A and B)
are considered, as discussed earlier.  The results are shown in Fig.~\ref{fig:Tchiwz}. 
The chargino mass sensitivity can be increased to $500$--$600\GeV$, while
discovery potential for neutralinos ranges from $150$ to almost $300\GeV$. 

\subsection{Chargino-Neutralino Production with Decays to a Higgs Boson}
\label{sec:tchiwh_projection}

\newcommand{\chipmo}{\ensuremath{\tilde{\chi}_{1}^{\pm}}}
\newcommand{\chitn}{\ensuremath{\tilde{\chi}_{2}^{0}}}
\newcommand{\lsp}{\ensuremath{\tilde{\chi}_{1}^{0}}}
\newcommand{\mchipmo}{\ensuremath{M_{\chipmo}}}
\newcommand{\mchitn}{\ensuremath{M_{\chitn}}}
\newcommand{\mlsp}{\ensuremath{M_{\lsp}}}
\newcommand{\mchi}{\ensuremath{M_{\tilde{\chi}}}}
\newcommand{\signal}{\ensuremath{ \chipmo\chitn \rightarrow (\PW \lsp)(\PH \lsp)}}
\newcommand{\wl}{\ensuremath{\PW}+light jets} 
\newcommand{\wbb}{\ensuremath{\PW+\bbbar}} 
\newcommand{\wzbb}{$\PW\Z\to\ell\nu\bbbar$}

\begin{figure}[htbp]
\begin{center}
  \subfigure[]{ \includegraphics[width=0.38\textwidth, viewport=1 -50 210 200]{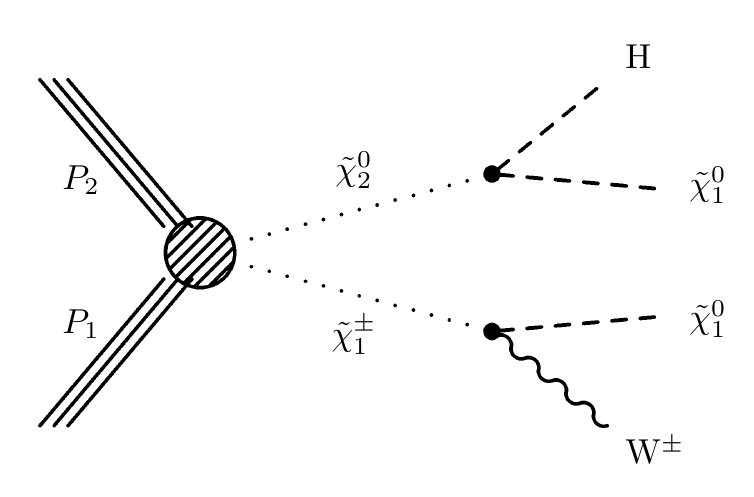} }
  \subfigure[]{ \includegraphics[width=0.58\textwidth]{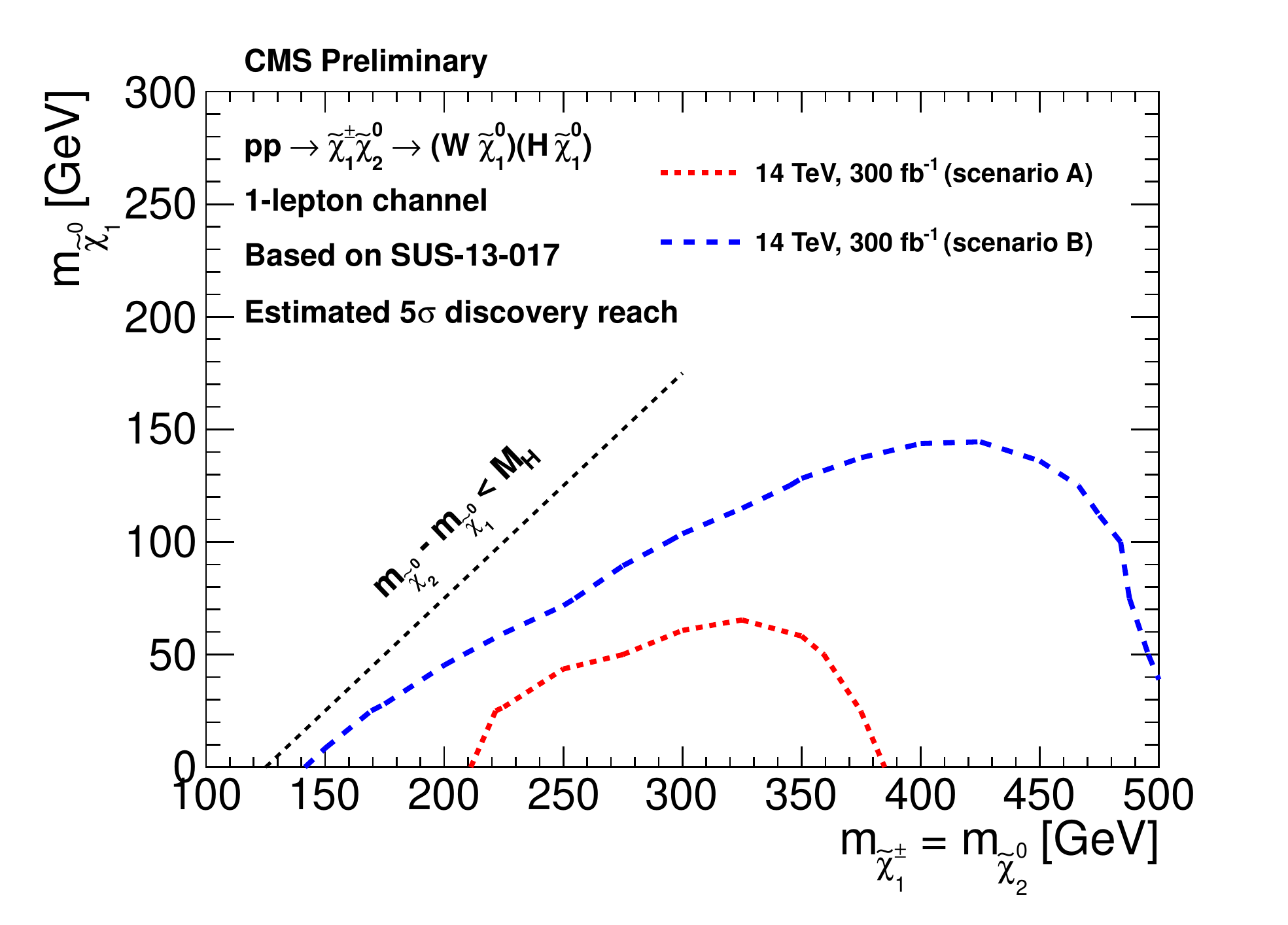} }
\caption{The simplified model topology for direct $\tilde{\chi}_1^{\pm}\tilde{\chi}_2^0$ production 
decaying to the $\PW\PH+\MET$ final state (a),
and the projected 5$\sigma$ discovery projections for this model (b).}
\label{fig:Tchiwh}
\end{center}
\end{figure}

In this section we also consider chargino-neutralino pair production with a signature 
that is similar to the one considered in Sec.~\ref{sec:tchiwz_projection}, except that
here the $\tilde{\chi}_2^0$ instead decays to a Higgs boson and the $\tilde{\chi}_1^0$ LSP.
Hence we target the process $\chipmo\chitn\to(\PW^{\pm}\lsp) (\PH \lsp)$ as indicated in Fig.~\ref{fig:Tchiwh}(a), 
and extrapolate the discovery reach based on the analysis of Ref.~\cite{SUS-13-017}.

The projections are based on the analysis in the single lepton final state, which targets the process
$\chipmo\chitn\to(\PW^{\pm}\lsp) (\PH \lsp) \to \ell\nu b\bar{b}+\MET$. The dominant background in
this search is from \ttbar production; \PW\ bosons produced in association with b-quarks are also relevant. SM backgrounds
are suppressed with requirements on \MET\ and related quantities, and we search for a peak in the
$m_{b\bar{b}}$ mass distribution consistent with $m_H=126$\GeV. For the projections, in the conservative scenario
we assume $\sigma_{\mathrm{syst}}=25$\% as in the current analysis, while in the optimistic scenario we assume a reduction
in the systematic uncertainty by a factor of 2.

The estimated 14 TeV discovery reach is shown in Fig.~\ref{fig:Tchiwh}. 
Sensitivity to charginos and neutralinos with masses up to $400$--$500$\GeV is achieved,
for LSP masses up to $60$--$150$\GeV. Note that realistic models contain a mixture of the decays
$\tilde{\chi}_2^0 \to \Z\lsp$ and $\tilde{\chi}_2^0 \to \PH\lsp$, so the sensitivity lies
between the projections in this section and those in Sec.~\ref{sec:tchiwz_projection}.

\section{Discovery Potential: Exotic New Particles}\label{exo}

\providecommand\Wprime{\PWpr\xspace}
\providecommand\Zprime{\cPZpr\xspace}
\providecommand\MT{\ensuremath{M_\mathrm{T}}\xspace}
\providecommand\MTlower{\ensuremath{M_\mathrm{T}^\mathrm{min}}\xspace}
\providecommand\ET{\ensuremath{E_\mathrm{T}}\xspace}
\providecommand\pT{\ensuremath{p_\mathrm{T}}\xspace}

\providecommand{\stau}{\ensuremath{\sTau_1}\xspace}
\providecommand{\smu}{\ensuremath{\PSgm_1}\xspace}
\providecommand{\sel}{\ensuremath{\PSe_1}\xspace}
\providecommand{\sto}{\ensuremath{\sTop_1}\xspace}
\providecommand{\ias}{\ensuremath{I_{as}}}
\providecommand{\iasp}{\ensuremath{I_{as}^\prime}}
\providecommand{\id}{\ensuremath{I_{d}}}
\providecommand{\ih}{\ensuremath{I_{h}}}
\providecommand{\dedx}{\ensuremath{\cmsSymbolFace{d}E\kern-0.12em/\kern-0.2em\cmsSymbolFace{d}x}\xspace}
\providecommand{\MeVcm}{\ensuremath{{\,\text{Me\hspace{-.08em}V\hspace{-0.16em}/\hspace{-0.08em}}\text{cm}}}\xspace}
\providecommand{\tof}{TOF}
\providecommand{\invbeta}{\ensuremath{1/\beta}\xspace}
\providecommand{\tkonly}{{tracker-only}}
\providecommand{\tktof}{{tracker+\tof}}
\providecommand{\multicharge}{{multiply charged}}
\providecommand{\muononly}{{muon-only}}
\providecommand{\fractionalcharge}{{fractionally charged}}

\providecommand{\mpt}{\,/\!\!\!\!p_{T}}
\providecommand{\ttbar}{{\mathrm{t}\bar{\mathrm{t}}}}
\providecommand{\TTbar}{{\mathrm{T}\bar{\mathrm{T}}}}

In this section the discovery potential for exotic signs of new physics with the 14~\TeV HL-LHC dataset at CMS is explored.  
The benchmark channels presented below include searches for additional gauge bosons (Z$^{\prime}$ and W$^{\prime}$), dark matter in 
the monolepton + MET channel, heavy stable charged particles, and vector-like top partners.

\subsection{Searches for Heavy Gauge Bosons Decaying to Lepton Pairs}

A search for additional heavy gauge bosons decaying to lepton pairs has been performed with the existing 7 and 8~\TeV
datasets~\cite{Chatrchyan:2012oaa}.  In order to project the discovery potential of this search to the HL-LHC scenarios, the
background and signal yields are predicted using generator level simulation parameterized by the efficiencies and resolutions
measured in the 8~\TeV data.  The POWHEG event generator and CT10 PDF sets were used to generate \ttbar and the dominant Drell-Yan
backgrounds, while WW events were generated using \PYTHIA.  Samples of Z$^{\prime}$ events were also
generated using \PYTHIA and
no interference effects were considered.

The same acceptance is assumed as in the 8~\TeV search.  In the electron channel, each electron is required to have $E_T>$35~\GeV and be reconstructed with $|\eta|<1.442$ (ECAL barrel region) or $1.56<|\eta|<2.5$ (ECAL endcap region).  At least one electron must be found in the barrel region.  Also studied is a case of reduced acceptance due to the degradation of the ECAL endcaps at high luminosity, where both electrons are required to be in the barrel region.  In the muon channel, both muons are required to have $p_T > 45$~\GeV; one muon must be within $|\eta| < 2.1$ and the other within $|\eta| < 2.4$.  
The effects of lepton isolation are simulated by requiring $\Delta R>0.8$ between the leptons and jets in \ttbar~background events.

Signal and background events are simulated at generator level and smeared to simulate the detector response.  The electron identification efficiency is taken to be 88\% per electron, from the 8~\TeV analysis.  The \pt of electrons within the ECAL barrel (endcap) acceptance is smeared by 0.8\% (1.6\%).  Very high energy deposits in a single ECAL crystal (above $\sim$1.7~\TeV in the barrel and above $\sim$3.0~\TeV in the endcap) will result in saturation of electronics readout.  While negligible at 8~\TeV, this effect will occur more frequently at 14~\TeV and is hence included in these projections.  The probability for an electron to saturate is predicted using the fraction of electron energy deposited in the leading crystal in fully simulated 8~\TeV electron events.  Saturating electrons have their energy smeared by 7\%, which is the expected resolution from~\cite{Ball:2007zza}.  The muon identification efficiency is 85\%, taken from the 8~\TeV analysis.  The di-muon mass is smeared according to a parameterization of the 8~\TeV di-muon mass resolution, which is approximately 10\% at 3~\TeV.  Charge misassignment is not included.  The effect of pile-up is observed to be negligible for high energy electrons and muons.

The dominant background for both channels is the Drell-Yan production of lepton pairs.  The background due to \ttbar is found to self-veto above $\sim$1~\TeV, where the boost of the top causes the lepton to fail isolation criteria due to the proximity of a b-jet.  The background due to WW is expected to be the dominant non-DY background above 1~\TeV, but is found to be small (1-2\% of the DY).  The photon-induced irreducible dilepton background was estimated using the FEWZ3 program \cite{Li:2012wna} and found to be around 5\% of the Drell-Yan background in the \TeV range relevant for this study, having a negligible impact on the Z$^{\prime}$ limits.  The background due to jets in the electron channel is small and difficult to measure at 8~\TeV and is not included in the study.

In order to derive the discovery cross section sensitivity, an empirical fit to both background and signal acceptance is performed as a function of the dilepton mass.  For discovery, it is required that the number of signal events in a mass window gives a p-value, calculated using Poisson statistics, less than than $3\times 10^{-7}$, with a minimum of 5 events required.  The mass window is defined such that it contains 95\% of the signal peak after resolution effects.  This strategy leads to conservative estimates at high luminosity for Z$^{\prime}$ production at low mass due to large background levels, but preserves discovery sensitivity at high mass where background is minimal.

\begin{figure}
\begin{center}
\includegraphics[width=0.53\linewidth]{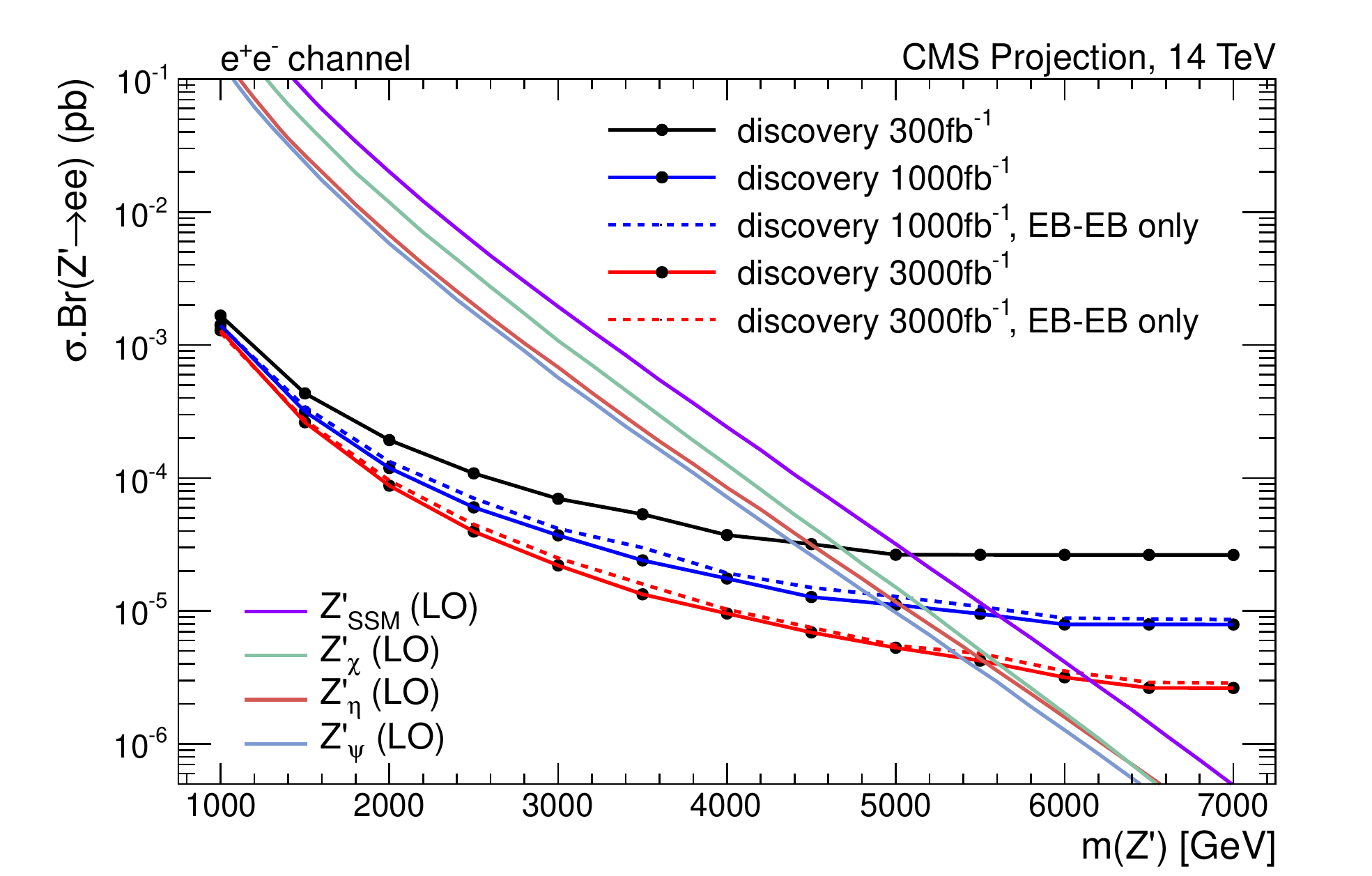}
\includegraphics[width=0.46\textwidth]{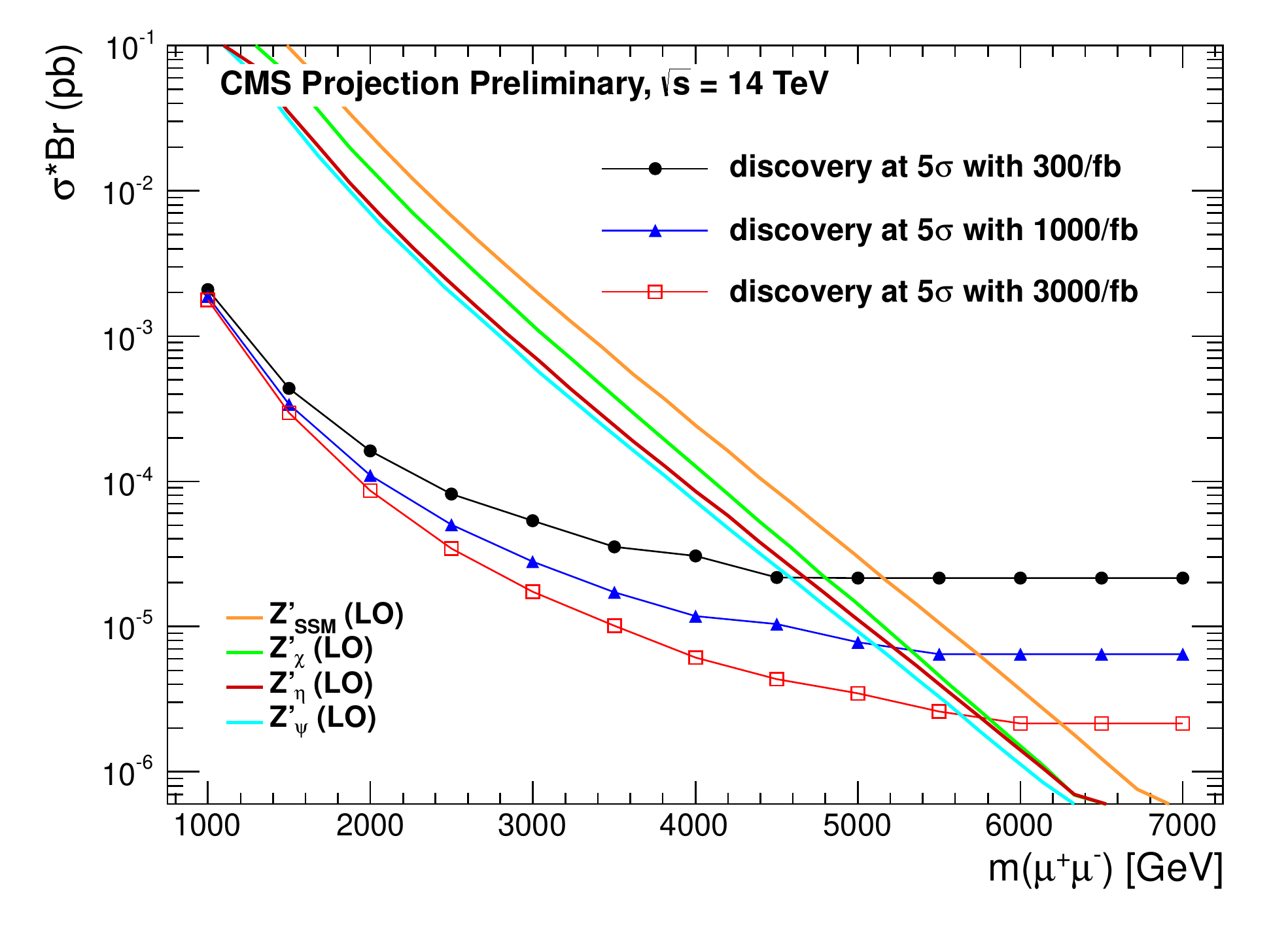}
\caption{The minimum cross section times branching ratio for discovery as function of dielectron (left) and dimuon (right) mass for various luminosity scenarios.  For the dielectron search, various luminosity and detector scenarios are considered, where the ``EB-EB only'' lines represent the reduced acceptance scenario in which electrons are reconstructed in the ECAL barrel only.}
\label{fig:ZprimelimitsE}
\end{center}
\end{figure}

The discovery reach in the electron and muon channels is shown in Fig.~\ref{fig:ZprimelimitsE}.  In both cases, the leading order cross section times branching ratio for various Z$^{\prime}$ models is also shown.  In the electron channel, a 5.1~\TeV Z$^{\prime}_{\mathrm SSM}$ in the sequential standard model (SSM) can be discovered with 300~fb$^{-1}$ of 14~\TeV data. A 5~\TeV Z$^{\prime}_{\eta}$ can be discovered with with 1000~fb$^{-1}$ of 14~\TeV data. In the muon channel, Z$^{\prime}_\psi$ with a mass of 5~\TeV can be discovered with approximately 900 fb$^{-1}$.  These results are in good agreement with estimates of discovery potential prior to LHC operations~\cite{Ball:2007zza}.

\subsection{Searches for Monoleptons+MET}

In searches for new physics involving a high \pt lepton ($\ell = e, \mu$) and missing energy, two different models are considered for extrapolation to HL-LHC: the SSM \Wprime~\cite{reference-model} and a dark matter effective theory~\cite{monoLepton,DMModelSummery2}.
In the SSM, the \Wprime boson is considered to be a heavy analog of the SM $\PW$ boson and thus can decay into a lepton and a neutrino, the latter giving rise to missing transverse energy as the observable detector signature.  The branching fraction is expected to be 8\% for each leptonic channel.  In the dark matter model, a pair of dark matter particles ($\chi$) are produced in association with a lepton and a neutrino deriving from an intermediate standard model $\PW$.  Depending on the couplings (vector or axial-vector type), a scenario with constructive ($\xi=-1$) or destructive ($\xi=+1$) interference would be possible.  Both signatures result in an excess of events in the transverse mass (\MT) spectrum.  

The estimate of discovery reach is based on the 8~\TeV search performed by CMS~\cite{EXO-12-060}.  The signal acceptance at 14~\TeV is assumed to be the same as at 8~\TeV, which for \Wprime masses ranging from 0.5~\TeV to 2.5~\TeV was found to be around 70\% with a variation of $\pm5\%$ in both channels, including 90\% geometrical acceptance.
The primary source of background is the off-peak, high transverse mass tail of the Standard Model $\PW \to \ell\nu$ decays. Other backgrounds are negligible at high \MT, which is the dominant region to set the upper limits on the model parameters.  The background predictions are based on simulations up to very high transverse masses.  Both signal and background are generated using \MADGRAPH 4.5.1.

The signal parameter in case of a discovery is determined using the profile likelihood method by generating toy experiments. To assume a discovery, the median likelihood is required to be less than $5\sigma$. The electron and muon channel are treated separately and their likelihoods are combined. 

\begin{figure}[hbtp]
\begin{center}
 \includegraphics[width=.5\textwidth]{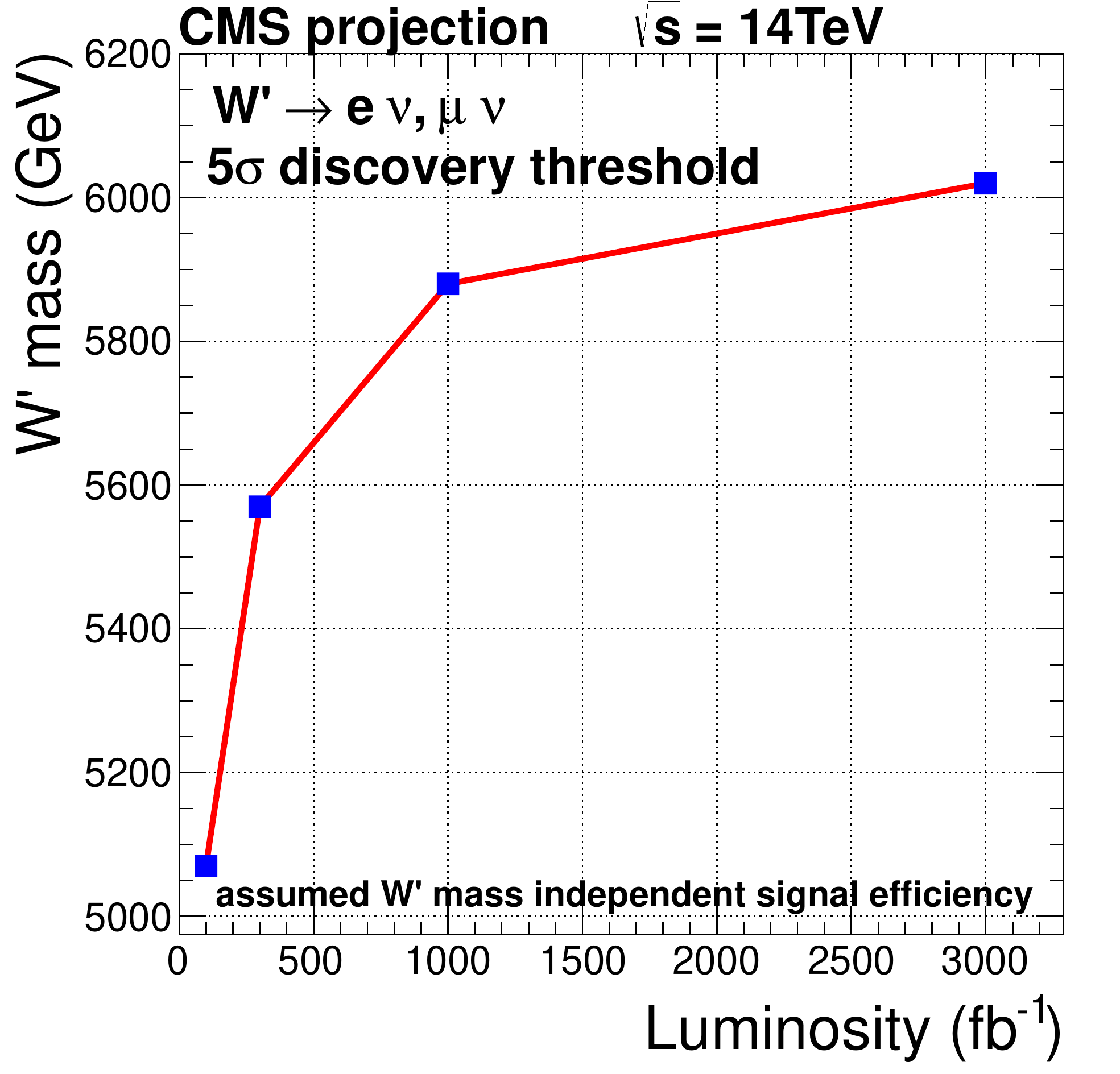}
\caption{Projection of the 5$\sigma$ discovery reach for $\sqrt{s} = 14$~\TeV for the 
         sequential standard model \Wprime.}
  \label{fig:wprimediscovery}
\end{center}
\end{figure}

The resulting discovery sensitivity on the \Wprime mass as a function of integrated luminosity is shown in Fig.~\ref{fig:wprimediscovery} for the combination of electron and muon channels. 
Given 3000\fbinv of data, it is possible to discover a \Wprime with a mass up to 6~\TeV. For high masses the sensitivity is affected by the center-of-mass energy due to the growing fraction of \Wprime bosons produced off-shell.  The extrapolation assumes that lepton reconstruction and, in particular, isolation efficiency are not affected by increased pile-up, based on the observation of flat efficiency in events from data with up to 50 vertices. The \Wprime cross sections are NNLO with a mass dependent k-factor.  

\begin{figure}[hbtp]
\begin{center}
 \includegraphics[width=.49\textwidth]{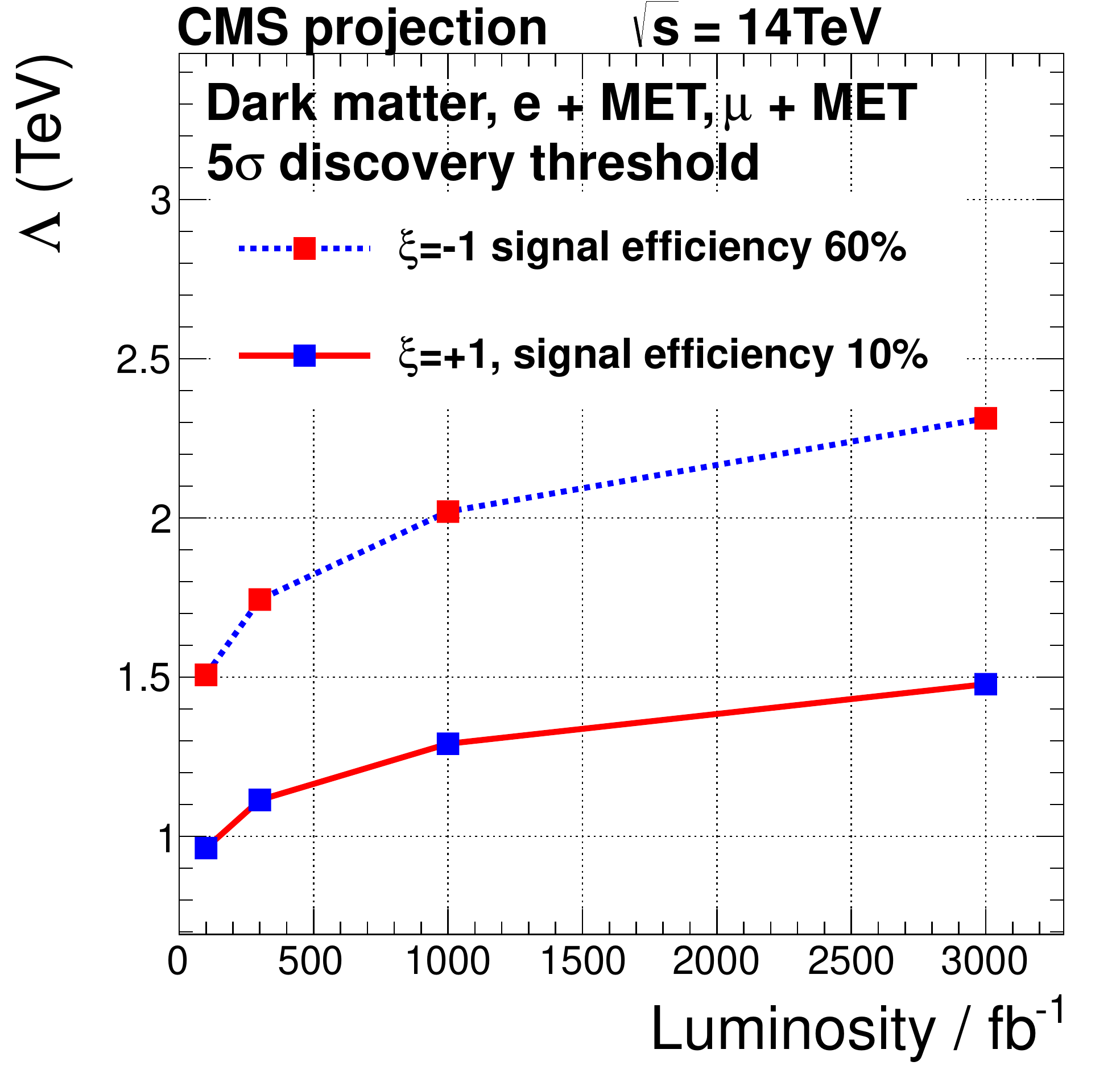}
 \includegraphics[width=.49\textwidth]{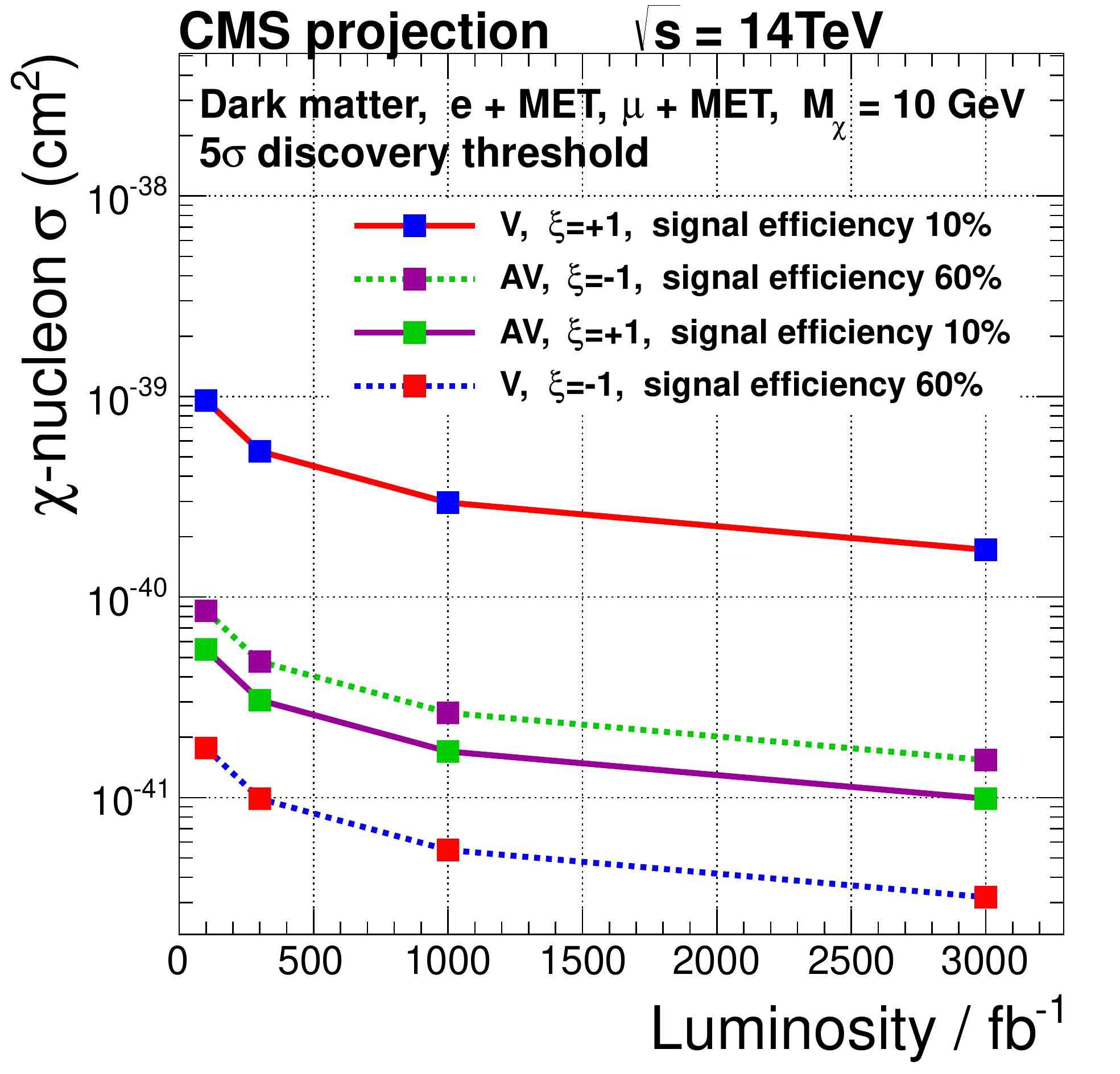}
\caption{Projection of the discovery reach on $\Lambda$ (left) and the dark matter-nucleon cross section (right) for the pair-produced dark matter model at $\sqrt{s} = 14$~\TeV and a variety of luminosity scenarios. The discovery threshold is $5\sigma$.}
  \label{fig:ppdmdiscovery}
\end{center}
\end{figure}

The same event selection optimized for SSM \Wprime can be used to search for pair-produced dark matter particles. Detailed studies at 8~\TeV using this method have shown the signal efficiency to be 60\% (10\%) in the case of constructive (destructive) interference~\cite{EXO-13-004}.  Applying this same procedure to the 14~\TeV lepton + \MET final state, the discovery reach relative to $\Lambda$, the scale of the effective interaction for associated dark matter pair production, is shown in Fig.~\ref{fig:ppdmdiscovery}. Signals with $\Lambda<1.4$~\TeV could be discovered in the case of destructive interference ($\xi = +1$) with an integrated luminosity of 3000~\fbinv.  For $\xi=-1$, values up to $\Lambda=2.3$~\TeV lie within the sensitivity of the experiment.  The discovery reach on the parameter $\Lambda$ can be translated to a nucleon cross section as shown in Fig.~\ref{fig:ppdmdiscovery}~(right)~for $M_\chi=10\GeV$ considering a vector or axial-vector coupling.

\subsection{Searches for Heavy Stable Charged Particles}

CMS has conducted searches for heavy stable charged particles (HSCP) produced in pp collisions at $\sqrt{s} = 7$ and 8\TeV, with integrated luminosities of 5.0\fbinv and 18.8\fbinv respectively, the results of which are presented in~\cite{Chatrchyan:2013oca}.  These searches present the most stringent limits to date on long-lived gluinos, scalar top quarks, and scalar $\tau$ leptons.  The signatures utilized include long time-of-flight to the outer muon system and anomalously large energy deposition in the inner tracker, and the existing results are presented for each separately and in combination.

The sensitivity of these searches in the HL-LHC era is projected by scaling the results of the 8~\TeV searches.  Unlike many conventional searches, where backgrounds arise from irreducible physical processes, the background to these searches comes primarily from instrumental effects.  It is therefore assumed that the backgrounds scale linearly with integrated luminosity, resulting in a constant signal over background ratio.  By scaling the signal yields linearly with integrated luminosity from the 8~\TeV result, a conservative assumption about the signal acceptance is introduced, since 14~\TeV kinematics are expected to yield increased acceptance.  Several changes are accounted for in LHC and detector operating conditions anticipated in the future.  First, since the LHC has operated at 50~ns bunch spacing to date, the 8~\TeV search was able to utilize a wide muon trigger time window, accepting candidates that arrive one LHC bunch-crossing after the collision.  The LHC is expected to run with 25~ns bunch-spacing from 2015 onwards, resulting in a reduced trigger time window, so the signal efficiency used in these projections has been adjusted, based on fully simulated 8~\TeV Monte-Carlo events.  Secondly, the current dE/dx measurement relies on analog readout of the CMS Tracker, which will almost certainly not be possible after the CMS Tracker is upgraded during LS3.  To account for this, the sensitivity with 3000~\fbinv is presented based on the combination of long time-of-flight and highly ionizing signatures, corresponding to an assumption that the dE/dx performance remains unchanged, and the sensitivity using the long time-of-flight signature alone, corresponding to an assumption that dE/dx measurements cannot be performed with the upgraded CMS Tracker.

\begin{figure}[ht]
 \begin{center}
  \includegraphics[clip=true, trim=0.0cm 0cm 2.3cm 0cm, width=0.48\linewidth]{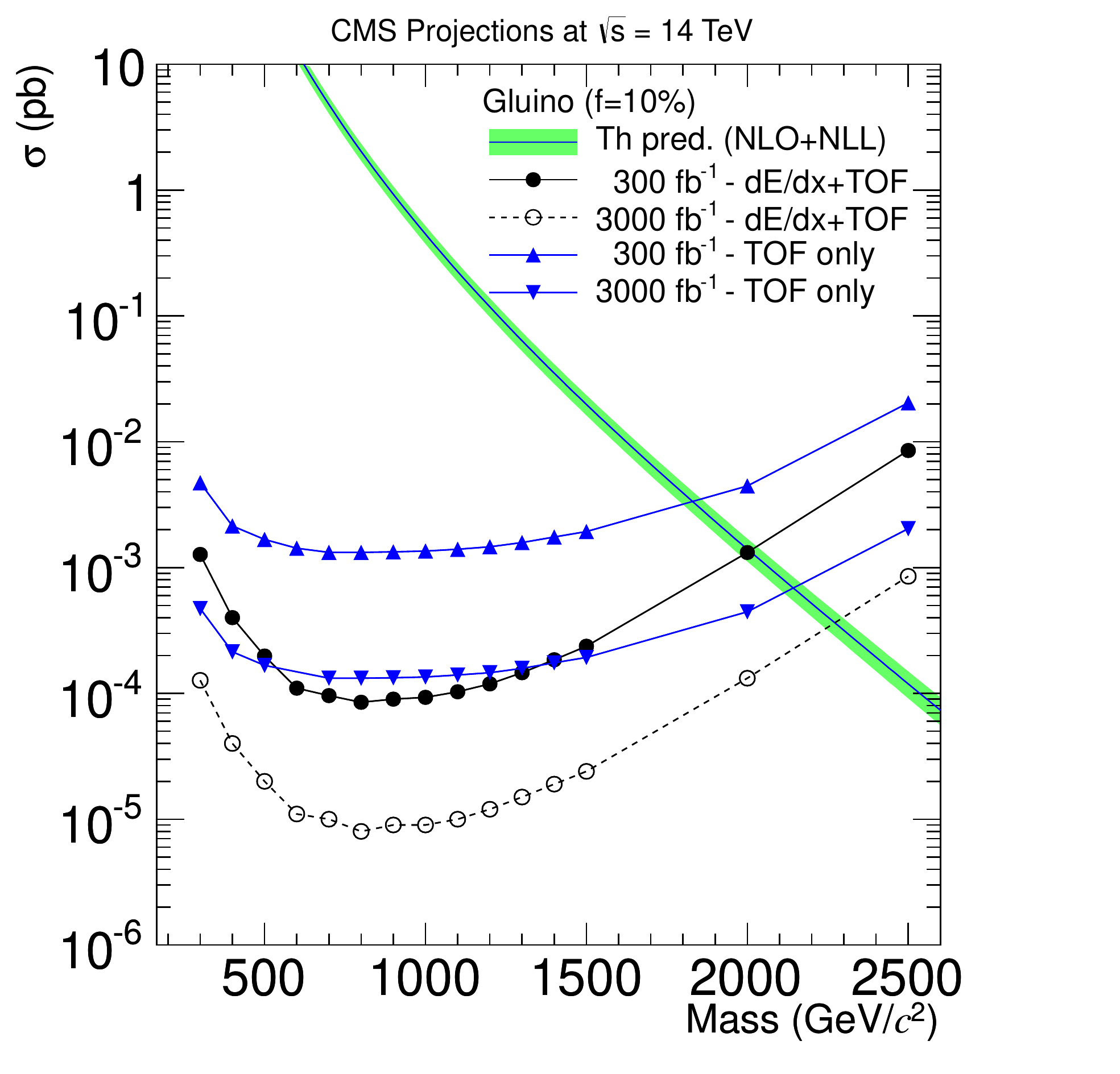}
  \includegraphics[clip=true, trim=0.0cm 0cm 2.3cm 0cm, width=0.48\linewidth]{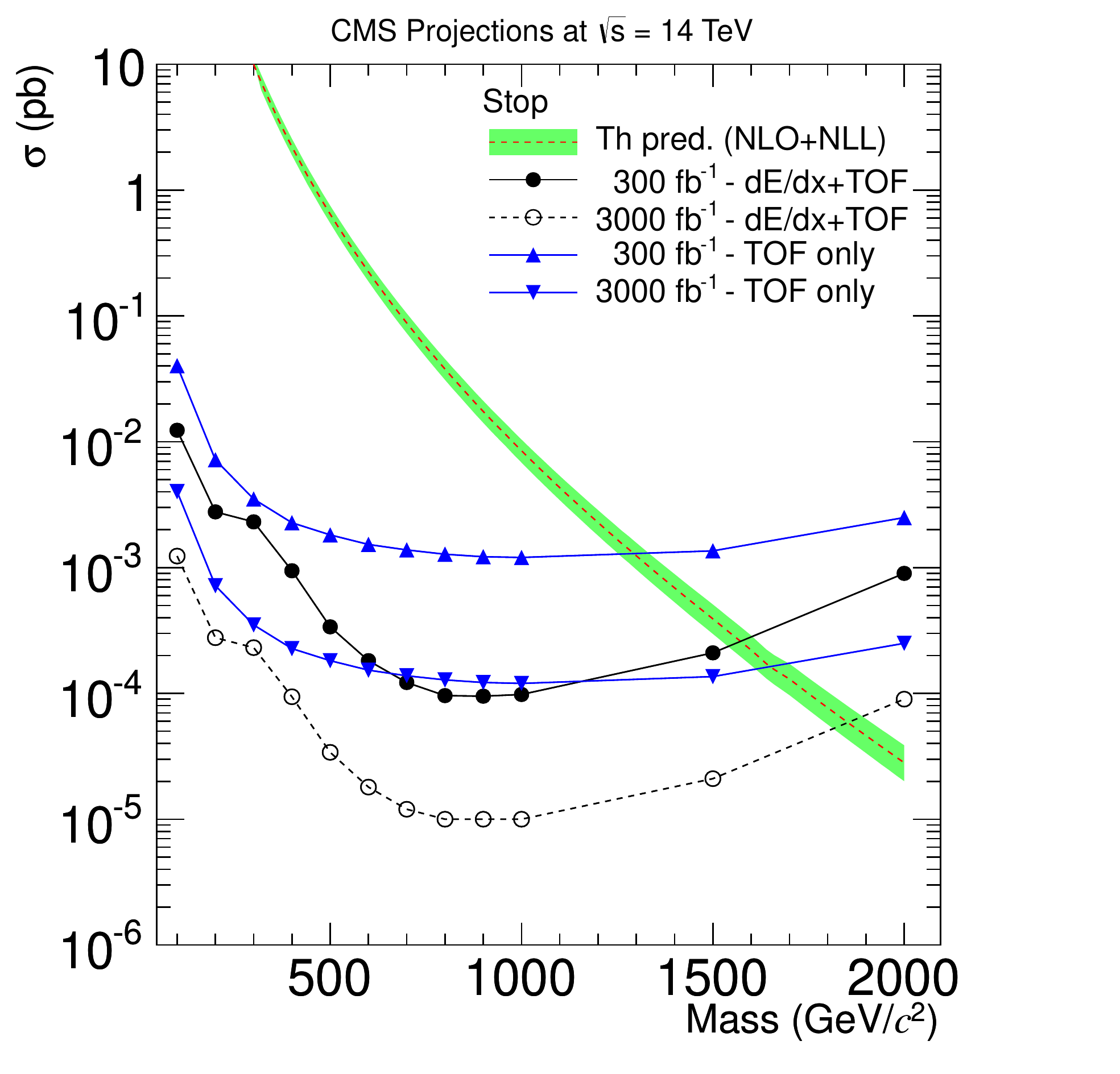}
 \end{center}
 \caption{ Minimum cross sections for an expected signal significance of 5 standard deviations.
   The signal models considered are the pair production of gluinos (left) and of stops (right).
   \label{fig:HSCPstopgluinoreach}}
\end{figure}

\begin{figure}[ht]
 \begin{center}
  \includegraphics[clip=true, trim=0.0cm 0cm 2.3cm 0cm, width=0.48\linewidth]{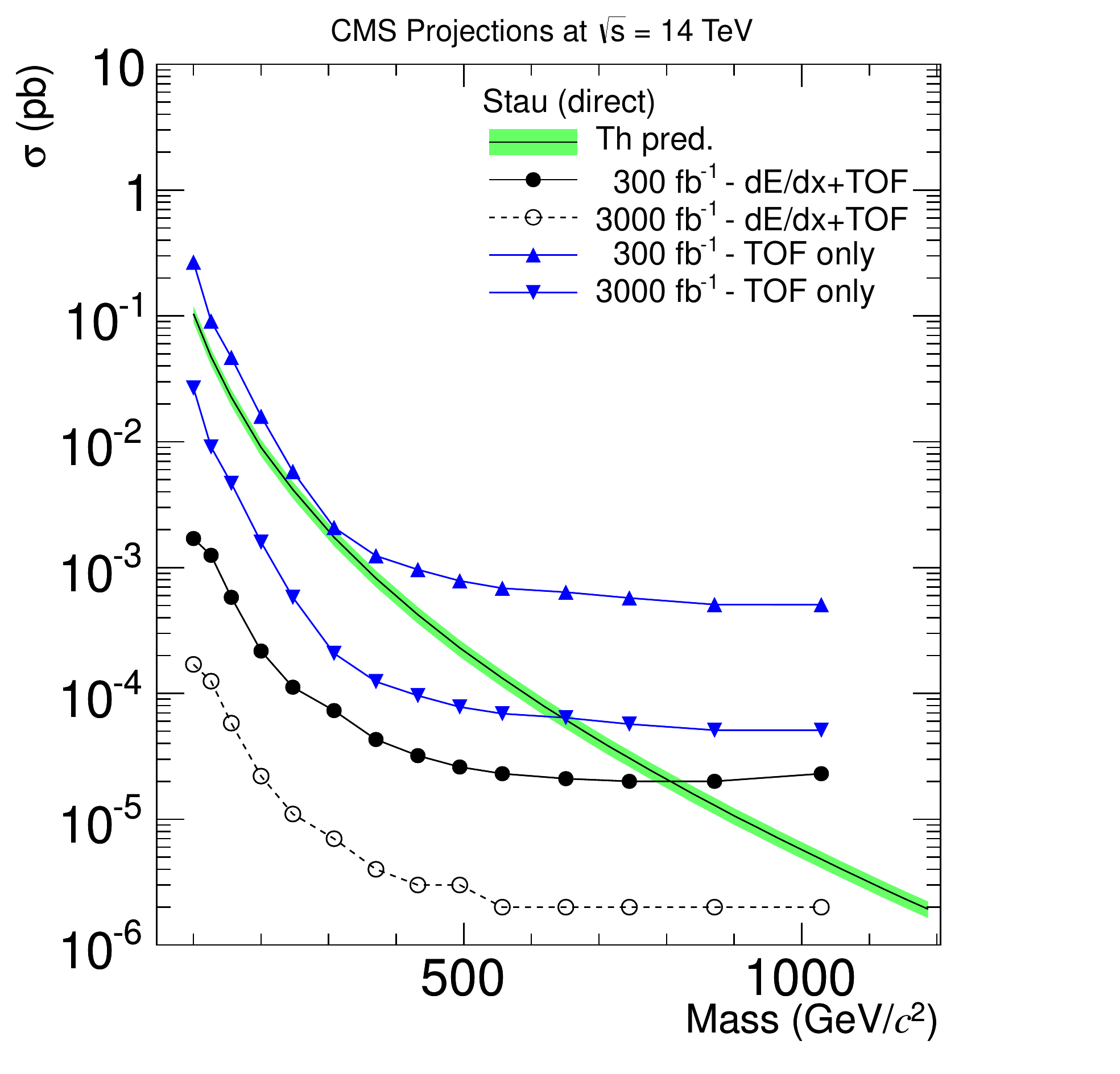}
  \includegraphics[clip=true, trim=0.0cm 0cm 2.3cm 0cm, width=0.48\linewidth]{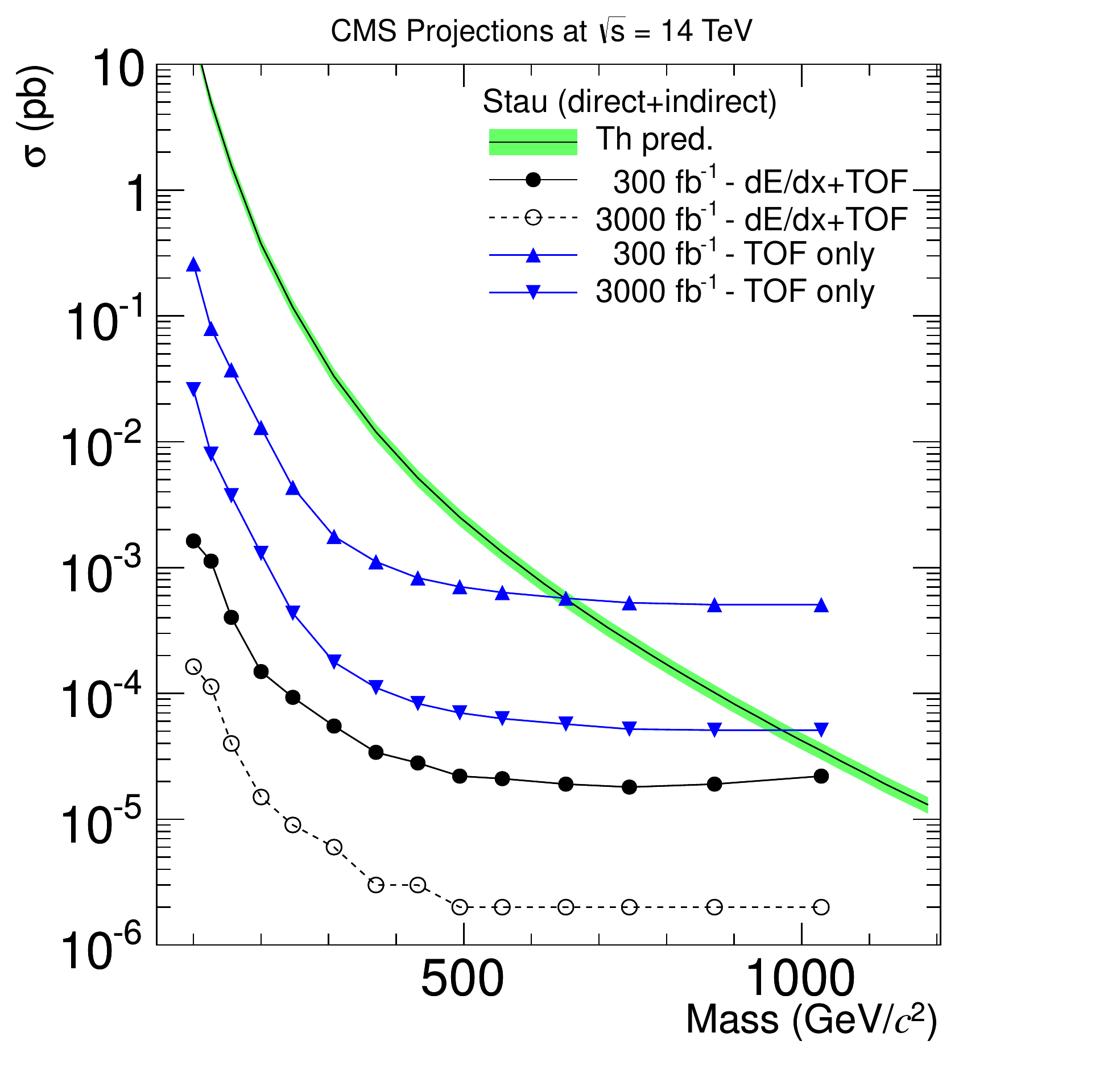}
 \end{center}
 \caption{ Minimum cross sections for an expected signal significance of 5 standard deviations.
   The signal models considered are the direct pair production of staus (left) and direct+indirect production of staus (right) in the context of GMSB.
   \label{fig:HSCPstaureach}}
\end{figure}

These assumptions allow us to rescale the results of ~\cite{Chatrchyan:2013oca} to both higher center of mass energy and integrated luminosity with little difficulty.  The results of this exercise are presented in terms of cross section reach defined as the cross section for which an observed signal is expected with a significance of at least 5 standard deviations (5$\sigma$). 
Figures \ref{fig:HSCPstopgluinoreach} and \ref{fig:HSCPstaureach} show the expected reach as a function of HSCP mass for hadron-like HSCP (stops and gluinos) and for lepton-like staus (direct and inclusive production), respectively.

The results show that the additional integrated luminosity will allow us to be sensitive to long-lived particles produced with a cross sections at least one order of magnitude lower than what has been excluded by~\cite{Chatrchyan:2013oca}.  It should be noted that the models considered in this search are simple benchmarks and the search for long-lived particles even in the already excluded mass range must be continued.  This is because the exclusion results rely entirely on theoretical cross section predictions made in the context of a given model (e.g., Split SUSY, GMSB), while the analysis itself is signature-based and mostly decoupled from any given theoretical model.  For example, it is known from past studies~\cite{CMS-PAS-EXO-08-003} that the sensitivity to lepton-like HSCPs in Universal Extra Dimension (UED) models is significantly less due to their lower production cross sections.  The cross section limits should therefore be pushed as low as possible regardless of the excluded mass range as interpreted in the context of a few popular benchmark models.

\subsection{Search for Heavy Vector-like Charge 2/3 Quarks}

Vector-like quarks differ from SM quarks in their electroweak couplings.  Whereas SM quarks have a V-A coupling to the W boson, i.e. their left and right-handed states couple differently to the W boson, vector-like quarks have only vector coupling to the W boson. One can thus write a mass term for them that does not violate gauge invariance without the need for a Yukawa coupling to the Higgs boson. Vector-like quarks are predicted, for example, by Little Higgs models~\cite{ArkaniHamed:2001nc, Schmaltz:2005ky}. They can cancel the diverging contributions of top quark loops to the Higgs boson mass offering an alternative solution to the hierarchy problem.

We search for a vector-like T quark with charge +2/3, which is pair produced together with its antiquark in proton-proton collisions through the strong interaction. Thus its production cross section can be calculated using perturbative QCD. 
The T quark can decay into three different final states: bW, tZ, or tH. If it is an electroweak singlet the branching fractions are predicted to be 50\% into bW and 25\% each into tZ and tH~\cite{delAguila:1989rq}. At low masses the tZ and tH modes are kinematically suppressed. All T quark decays produce final states with b quarks and W bosons.  Signal events therefore have large numbers of jets from b-quarks and hadronic W, Z, or H boson decays. For large T quark masses it becomes likely that the jets from one or more boson decay are not resolved which gives rise to jets that have substructure and a large invariant mass. 

This sensitivity is based on an estimate on the T quark search carried out by CMS based on 19.6~\fbinv of pp collision data collected at a center of mass energy of 8~\TeV~\cite{CMS-PAS-B2G-12-015}.  This search considers eight channels which differ in their selection criteria.  All require at least one electron or muon, together with a number of jets, which may be identified as originating from a b-quark or a boosted W or Z.
The same selections are used for the projection to HL-LHC, with one simplification.  For the single lepton channels, the 8~\TeV results are based on the full spectrum of a boosted decision tree discriminant.  Here, the same BDT discriminant is used, but simply accept events above a threshold, that was optimized for the expected significance.

In order to compute the expected sensitivity for pp-collisions at $\sqrt{s}=14$~TeV, the same selection efficiencies are used for signal and background as for the CMS analysis for $\sqrt{s}=8$~TeV. The signal yields are scaled using the calculated cross section from HATHOR. The backgrounds are scaled by the ratio of their NLO cross section. For the $W$+jets background this is 1.8, for the $\ttbar$ background it is 3.9, for the $\ttbar$W background it is 3.3, and for the $\ttbar$Z background it is 4.8. All other backgrounds are scaled by a factor 3. All yields are scaled proportionally to integrated luminosity assumed. 																							

A simplified treatment of systematic uncertainties is applied. A flat 30\% uncertainty is applied to the background yields in the multilepton channels. In the single lepton channels a 10\% uncertainty is assigned to the $\ttbar$ background and 50\% uncertainty to all other backgrounds. When estimating the sensitivity the background estimate is increased by the size of the uncertainties. 

To determine the sensitivity, 10000 random pseudoexperiments are generated based on the expected number of signal and background events in each channel. For each channel $i$ in each pseudoexperiment the p-value $p_i$ is determined for the observed number of events under the background only hypothesis. The combined p-value of all channels is then computed,
$P=k\sum_{i=0}^{n-1}(-\ln k^i/i!)$, where  
$k=\prod_{i=1}^n p_i$ 
and $n$ is the number of channels to be combined.

\begin{figure}[hbtp]
\begin{center}
\includegraphics[width=0.49\textwidth]{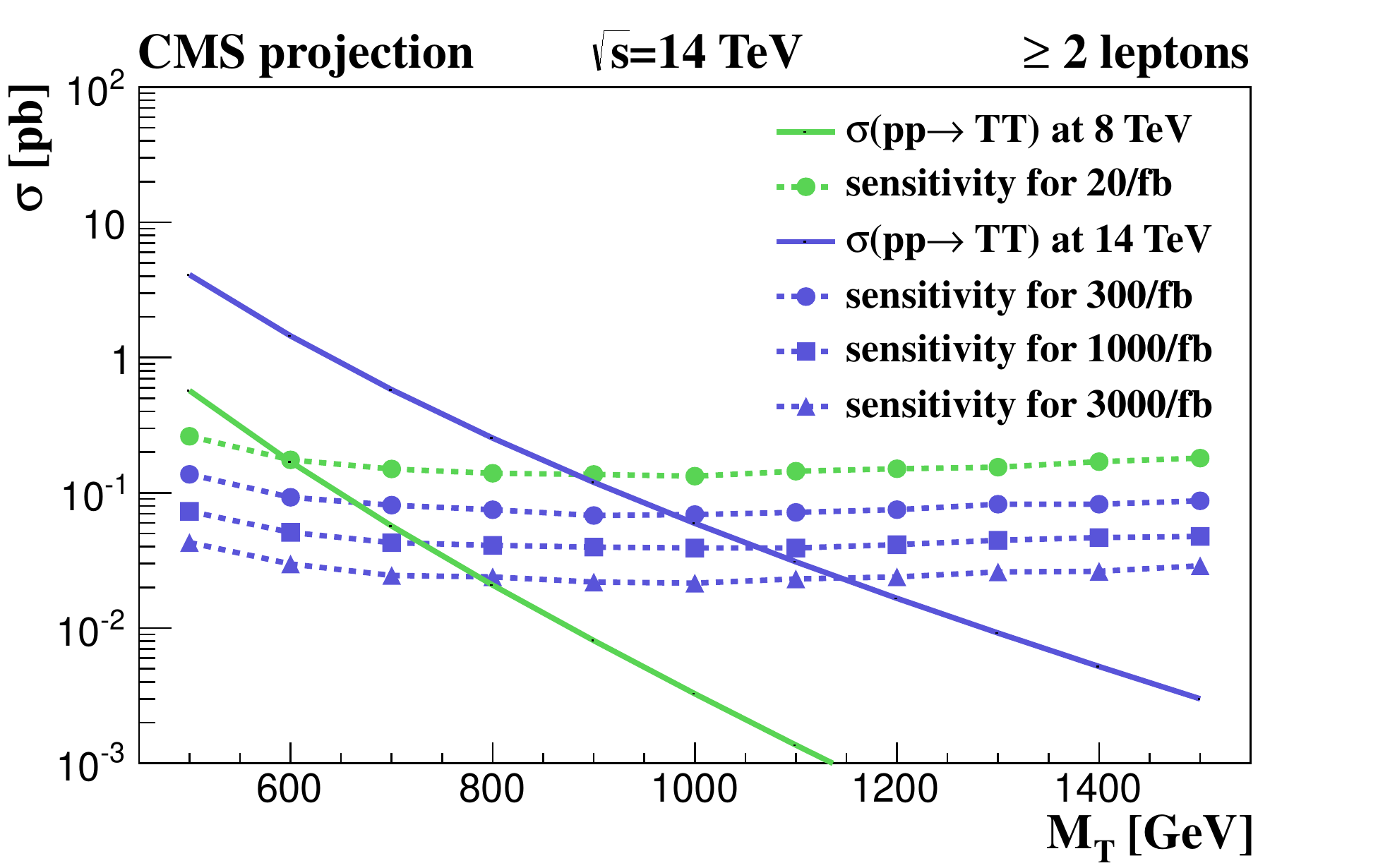}
\includegraphics[width=0.49\textwidth]{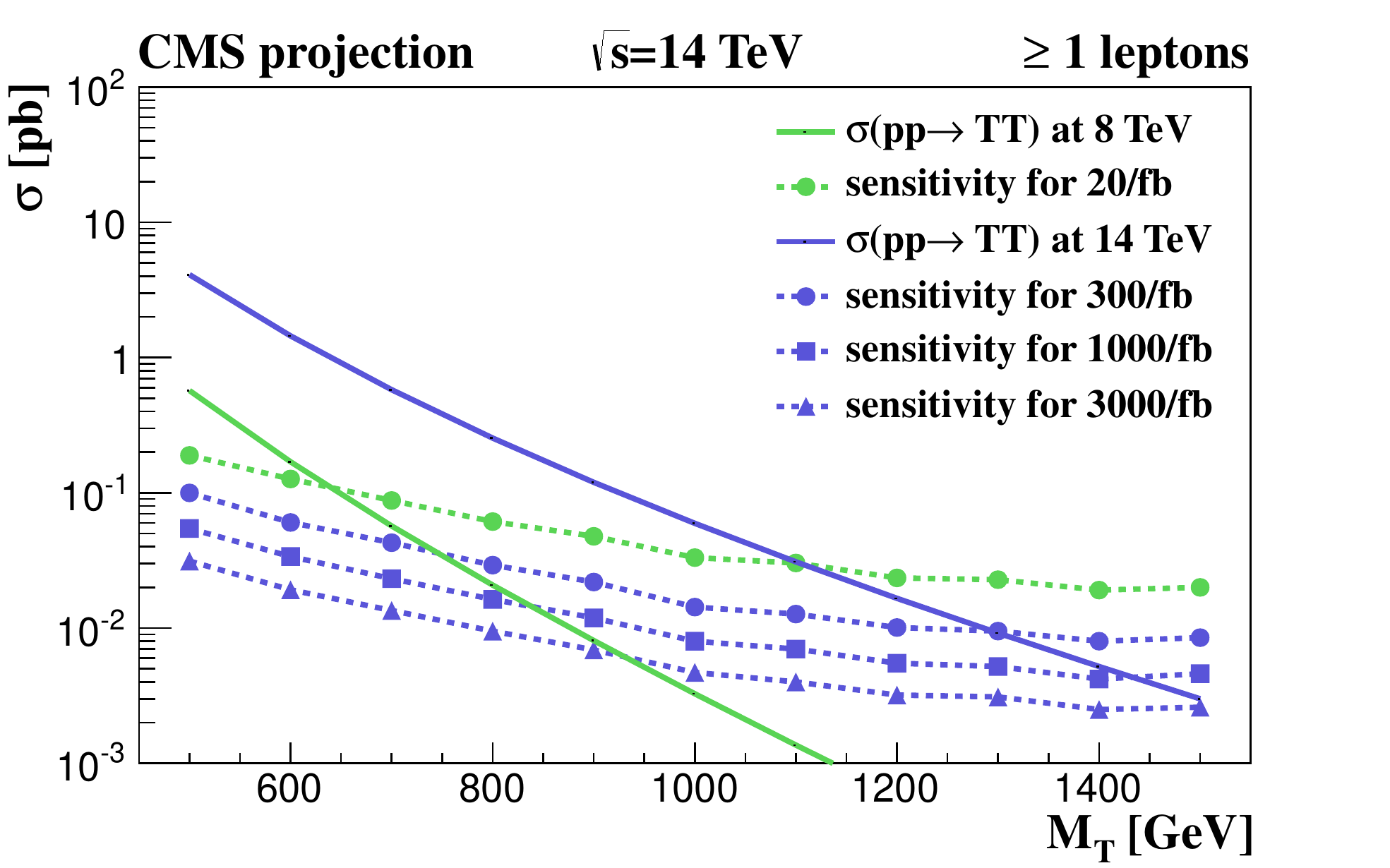}
\caption{Expected sensitivity for a T quark pair production signal in the multilepton channels only
(left), and in all channels combined (right).}
\label{fig:sensll}
\end{center}
\end{figure}

Figure~\ref{fig:sensll} shows plots of the discoverable T quark pair production cross section as a function of T quark mass, for the nominal branching fractions of 50\%/25\%/25\% to bW/tZ/tH. The sensitivity is not expected to vary substantially if the branching fractions deviate from the nominal values. For the CMS analysis at $\sqrt{s}=8\TeV$ the mass limits vary inside a $100\GeV$ interval for any combination of branching fractions.  At $14\TeV$, T quarks with mass below $1\TeV$ could be discovered with $300\fbinv$ of data, with the discovery reach extending to nearly $1.2\TeV$ with $3000\fbinv$ accumulated.

\section{Top Quark Physics}\label{top}

The copious production of top quarks at the LHC, together with the
excellent performance and understanding of the CMS detector, allows for
studies of top quark properties and the mechanism of their production and decay
with unprecedented detail and precision. Already with the 7 TeV LHC dataset, CMS has caught
up with the Tevatron and surpassed both D0 and CDF in precision for key measurements
such as that of the top quark mass where CMS has now achieved the most precise
single results in each of the $t \bar{t}$ decay modes. Similarly, the \ttbar and
single-top production cross sections have been determined with a relative
precision that is challenging the theoretical predictions in differential cross section measurements.

\begin{figure}[htb]
 \begin{center}
  \includegraphics[width=9 cm]{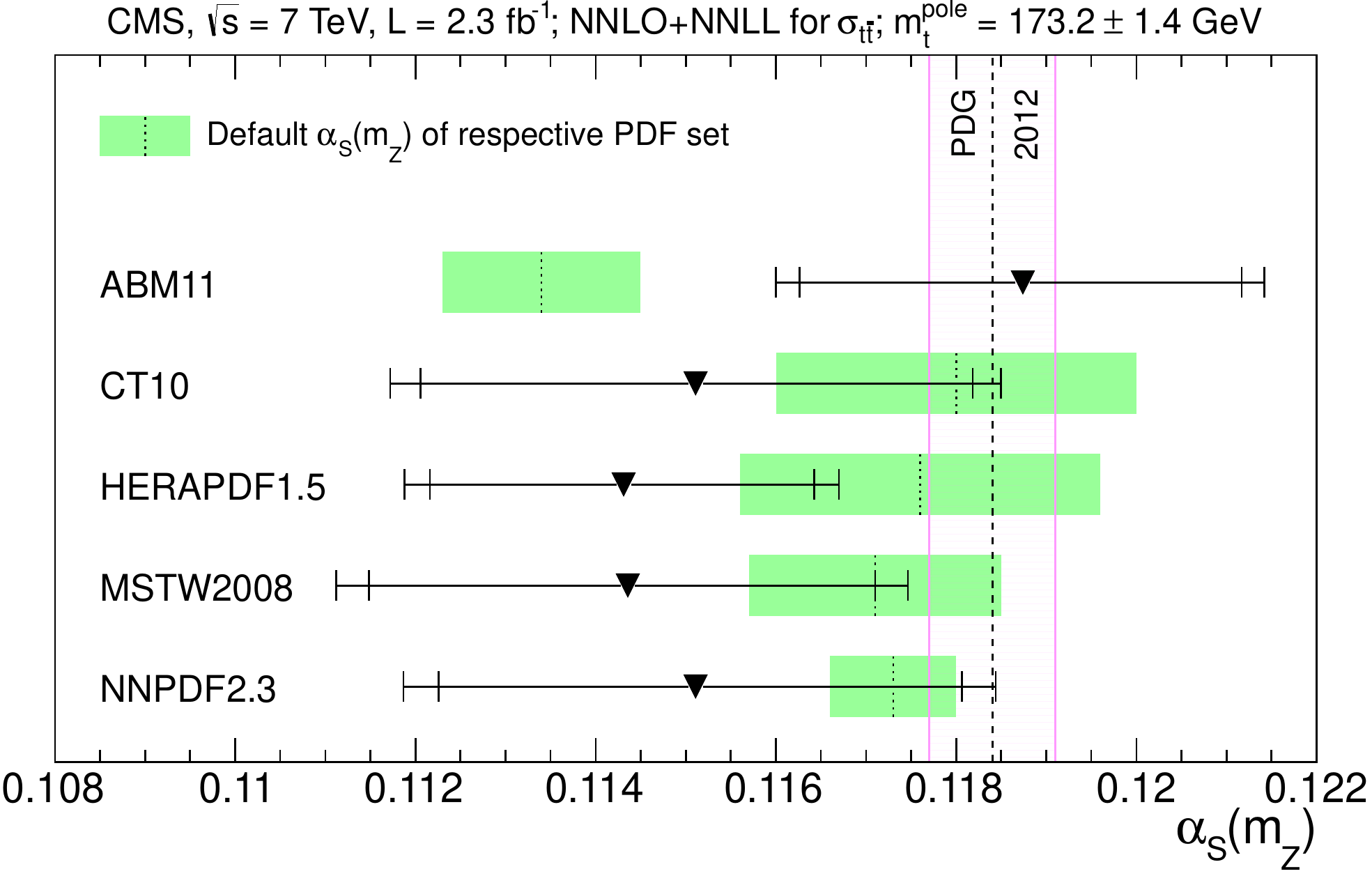}
  \includegraphics[width=6 cm]{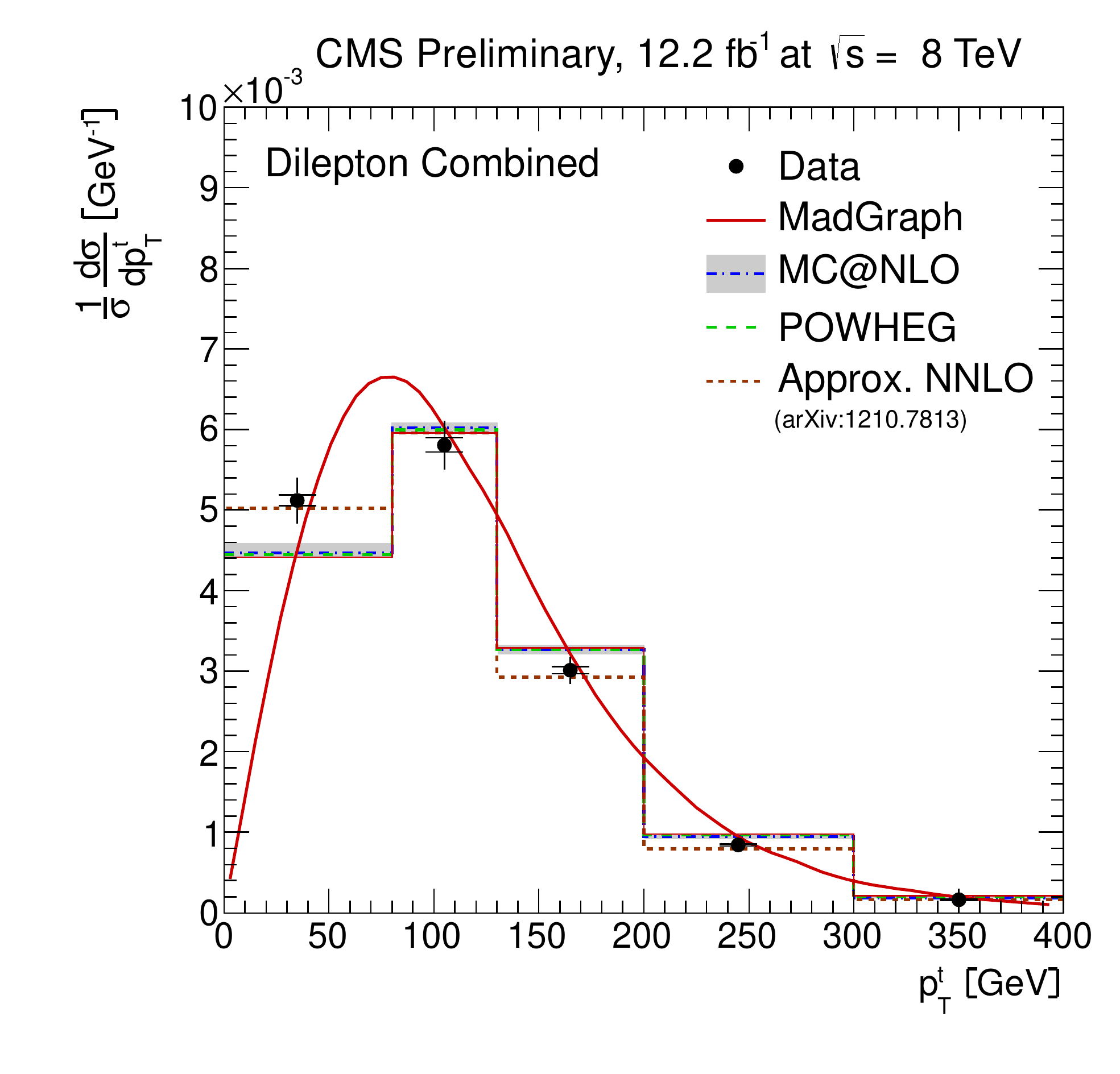}
 \end{center}
 \caption{\label{fig:alphasttbar}
    Left: Results obtained for $\alpha_S (m_{\rm Z})$ from
    the measured \ttbar cross section together with the prediction at NNLO+NNLL using different NNLO PDF sets.
    The inner error bars include the uncertainties on the measured cross section and on
    the LHC beam energy as well as the PDF and scale uncertainties on the predicted cross section.
    The outer error bars additionally account for the uncertainty on $m_{\rm top}$.
    For comparison, the latest $\alpha_S (m_{\rm Z})$ world average with its uncertainty is shown as a hatched band.
    For each PDF set, the default $\alpha_S (m_{\rm Z})$ value and its uncertainty are indicated using a dotted line and a
    shaded band (from~\cite{Mitov}). Right: Normalised differential \ttbar production cross section as
    a function of the $\pt^{\text{t}}$ of the top quarks. The inner (outer) error bars indicate the
    statistical (combined statistical and systematic) uncertainty. The measurements are compared to
    predictions from \MADGRAPH+\PYTHIA, POWHEG+\PYTHIA, MC@NLO+HERWIG, and to an approximate NNLO calculation.
    The \MADGRAPH+\PYTHIA prediction is shown both as a curve and as a binned histogram
    (from~\cite{CMS-PAS-TOP-12-028}).}
\end{figure}

With the addition of $8\TeV$ data, CMS top physics measurements will break new
ground across the entire program. Many goals here range from measuring cross
sections and rare production processes, to a further improvement in the mass
determination, the exploration of its spin and decay properties, the $V_{tb}$
coupling, the refinement of the strategies to use the top as a search tool, etc.
In the process of covering this, there are many bread and butter measurements
that will be obtained. One example is the determination of $\alpha_s$ using the
full NNLO+NNLL calculation of the \ttbar cross section~\cite{Mitov}, for different
PDF sets, as shown in Fig.~\ref{fig:alphasttbar}~(left). Preliminary results at
$8\TeV$ confirm earlier observations at 7 TeV~\cite{CMS-PAPERS-TOP-11-013} revealing
clear differences between data and NLO MC predictions in key differential kinematic
distributions such as the transverse momentum distribution of top quarks in
\ttbar events, shown in Fig.~\ref{fig:alphasttbar}~(right). Full NNLO differential
calculations, that should become available in the next years, may resolve some of
the current discrepancies. In the meantime, detailed study of the pre-LS1 dataset
continues to constrain (gluon) PDFs, and improve understanding of the modeling of
additional jets from ISR and FSR in ttbar events. Improved theoretical tools,
including NLO parton-shower matched simulation, are being commissioned and are
expected to play an important role. These new tools and studies are crucial to
lay a solid foundation for post-LS1 analysis of $14\TeV$ data.

With $300\fbinv$ at $14\TeV$ CMS will record some 50 million \ttbar events in the
lepton+jets channel, and about 10 million events in the dilepton channel.  In
addition, some 15 million events of single-top production can be collected in
the various channels with leptonic triggers.

With these large samples we can study very rare top decays like those induced
by flavor-changing neutral currents (FCNC). These decays occur in the SM only in
quantum-loop corrections with tiny branching fractions (10$^{-14}$) and their
observation would be a clear signal of new physics. There are models of new
physics that predict branching fractions as high as
$10^{-4}$~\cite{AguilarSaavedra:2000db,Lu:2003yr}. As an example, an extrapolation
from the current search for t $\rightarrow$ Zq decays~\cite{CMS-PAPERS-TOP-11-028}
yields sensitivities around $10^{-5}$, in the interesting range for
SUSY~\cite{Yang:1998yr}. Sensitivities in the range $10^{-5}$--$10^{-4}$ (and higher
with HL-LHC) are expected for other FCNC decays $q\gamma$ and
$qg$~\cite{GianottiManganoVirdee}. High luminosity is also necessary for precise
studies of the $tWb$ vertex which can be performed in terms of an effective
Lagrangian~\cite{Zhang:2011yr} requiring a common analysis of \ttbar and single-top
production.

The study of the associated production of top with $\gamma$, W  and Z  will give
access to the top coupling to bosons. First evidence of ttZ and ttW production has
been recently observed~\cite{CMS-PAPERS-TOP-12-014}, opening the road to precise
measurements in this area. A more difficult analysis is the study of the ttH
associated production where a very good control of the background will be needed
to extract the coupling with precision.
In addition, it is key to be able
to keep systematic uncertainties under control, which would for example require
excellent understanding of the b-tagging performance with the upgraded CMS detector.
As shown in Sec.~\ref{higgs}, with the HL-LHC a measurement of the top-Yukawa coupling
with a precision better than $10\%$ is expected.

New physics coupled to the third generation can affect \ttbar production and give
rise to bumps and distortions in the \ttbar invariant mass spectrum. High statistics
enhances the sensitivity to distortions to the expected shape, and will extend the
range to higher mass resonances. The study of asymmetries is particularly promising
for high-luminosity measurements, since many systematic effects cancel in the
ratios. New physics in \ttbar production can be unvealed by studying charge
asymmetries and spin correlations~\cite{Krohn,theoretical,SaavedraJuste}.

The very large statistics sample will open also the possibility to exploit
new methods to extract more accurate values of the top mass with exclusive
decays like t$\rightarrow$b$\rightarrow\JPsi$, or exploiting the shape of the
lepton spectrum.  Another interesting technique which requires high statistics is
the measurement of the top mass using the B hadron decay
length~\cite{CMS-PAS-TOP-12-030}. A new development pioneered by CMS is the
measurement of the top mass as a function of kinematic properties of the ttbar
events~\cite{CMS-PAS-TOP-12-029}, and the measurement of the difference between
the top quark and anti-quarks~\cite{CMS-PAPERS-TOP-11-019,CMS-PAS-TOP-12-031}.
In addition to a test of CPT invariance the latter measurement is a test of soft
non-perturbative QCD effects that may affect top quark and anti-quark decays
differently. With further study and sufficiently large statistics these methods
can shed light on the dependency of the top mass on color reconnection and other
QCD effects. An improved understanding of the interpretation of the top mass
measurements is particularly important in view of a very high precision
measurement of the W mass, and since the Higgs boson mass is now known.

\section{Electroweak Physics}\label{ewk}

The goals of the CMS electroweak program are at least threefold:

\begin{itemize}
\item To test the standard model theory of electroweak symmetry
  breaking at the TeV scale via a comprehensive portfolio of
  (multi)boson production measurements;
\item To improve upon traditional precision electroweak observables;
\item To produce precision PDF constraints and tests of perturbative
  QCD/electroweak predictions.
\end{itemize}

\subsection{Multiboson production}

The observed Higgs boson is a very attractive explanation for mass and
electroweak symmetry breaking, however what is not yet known to any
precision is whether it plays the desired role of completely restoring
unitarity to the gauge-boson interaction sector of the SM, or whether
other physics is also participating.  The classic test is to measure
$W_L-W_L$ scattering via vector-boson fusion (VBF) production of WW
pairs. In addition, VV scattering is potentially linked also, in the
context of electroweak-gauge-invariant effective field theory, with
other triple and quartic gauge interactions (TGC and QGC), and hence
as many sensitive multiboson final states as possible should be
studied.  Examples of this rich sector include:
\begin{itemize}
\item Differential cross sections of inclusive diboson production, in all 
combinations (especially at the highest \pt or mass);
\item VBF diboson production (a recent search for exclusive
  photoproduction of WW~\cite{Chatrchyan:2013foa} is already an example of this which probes the
  WW$\gamma\gamma$ QGC);
\item VBF single boson production (already examined for Z production at 7 TeV~\cite{Chatrchyan:2013jya}) ;
\item triboson production (not yet observed at the LHC).
\end{itemize}

These measurements typically involve processes with small cross
sections, and the most valuable information is contained in the
high-mass, high-momentum tails of their distributions.  Their study
will therefore be relevant throughout the $14\TeV$ era and at the
highest integrated luminosities.  They will benefit from detector
upgrades in similar fashion to the Higgs program, which examines most
all of the same final states.  Of particular value will be VBF dijet
triggering and reconstruction, crucial to isolating VV scattering
events.  Another important upgrade impact will be preserving the
capability, through precision tracking and higher-granularity
calorimetry, of the analysis of jet substructure.  At the TeV scale,
hadronic decays of bosons are exhibited in the detector as dijets
merged into a single jet.  Analysis at the particle level of these
jets (``W/Z-tagging'') is critical to identify VV (or VVV) candidates.

\subsection{Precision constraints of the electroweak theory}

In the domain of precision electroweak data, the statistical power is
clear: CMS has already recorded W samples far in excess of the
Tevatron, and Z dilepton samples in excess of LEP 1. If systematic
uncertainties can be appropriately constrained, then progress can be
made in the understanding of the W mass (through kinematic
distributions of leptonic W decays) and the effective weak mixing
angle (through angular distributions of dileptonic Z decays).

In the case of the W mass, the appropriate statistical power is known
to be present in the Run 1 data alone.  However the systematic
limitations from PDFs (and possibly the detector) may only find
solution in higher statistics samples and an upgraded detector in Run
2.  For the weak-mixing angle, it is known from the first successful
CMS measurement~\cite{Chatrchyan:2011ya} that the limitations are from statistics, PDFs, and
tracker alignment.  With the Run 2 data the statistical uncertainty
will be comparable to the world-average uncertainty; it then remains
to improve tracker alignment and PDFs, again exploiting higher Run 2
statistics, and the improved capabilities of tracking with pixel
upgrades.

\subsection{Constraints of PDFs and validation of perturbative predictions}

Finally, in the domain of precision tests of PDFs and perturbative
predictions, there are several samples of electroweak boson production
relevant to validating our understanding of standard model predictions
generally:

\begin{itemize}
\item Precision understanding of W or Z production at the LHC requires
  improved determination of gluon, strange, and charm quark PDFs, as
  they play a much stronger role at LHC energies than at the Tevatron.
  In addition, further progress must be made with valence quarks, and
  over a broader range in Bj\"{o}rken $x$.  Similarly to the Tevatron,
  however, W and Z samples can be boot-strapped to measure these PDFs
  in situ. For example, the recent CMS W$+$c cross section measurement~\cite{CMS-SMP-12-002}
  provides a strange PDF constraint which is already starting to be
  competitive with fixed-target neutrino data, and the CMS W lepton
  charge asymmetry measurements~\cite{Chatrchyan:2012xt, CMS-SMP-12-021}already over-constrain the u/d PDF
  ratio compared to pre-LHC PDFs.

\item A new generation of next-to-leading-order QCD, parton-shower
  matched (NLO+PS) simulations are now being commissioned (Sherpa 2.0,
  aMC@NLO, POWHEG), which give true NLO+PS-matched predictions for
  $2\rightarrow 3,4,5,6$-body processes; they will be crucial to
  understand 6-fermion final states like VV scattering. In the coming
  years, next-to-next-to-leading order (NNLO) QCD predictions for
  these processes are also expected to mature, demanding another round
  of confrontation with experiment.

\item At the TeV scale, radiative and loop corrections due to W, Z, or
  photon emission leads to next-to-leading order electroweak
  corrections comparable to NLO QCD.  These effects must be isolated
  (typically in the highest mass Drell-Yan or diboson production
  differential cross sections), measured, and compared with NLO
  electroweak (possibly in concert with NLO or NNLO QCD corrections)
  predictions where available.
\end{itemize}

This is a broad and ongoing program of cross section measurement which
evolves with beam energy and luminosity, as the relevant higher-order
effects are typically strongly energy-dependent.  To understand the
heavy-flavor PDFs, a dedicated triggering strategy must be adopted,
with detector upgrade requirements similar to the $H\rightarrow bb$
analysis.

\section{Heavy Ion Physics}\label{hin}

The primary goal of relativistic heavy-ion physics is to study the phases of 
nuclear matter and the transitions between them. Of particular interest is the 
characterization of the extremely high energy-density phase, the quark-gluon 
plasma (QGP), in which the quarks and gluons comprising the atomic nuclei are 
deconfined and chiral symmetry is restored. Owing to its high center-of-mass
 energy and high luminosity, the LHC is in a unique position to explore the 
properties of the QGP using rare processes, such as the production of heavy 
quarkonium states, jets of different flavors, electroweak bosons, and the
 correlations between these rare probes of the medium produced in the nuclear
 collisions. CMS is ideally suited to carry out this program, due to its large 
acceptance, high-rate triggering and data acquisition capabilities.
The high luminosity regime of the LHC gives the opportunity to explore the QGP phase with rare probes and the Heavy Ion community 
is proposing to collect 10~nb$^{-1}$ of  lead-lead (PbPb) collisions. This number is used in this document to evaluate  
the physics prospects of the CMS heavy-ion program.

\subsection{Proposed strategy}

By the end of LS2, CMS will have 
completed its first upgrade phase including a new pixel vertex detector with 
larger redundancy, an upgraded trigger, an extension of the forward muon 
system and a refurbishment of the hadron calorimeter electronics 
including the replacement of the PMTs of the forward hadron calorimeter. 
The selectivity of high-\pt jets at L1 will be augmented by implementing 
a subtraction for the underlying event and the DAQ upgrade will allow to increase 
the bandwidth of the tracker detector.
The combination of these upgrades will raise the maximum PbPb interaction rate 
that can be sampled by CMS detector from about 5~kHz, as observed in Run 1 to 
above 50~kHz. This will allow the CMS heavy-ion program to fully exploit the 
high luminosity heavy-ion running, after the LHC collimator upgrades are 
completed in LS2. 

During LS3, CMS will upgrade the inner tracker and will make many other 
changes to sub-detectors, trigger and data acquisition systems. These 
improvements will further augment the capabilities of the CMS experiment 
to study heavy-ion collisions by increasing the track reconstruction 
efficiencies and acceptance among other improvements.

\subsection{Plans for heavy-ion physics in the HL-LHC era }

In case of a successful dispersion collimator installation in the LHC, and 
stochastic cooling tests in the PS/SPS in the LS2, the LHC is expected to reach its 
currently maximum foreseen instantaneous PbPb luminosity 
of $\approx 6 \times 10^{-27}$~cm$^{-2}$~s$^{-1}$, providing a 50~kHz collision 
rate at the nominal PbPb center-of-mass energy of 5.5~TeV. With these running 
conditions the heavy-ion community is proposing to collect a total of 10nb$^{-1}$, 
which is approximately 60 times more than what is presently available, in terms of number of events sampled, with the additional benefit of a higher center-of-mass energy, which in turn leads to higher production cross sections of hard probes.

This increase in collision energy and luminosity will again necessitate recording
reference pp and pPb data, with a center-of-mass energy and statistics 
corresponding to the ones of the PbPb dataset.

One of the most important questions is to precisely quantify the parton energy 
loss in the hot and strongly interacting medium produced in PbPb collisions.
Of particular interest are the parton flavor dependence and the path length 
dependence of this phenomenon, which was for the first time directly observed
as an imbalance of the energies of back-to-back jets~\cite{Chatrchyan:2011sx,Chatrchyan:2012nia}. Such an imbalance is also observed in isolated-photon+jet 
pairs~\cite{Chatrchyan:2012gt}. Expanding these initial observations to 
precision measurements, especially measurements of differential parton flavor or 
in the azimuthal angle relative to the reaction plane, to study the path length dependence
 will require significantly higher event rates.

Table \ref{pbpb_statistics} illustrates the expected number of 
events containing the various physics signatures of interest for high-precision parton 
energy loss studies in a 10~nb$^{-1}$ PbPb collision sample. The estimated numbers are 
based on the observed rates in the Run 1 dataset. 

\begin{table}[h!]
\begin{center} 
\caption{Expected hard probe event rates in PbPb collisions at the LHC (2015-2017) and HL-LHC (2019-2025) heavy-ion running. The estimate is derived from the data collected in Run 1 of the LHC.} 
\label{pbpb_statistics} 
\begin{tabular}{|c|c|c|c|c|}
\hline
                                                             & $2010$--$2013$                      & $2015$--$2017$                   & $2019$--$2025$ \\
                                                             & 2.76 TeV 160${\rm \mu b}^{-1}$ & 5.5 TeV 0.3-1.5${ \rm nb}^{-1}$ & 5.5 TeV$\sim 10{ \rm nb}^{-1}$  \\
\hline
Jet $p_T$ reach (GeV/$c$)                                    & $\sim 300$                     & $\sim 500$                  & $ \sim 1000$ \\
Track $p_T$ reach (GeV/$c$)                                  & $\sim 100$                     & $\sim 160$                  & $ \sim 300$ \\
\hline
Dijet ($p_{T,1} > 120 {\rm GeV}/c$)                          & $50k$                          & $\rm \sim 300k-1.5M$        & $\rm  \sim 10M$ \\
b-jet ($p_{T} > 120 {\rm GeV}/c$)                            & $\sim 500$                     & $\rm \sim 4k-21k$           & $\rm  \sim 140k$ \\
Isolated $\gamma$ ($p^{\gamma}_{T} > 60 {\rm GeV}/c$)    & $\rm \sim 1.5k$                & $\rm \sim 9k-45k$           & $\rm  \sim 300k$ \\
Isolated $\gamma$ ($p^{\gamma}_{T} > 120 {\rm GeV}/c$)   & $-$                            & $\rm \sim 300-1.5k$         & $\rm  \sim 10k$ \\
W ($p^{\rm W}_{T} > 50 {\rm GeV}/c$)                   & $\sim 100$                     & $\rm \sim 600-3k$           & $\rm  \sim 20k$ \\
Z ($p^{\rm W}_{T} > 50 {\rm GeV}/c$)                   & $\sim 10$                      & $\sim 60-300$               & $\rm  \sim 2k$ \\
$ {\rm t}\bar{\rm t}\rightarrow {\rm l^+l^-}{\rm b}\bar{\rm b}$ MET                     & $\rm \sim O(10)$               & $\sim 100-200$            & $ \sim 600$ \\
\hline
\end{tabular}
\end{center} 
\end{table}

A first set of studies of the quark flavor dependence of parton energy loss using jets 
arising from bottom quark fragmentation or three jet events will become feasible
when the LHC energy is raised to $14\TeV$, as illustrated by the corresponding studies 
presented below. 
These analyses were conducted to study the performance impact of the L1 calorimeter 
trigger upgrade on CMS heavy ion data taking.

The ideal measurement in the b-jet channel would be a measurement of the
dijet asymmetry for doubly tagged b jets, where we expect systematic 
uncertainties to be small, and to mostly cancel with respect to the corresponding 
light-quark jet measurements.  
The rate of doubly tagged b jets has been estimated based on the number
of inclusive dijets in the 2011 sample. The b-jet-to-inclusive-jet ratio was measured 
to be approximately 0.03 in pp collisions at 7 TeV, as well as in 2.76 TeV PbPb 
collisions, with significantly larger uncertainties in the latter case.  
Of these b jets only about 20\% will be produced back-to-back with another b jet, 
in the so-called flavor-creation mode.  Using a simple secondary-vertex
tagger, one can achieve about 50\% tagging efficiency in PbPb.  
For doubly tagged jets, then, one only obtains a tagging efficiency of 25\%,
but with a purity close to unity. Assuming $x_{\mathrm T}$ scaling with an exponent 
of $n = 4.5$, the yield of jets at fixed \pt\ increases by a factor of 5 for the 
increased collision energy expected in 2015.  

\begin{figure}[!ht]
\begin{center}
\includegraphics[width=.60\textwidth]{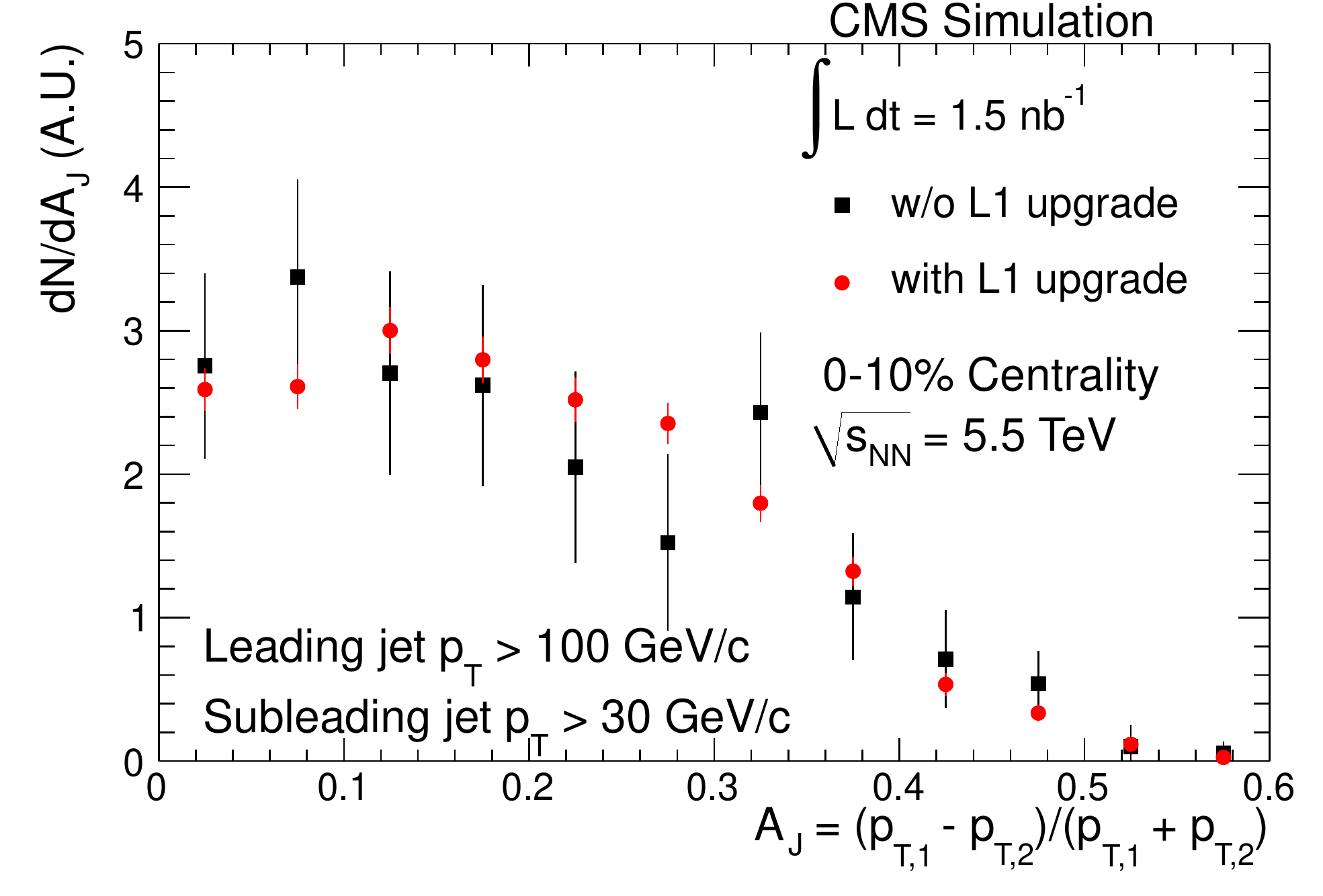}
\caption{$A_{J}$ distribution for doubly tagged b jets of $p_{T,1}$ $> $ 100
GeV/c, $p_{T,2} > $ 30 GeV/c and $|\eta| < 2$ 
in the ten percent most central collisions expected in the 2015 PbPb Run.}
\label{fig:aj_2015}
\end{center}
\end{figure}

The dijet asymmetry $A_{J}$ is defined as the
difference between the leading and subleading jet transverse momenta,
divided by their sum. The $A_{J}$ distribution for doubly tagged b jets is 
estimated from the inclusive jet $A_{J}$ distribution, scaling the uncertainties 
to those expected from $1.5\ {\rm nb^{-1}}$ of data at
5.5 TeV, with a tagging efficiency of 25\%.  This distribution is shown in
Fig.~\ref{fig:aj_2015} for the $10\%$ most-central PbPb events.  
The kinematic selection on the leading and sub-leading jets are \pt $>$ 100 GeV/c
requires $\pt> 30\GeV/c$, respectively, for jets in $|\eta| < 2$.  

\begin{figure}[!ht]
\begin{center}
\includegraphics[width=.60\textwidth]{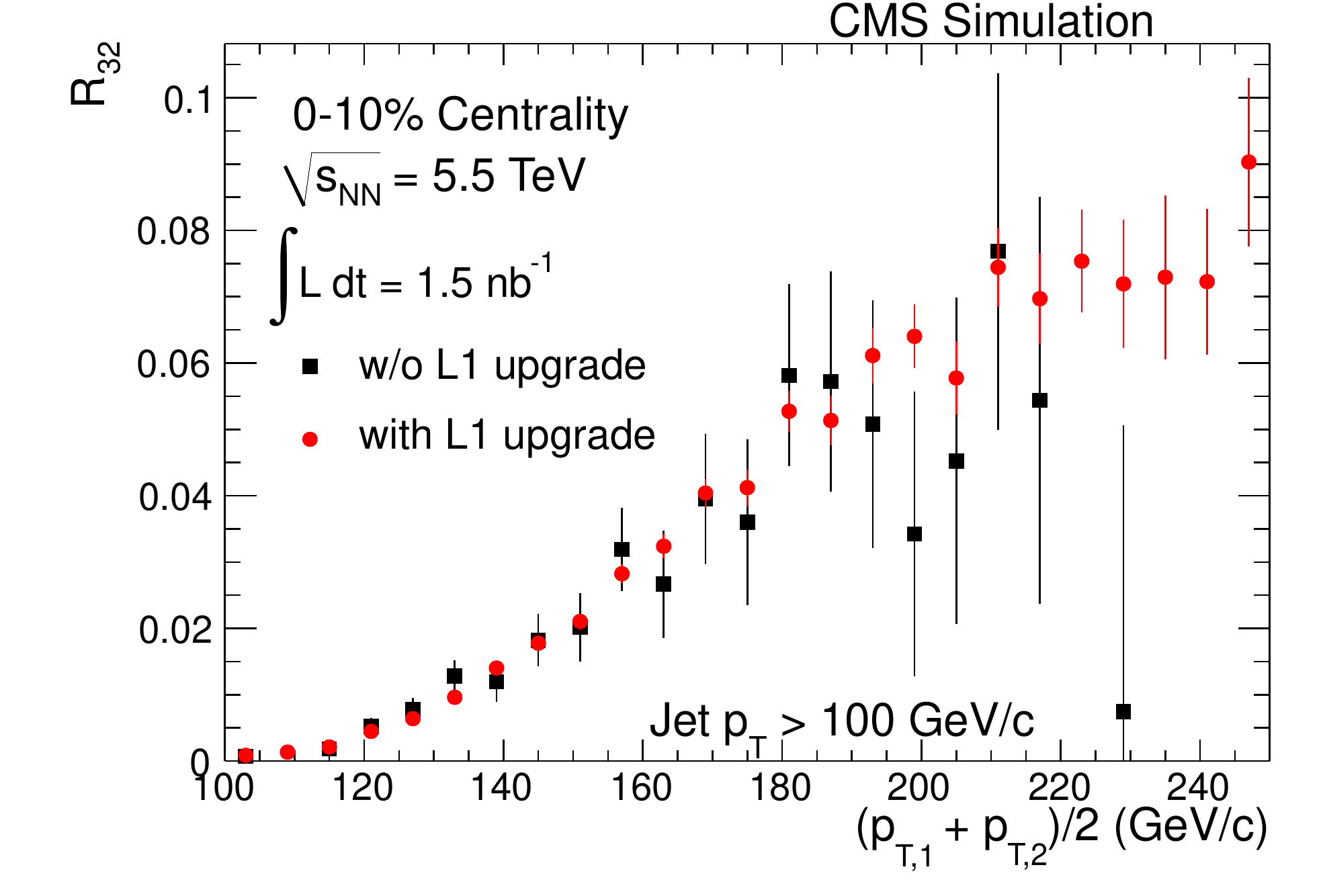}
\caption{The ratio of 3-jet to 2-jet events ($R_{32}$) as a function of the
average \pt\ of the two leading jets for \pt\ $> $ 100 GeV/c 
in the $10\%$ most-central collisions expected in the 2015 PbPb Run.}
\label{fig:r32_2015}
\end{center}
\end{figure}

Events with three or more jets in the final state originate from hard gluon 
radiation and other higher-order QCD processes. A measurement of the inclusive 
3-jet to 2-jet cross section ratio ($R_{32}$) is an interesting testing ground of 
pQCD, with possible modification of parton-shower and gluon jet quenching in QGP, 
because major systematic uncertainties such as jet energy scale, reconstruction 
efficiency and integrated luminosity largely cancel. The expected number of 3-jet 
events at $5.5\TeV$ is estimated based on the observed statistics in the 2011 data 
sample, in which we recorded about 106 3-jet events and 8225 dijet events, with 
all jets having $\pt > 100\GeV/c$.  
The ratio from PYTHIA events, with uncertainties scaled to the expected
2015 statistics, is shown in Fig.~\ref{fig:r32_2015}. 

While the measurements detailed above will become accessible already based om 
the 1.5~nb$-1$ expected to be recorded in the next run of the LHC, 
extracting the full path length dependence of parton energy loss in these channels 
by an analysis differential in the angle with respect to the reaction plane will 
require another increase of event rate by a factor of $5$--$10$, which will be achieved 
in the High Luminosity Ion running period of the LHC.

Other important channels to study parton energy loss, which should be possible at
the HL-LHC are the $\gamma$+jet and Z+jet energy balance. Comparing Z+jet 
and $\gamma$+jet observables to inclusive jets will allow to start separating 
quark jets from gluon jets. The collected number of $\gamma$+jet events will 
be sufficient to study the jet quenching as a function of the reaction plane. 

The suppression of quarkonium states, such as the Y family members, is another interesting 
signal of the QCD phase transition occurring in heavy-ion collisions. The dependence of 
this suppression on the collision centrality is especially interesting, where the 
measurement is severely limited by the number of peripheral collisions. A large 
statistics dataset would allow CMS to precisely map out the centrality dependence 
as well as conduct a more differential, reaction-plane dependent study.

The list of physics studies that can be performed with a 10~nb$^{-1}$ data sample 
will include:
\begin{itemize}
\item Detailed measurements of multijet correlations, shedding light on gluon versus quark jet quenching;
\item Differential studies of photon+jet correlations as a function of photon $p_{\rm T}$, event centrality or the reaction plane orientation;
\item Detailed measurements of $Z$+jet correlations, with up to 3000 and 600 $Z$ bosons having transverse momentum above 50 and 100~GeV/$c$, respectively;
\item Differential measurement, such as azimuthal correlations for the rarest probes ($\psi(2S)$);
\item Multiobject correlations;
\item Jet quenching studies up to the TeV scale;
\item Precision measurements of the quarkonium suppression pattern.
\end{itemize}

The physics channels listed above are probably the most interesting topics in heavy-ion
 physics at the LHC today and in the future, and the CMS apparatus is especially well 
suited to study those, thanks to the excellent and flexible trigger, to the extended and 
precise muon system, to the high-coverage calorimetry, to the photon isolation, vertexing, 
and precise jet energy measurement capabilities. The high luminosity era of the LHC 
will mark the beginning of high-precision studies of the parton energy loss mechanism 
and the dynamics of the medium created in heavy-ion collisions.

\section{Conclusions}\label{conclusion}

The discovery of a Higgs boson last summer invigorated the field of particle physics and offered new insights
as to where the next discoveries may occur. We believe that building on the success of this discovery by detailed
characterization of the newly observed particle, elucidating the nature of EWSB mechanism, and continuing searches 
for physics beyond the standard model should therefore be the highest priority for high-energy physics in the next 
decade. One of the most important components of this quest is the exploitation of the full potential of the upgraded 
LHC. The HL-LHC upgrade will contribute greatly to our understanding of Nature and will allow us to carry out the 
above ambitious goals, along with a host of electroweak precision measurements, which will extend our sensitivity 
to new physics models.

The characterization of the newly discovered 125 GeV boson by precision measurements of its mass and tree-level couplings to
fermions, W and Z bosons, as well as self-coupling, at the HL- LHC will allow us to prove that it is the SM Higgs boson or, if
it is not, to uncover the true nature of the observed particle. In addition, precision measurement of the couplings of the
Higgs to photons and gluons via quantum loops will provide sensitive probes for possible new physics beyond the SM.
Continuation of searches for SUSY with massive squarks and gluinos, as well as for the superpartners of the third-generation
quarks and electroweak bosons will either result in the finding of a ``natural" solution to the hierarchy problem of the Standard
Model or proving that this model is ultimately fine-tuned.

The discovery of the new boson was an immensely exciting and important event in humankind's quest for the fundamental laws of
physics. There is every reason to believe that more discoveries await us at the LHC. The rich physics program described above
provides overwhelming justification for the upgrades of the CMS detector needed to exploit this magnificent and unique
opportunity.

\bibliography{auto_generated}   

\end{document}